\documentclass[manuscript]{acmart}

\setcopyright{rightsretained}
\acmBooktitle{ACM Comput. Surv.}
\acmDOI{}
\acmISBN{}
\copyrightyear{2023}
\acmYear{2023}
\acmDOI{}

\usepackage{booktabs}
\usepackage{lettrine}
\usepackage{color}
\usepackage{longtable}

\usepackage{hyperref}

\newcommand\marios[1]{\textcolor{black}{#1}}
\newcommand\MM[1]{\textcolor{black}{#1}}

\usepackage{tikz}
\newcommand*\fullcircle{\tikz\filldraw[gray, thick] (0,0) circle (.5ex);}

\begin{document}

\title{A Systematic Literature Review of Human-Centered, Ethical, and Responsible AI}

\author{Mohammad Tahaei}
\orcid{0000-0001-9666-2663}
\affiliation{
  \institution{Nokia Bell Labs}
  \streetaddress{21 J.J. Thomson Avenue}
  \postcode{CB3 0FA}
  \city{Cambridge}
  \country{UK}
}
\email{mohammad.tahaei@nokia-bell-labs.com}

\author{Marios Constantinides}
\orcid{0000-0003-1454-0641}
\affiliation{
  \institution{Nokia Bell Labs}
  \city{Cambridge}
  \country{UK}
}
\email{marios.constantinides@nokia-bell-labs.com}

\author{Daniele Quercia}
\orcid{0000-0001-9461-5804}
\affiliation{
  \institution{Nokia Bell Labs}
  \city{Cambridge}
  \country{UK}
}
\email{daniele.quercia@nokia-bell-labs.com}

\author{Michael Muller}
\orcid{0000-0001-7860-163X}
\affiliation{
    \institution{IBM Research AI}
    \city{Cambridge}
    \state{MA}
    \country{USA}
}
\email{michael_muller@us.ibm.com}

\begin{abstract}
As Artificial Intelligence (AI) continues to advance rapidly, it becomes increasingly important to consider AI's ethical and societal implications. In this paper, we present a bottom-up mapping of the current state of research at the intersection of Human-Centered AI, Ethical, and Responsible AI (HCER-AI) by thematically reviewing and analyzing 164 research papers from leading conferences in ethical, social, and human factors of AI: \emph{AIES}, \emph{CHI}, \emph{CSCW}, and \emph{FAccT}. The ongoing research in HCER-AI places emphasis on governance, fairness, and explainability. These conferences, however, concentrate on specific themes rather than encompassing all aspects. While AIES has fewer papers on HCER-AI, it emphasizes governance and rarely publishes papers about privacy, security, and human flourishing. FAccT publishes more on governance and lacks papers on privacy, security, and human flourishing. CHI and CSCW, as more established conferences, have a broader research portfolio. We find that the current emphasis on governance and fairness in AI research may not adequately address the potential unforeseen and unknown implications of AI. Therefore, we recommend that future research should expand its scope and diversify resources to prepare for these potential consequences. This could involve exploring additional areas such as privacy, security, human flourishing, and explainability.
\end{abstract}



\maketitle

\section{Introduction}
\label{sec:intro}
Artificial Intelligence (AI) is a rapidly growing field with tremendous potential to transform our lives on an unprecedented scale through a new technological revolution~\cite{kissinger2021age}. It can advance the global economy and contribute to human flourishing~\cite{laker2022artificial, lee2022ai}. However, like every significant technological revolution, ethical and societal risks such as invasion of privacy or identity theft in social media~\cite{fowler2021there, press2021judge} are taking center stage. Risks associated with AI are no exception. If left unattended, AI can negatively impact specific populations, perpetuate historical injustices, or, even worse, amplify them~\cite{baeza_yates2018bias, bloomberg2023humans}. These risks include misclassifying people of color~\cite{buolamwini2018gendera, bloomberg2023humans}; exacerbating economic disadvantage by denying bank loans through digital redlining~\cite{hertzberg2010information, kizilaslan2017can}; reducing medical care for economically disadvantaged people based on their prior medical spending~\cite{arthur2008racial, jones2020covid, wilkinson2020which}; denying bail to Queer, Trans, Black, Indigenous, or People of Color (QTBIPOC)\cite{silva2018algorithms}; imposing longer prison sentences on QTBIPOC people\cite{lyn2020risky}; and removing children from QTBIPOC families into foster care~\cite{leckning2021patterns, saxena2020human}. In addition to the considerable attention given to biases within data sources, recent work examines human ``wrangling'' activities necessary to make so-called ``raw'' data fit-for-purpose. However, these changes to datasets and models during AI development introduce biases as a consequence~\cite{muller2022forgetting, pine2015politics, feinberg2017design, passi2017data, mentis2016crafting}. These examples remind us that people build AI systems and that social biases and prejudices may be included---intentionally or inadvertently~\cite{aragon2022human, muller2022forgetting}.

The scientific community has recently made attempts to address these pressing issues. Many research conferences, journals, and groups have begun to argue for responsible AI development, bringing to the table issues including but not limited to fairness, explainability, and privacy in AI~\cite{e2022responsible, t2022responsible, c2022pwc, labs2022responsible, arrieta2020explainable, co_operation2019recommendation, standards2023ai}, and centering AI around humans~\cite{aragon2022human, chancellor2023practices, ehsan2021expanding}. These topics are discussed in research conferences with a long history of advocating for human-centered design, such as the Conference on Human Factors in Computing Systems (CHI) and the Conference on Computer-Supported Cooperative Work and Social Computing (CSCW), and in newer conferences like the Conference on AI, Ethics, and Society (AIES), and the Conference on Fairness, Accountability, and Transparency (FAccT), which were established to cover topics related to the ethics of AI. In response to the large and growing interest in AI, there have been several survey and review papers mapping out the AI research landscape~\cite{capel2023what, wong2023seeing}, such as those surveying responsible development of AI in healthcare~\cite{siala2022shifting}, biases and fairness in AI~\cite{sun2019mitigating, pessach2022review, orphanou2022mitigating}, and explainable AI~\cite{abdul2018trends, bertrand2022how, arrieta2020explainable, angelov2021explainable}. However, a comprehensive overview at the intersection of \emph{human-centered AI}, \emph{ethical AI}, and \emph{responsible AI} is lacking. The intersection of Human-Centered, Ethical, and Responsible AI (HCER-AI) is significant because it presents a complex challenge of addressing multiple aspects simultaneously (see Section~\ref{sec:method} for how we define these terms). Each aspect---human-centeredness, ethical considerations, and responsible AI---cannot be achieved in isolation; they are interdependent. This interdependence underscores the crucial need to thoroughly understand and explore this intersection to create AI systems that are comprehensive and aligned with societal values. We aim to fill this gap by surveying the current state of research in HCER-AI in four leading research conferences on these topics. We formulated four Research Questions (RQs):

\begin{enumerate}
\item[\textbf{RQ\textsubscript{1}}:] What is the state of research in HCER-AI in the four research conferences that cover topics related to HCER-AI?
\item[\textbf{RQ\textsubscript{2}:}] What research methods are used in HCER-AI studies?
\item[\textbf{RQ\textsubscript{3}:}] What are the research gaps in HCER-AI?
\end{enumerate}
To further assess the current state of research in HCER-AI and compare it with potential future implementations, as indicated by patents, we conclude with the following question:

\begin{enumerate}
\item[\textbf{RQ\textsubscript{4}:}] What is the landscape of patent applications in HCER-AI?
\end{enumerate}

To answer our RQs, we reviewed and thematically analyzed \textbf{164} research papers from the proceedings of AIES, CHI, CSCW, and FAccT related to HCER-AI (Section~\ref{sec:method}). We found that the landscape of HCER-AI covers six primary themes: \emph{governance}, \emph{fairness}, \emph{explainability}, \emph{human flourishing}, \emph{privacy}, and \emph{security}, with a heavy emphasis on the first three themes (Section~\ref{sec:themes}). These results echo prior empirical studies with different groups of people, highlighting a gap between public and AI research concerns~\cite{jakesch2022how}. For instance, while there is a critical mass of research on fairness, average users may care more about their safety and privacy~\cite{jakesch2022how}. We also found that conferences have distinct areas of focus within HCER-AI. For example, AIES and FAccT publish more papers on governance and fairness, while CHI and CSCW cover a broader range of topics, including explainability and human flourishing. Our results suggest that future research should take a broader view of AI and diversify resources beyond governance and fairness to prepare for AI's unexpected and unknown ramifications.

As part of our \textit{posteriori} analysis, we used AI-assisted summarization to answer our RQs using the abstracts of the papers as input to ChatGPT 4.0, following the recent emerging literature on using AI for qualitative analysis~\cite{byun2023dispensing, abram2020methods}. We compared its findings with our manual analysis (Section~\ref{sec:results-chatgpt}). Although ChatGPT provided an overview of the dataset, it is important to highlight that its classification was not specifically tailored to the research areas currently under investigation. Instead, it provided a broad classification intended for general public understanding, rather than specialized research purposes. In the future, AI-powered research tools like ChatGPT could be a helpful supplementary resource for gaining new insights. Further examination is needed to determine how AI can be effectively utilized in literature reviews.

Materials concerning our classification and analysis will be available on our project's website, which can be found at \href{https://social-dynamics.net/RAI-Review}{social-dynamics.net/RAI-Review}.

\section{Related Work}
To situate our review within the broader human-centered, ethical, and responsible AI literature, we next discuss previous surveys that focused on human-centered AI, specific concerns with AI such as fairness and explainability, and surveys dedicated to specific domains of AI.

Developing AI involves socio-cultural and technical factors, with scholars arguing for developing human-centered AI in the past few years~\cite{aragon2022human, shneiderman2022human, bingley2023where,ehsan2021operationalizing}. In particular, there is an emphasis on the challenges of AI integration into socio-technical processes to preserve human autonomy and control, as well as the impacts of AI deployment and applications on society, organizations, and individuals~\cite{boyarskaya2020overcoming, whittlestone2019role}. Scholars also argue that understanding socio-technical and environmental factors can surface why and how an AI may become human-centered~\cite{shneiderman2020bridging, ehsan2021expanding, liao2021human, muller2022neurips}.

Despite its recent surge, bias in AI has been a longstanding topic of interest among academic circles~\cite{caton2020fairness}. Several surveys discussed data and model biases across domains, including social data biases that stem from user-generated context and behavioral traces~\cite{olteanu2019social}, gender bias in natural language processing~\cite{sun2019mitigating}, and fairness metrics with potential mitigation strategies~\cite{caton2020fairness, pessach2022review, orphanou2022mitigating}. Others investigated emerging research trends in responsible AI, such as fair adversarial learning, fair word embeddings, fair recommender systems, and fair visual description~\cite{pessach2022review}. 

As AI comes with the promise of advancing our economy and augmenting our lives, it becomes increasingly essential for people to understand AI's implications and remain in control~\cite{arrieta2020explainable}. To this end, much of the previous literature has been dedicated to discussing explainability in AI. For example, a survey of 289 core papers on explanations and explainable systems and more than 12,000 citing papers represents the diverse research landscape of AI, such as algorithmic accountability, interpretable machine learning, context awareness, cognitive psychology, and software learnability~\cite{abdul2018trends}. More recently, \citet{angelov2021explainable} provided an overview of AI explainability techniques considering recent advancements in machine learning and deep learning. Methods included featured-oriented ones such as SHapley Additive exPlanation (SHAP)~\cite{lundberg2017unified}, and surrogate models such as Local interpretable model-agnostic explanations (LIME)~\cite{dieber2020why}. Another line of work reviewed specific domains (e.g., recommender systems, social networks, and healthcare). These reviews cover notions of fairness specifically tailored for recommender systems~\cite{li2022fairness}, machine learning fairness in the domain of public and population health~\cite{mhasawade2021machine}, and algorithmic fairness in computational medicine~\cite{xu2022algorithmic}.

The closest survey to ours is \citet{capel2023what}'s survey, published at CHI 2023, where the authors focus on the ``human-centered AI'' as the primary search term. The landscape mapping resulted in four major themes: explainable and interpretable AI, human-centered approaches for designing and evaluating AI, human-AI teaming, and ethical AI. Our survey extends this survey beyond human-centered AI by adding two keywords (ethical AI and responsible AI). Adding these keywords has significance because it presents the complex challenge of addressing multiple aspects simultaneously. Each aspect---human-centeredness, ethical considerations, and responsible AI---cannot be achieved in isolation; they are interdependent. This interdependence underscores the crucial need to thoroughly comprehend and explore this intersection in order to create AI systems that are comprehensive and aligned with societal values. Our work also shows the differences and similarities between the research communities, which are not covered in the prior work.

While previous surveys mapped out parts of research in human-centered AI, ethical AI, and responsible AI, these mappings do not fully cover the intersection of the three, underscoring the need to explore it systematically. We also provide a lens into different research practices across the research communities as a new angle to the prior surveys. Therefore, this review aims to fill this gap and contribute to the ongoing discourse about HCER-AI.

\section{Method}
\label{sec:method}
\begin{figure}
\centering
\includegraphics[width=0.6\linewidth]{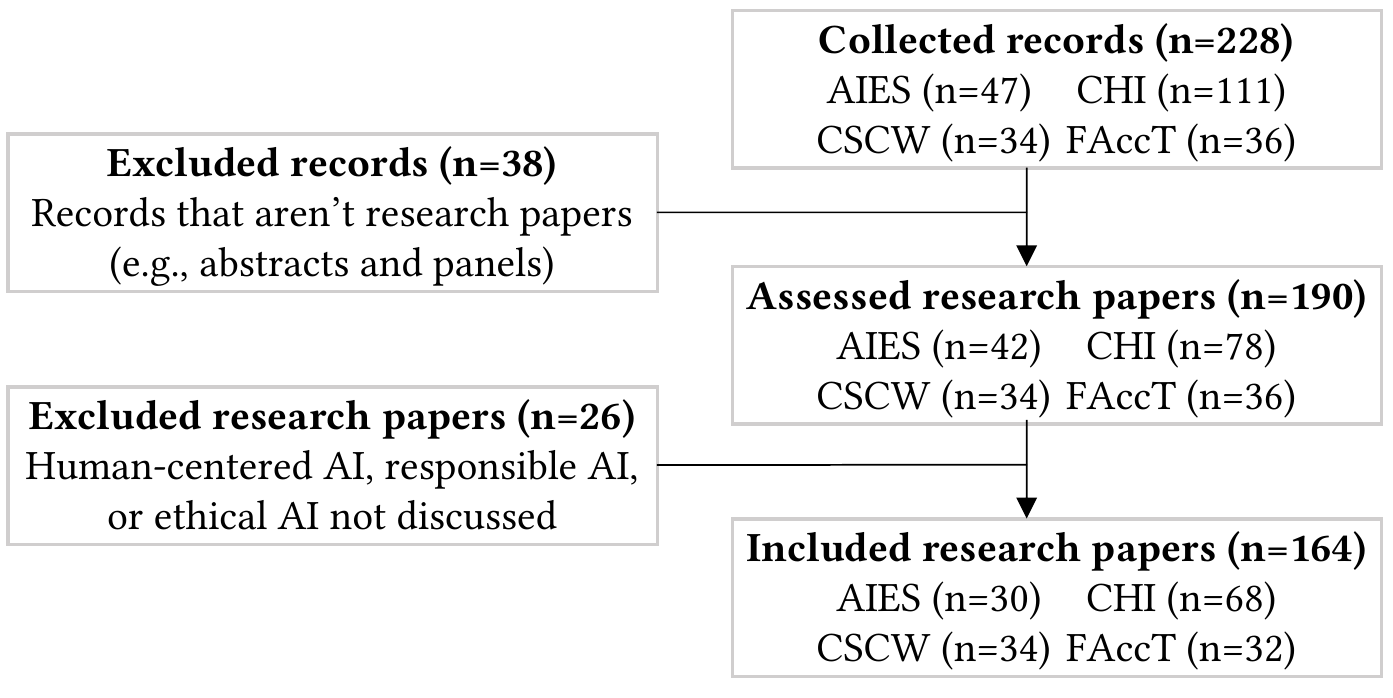}
\caption{An overview of our research method. We started with 228 records from the ACM Digital Library. After assessing quality and eligibility based on our inclusion criteria, we included and analyzed 164 research papers.}
\label{fig:prisma}
\end{figure}

To answer our RQs (\S\ref{sec:intro}), we surveyed the HCER-AI literature in AIES, CHI, CSCW, and FAccT following the PRISMA 2020 statement for conducting systematic literature reviews~\cite{page2021prisma}. Figure~\ref{fig:prisma} shows the flowchart of our method.

\subsection{Positionality Statement}
\label{sec:positionality}
Understanding researcher positionality is essential for demystifying our lens on data collection and analysis~\cite{frluckaj2022gender, havens2020situated}. We situate this paper in two Western countries in the 21\textsuperscript{st} century, writing as authors primarily working as academic and industry researchers. We identify as males from the Middle East, Southern Europe, and North America with diverse ethnic and religious backgrounds. Our shared research backgrounds include human-centered computing, privacy, security, software engineering, AI, social and ubiquitous computing, urbanism, and conducting systematic literature reviews. While we aimed at a bottom-up literature analysis, our results are constructs of our expertise and understanding of the landscape, as well as our positionality and cultural background. At least two authors were involved in all research steps to diversify the biases that attend any humans' judgments in analyzing the papers and reporting the results.

\subsection{Eligibility and Inclusion Criteria}
\label{sec:eligibility}
For this systematic review and our scoping, papers qualified for the analysis if their contributions pertained to three main areas: \emph{human-centered AI}, \emph{responsible AI}, and \emph{ethical AI}. By searching for keywords, we relied on the paper authors' use of terms related to our scoping. Our interpretation of AI is based on the definition of the National Institute of Standards and Technology (NIST), published in 2023: ``as an engineered or machine-based system that can, for a given set of objectives, generate outputs such as predictions, recommendations, or decisions influencing real or virtual environments. AI systems are designed to operate with varying levels of autonomy.''\cite{standards2023ai} In the next section (Section~\ref{sec:data-collection}), we describe our data collection method.

\subsection{Data Collection}
\label{sec:data-collection}
We decided to base our review on four research conferences: two well-established venues with a wide net for human-computer interaction research and two recently established venues with a narrow emphasis on AI and its implications (excerpts taken from conferences' websites):

\begin{itemize}
    \item AIES (established in 2018): ``our goal is to encourage talented scholars in these and related fields to submit their best work related to morality, law, policy, psychology, the other social sciences, and AI.'' 
    \item CHI (established in 1982): ``annually brings together researchers and practitioners from all over the world and from diverse cultures, backgrounds, and positionalities, who have as an overarching goal to make the world a better place with interactive digital technologies.''
    \item CSCW (established in 1986): ``is the premier international venue for research in the design and use of technologies that affect groups, organizations, communities, and networks.'' 
    \item FAccT (established in 2019; previously known as FAT*): ``a computer science conference with a cross-disciplinary focus that brings together researchers and practitioners interested in fairness, accountability, and transparency in socio-technical systems.''
\end{itemize}

We collected 228 records from the ACM Digital Library from these four conferences in May 2023. We queried for \texttt{``human-centered AI''} \texttt{OR} \texttt{``human-centered artificial intelligence''} \texttt{OR} \texttt{``ethical artificial intelligence''} \texttt{OR} \texttt{``ethical AI''} \texttt{OR} \texttt{``responsible AI''} \texttt{OR} \texttt{``responsible artificial intelligence''} within anywhere in the proceedings of AIES, CHI, CSCW, and FAccT, including only full research papers using ACM's filters (excluding other materials such as abstracts, panels, and tutorials). We chose these keywords because human-centered AI, ethical AI, responsible AI are widely used in academia and industry; examples include prior literature~\cite{capel2023what, shneiderman2022human} and guidelines produced by large technology companies (e.g., Google, Microsoft, and IBM use responsible AI to address ethical concerns and risks with AI~\cite{e2022responsible, t2022responsible, c2022pwc, labs2022responsible, arrieta2020explainable, co_operation2019recommendation, tahaei2023human}). We define human-centered AI as a paradigm that advocates for including people's requirements and needs in designing, developing, and deploying AI systems. Ethical AI and responsible AI refer to paradigms that emphasize the inclusion of ethical values in the design, development, and deployment of AI, as well as the responsibility to consider the risks and consequences on individuals and societies. Our review aims to explore the intersection of these research areas. We explore the topics related to the impact of AI to inform the building of AI systems that prioritize the needs and requirements of people and society while mitigating potential harms through a human-centered approach.

The above process resulted in 190 full research papers that two authors read to assess their relevance and eligibility (Section~\ref{sec:eligibility}). They examined papers discussing human-centered AI, ethical AI, and responsible AI. They excluded papers that mention the keywords in the references or acknowledgments without discussing them in the main text and papers that did not cover our topics of interest (human-centered AI, ethical AI, and responsible AI). After discussing and resolving disagreements, 26 of these papers were excluded. Therefore, our review and the rest of this paper are based on a final set of \textbf{164} research papers from AIES (n=30), CHI (n=68), CSCW (n=34), and FAccT (n=32).

We were also interested in discovering patents related to our keywords. Patents can show the practical applications of an invention and how it propagates in the industry and products~\cite{cao2023breaking}. HCI research, in particular, significantly impacts patents compared to other computer science research areas~\cite{cao2023breaking}. Therefore, we ran the same query we executed on ACM on the United States Patent and Trademark Office search tool~\cite{patents2023} and retrieved 67 results. One author manually reviewed these patents and removed 30 because they were unrelated and only appeared in the results due to a keyword (e.g., patents with ``responsible for AI'' were also included). This process left us with 37 patents.

\subsection{Data Analysis}
Following a thematic analysis method~\cite{braun2006using, lazar2017chaptera}, the first two authors split the papers between themselves. After discussions, they decided to open code each paper with the research methods used, primary contributions, human aspects, responsible aspects, and sample description (any information about the population or dataset of the paper). They used the paper authors' words to code each paper (also known as ``in vivo coding''~\cite{lazar2017chapterb}). When in doubt, they flagged a paper for further discussion (n=27). In a joint session, they used thematic analysis to create themes from the codes and resolved disagreements through discussions. They did not use a set of predefined themes but came up with the results using a bottom-up approach. In our analysis, a paper may appear in multiple codes---codes are not mutually exclusive. Sections~\ref{sec:research-methods} and \ref{sec:themes} are derived from this analysis. Although our findings are primarily based on this analysis and review, as authors actively engaged in this field, we have also utilized other papers and online resources to enhance our discussion beyond the reviewed papers, especially when addressing future directions. For a comprehensive list of the reviewed papers, refer to Table~\ref{tab:all-papers} in the Appendix.

In a joint session, two authors read the title and abstract of the remaining 37 patents and built an affinity diagram. During this process, they removed another 9 patents because they were unrelated and did not discuss our topics of interest as their primary claims. The final set of patents totaled \textbf{28}. Section~\ref{sec:patents-results} is based on the findings of this analysis.

\subsection{Limitations}
\label{sec:limitations}
We acknowledge that our literature search may not be comprehensive and exhaustive. However, covering AIES, CHI, CSCW, and FAccT as prominent venues for research in AI, Human-Computer Interaction (HCI), ethical, and responsible AI gave us insights into the state of the art of HCER-AI. Future research may also build on our work and expand it to other academic and industry literature. We also used keywords instead of reviewing a list of papers and deciding which paper to include or exclude based on our judgment (a typical approach in running systematic literature reviews, for example, \citet{holl_ander2021taxonomy} and \citet{bergram2022digital}). Through this approach, we collected papers related to HCER-AI from the authors' perspective rather than using our own judgment of whether a paper is related to HCER-AI. Future research may take a representative sample of papers from research publications, conduct a bottom-up analysis, and compare results with ours.

\section{Results: Research Methods of HCER-AI}
\label{sec:research-methods}
\begin{figure*}
\centering
\includegraphics[width=\linewidth]{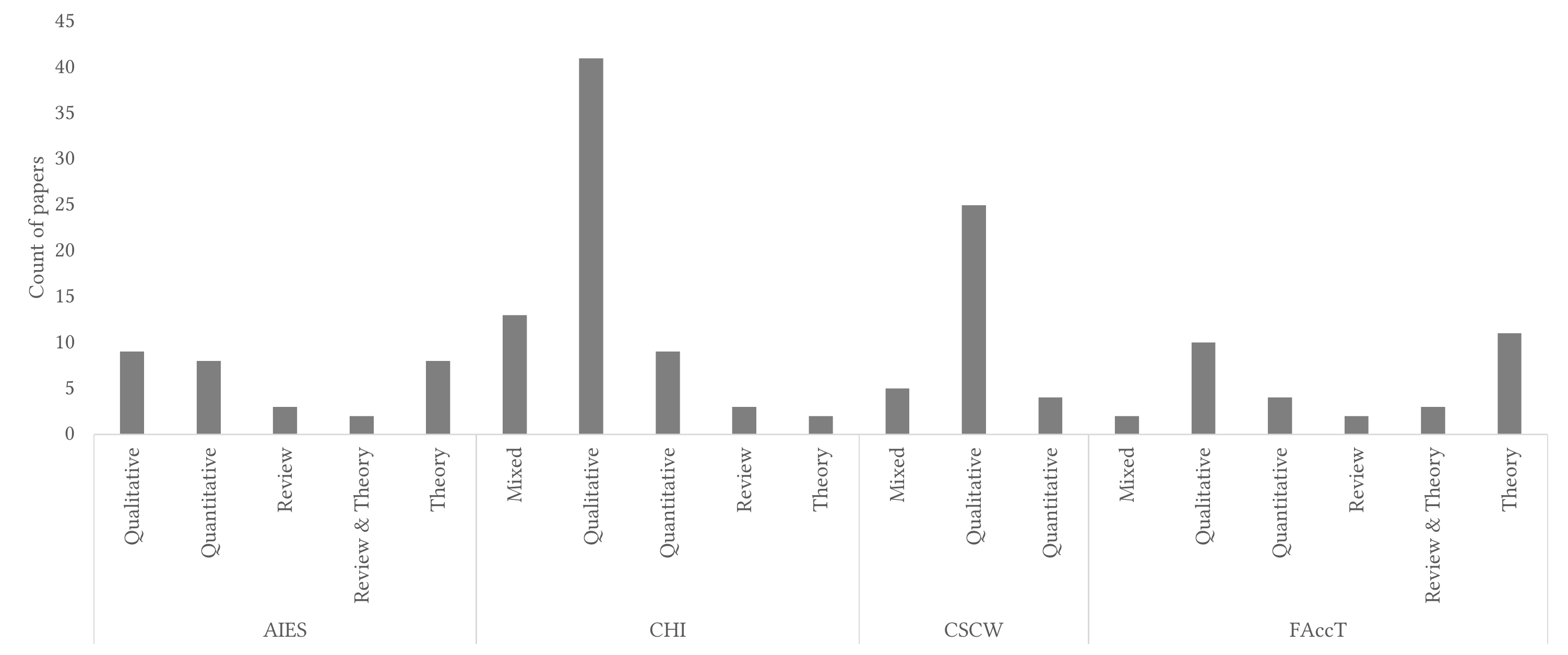}
\caption{Count of HCER-AI papers per research method (n=164).}
\label{fig:papers-methods}
\end{figure*}

In the 164 papers reviewed, 85 were qualitative (e.g., user studies for system design and evaluation, interviews, and workshops), 25 were quantitative (e.g., survey and log analysis), 21 were theoretical (e.g., essays and framework proposal), 20 were mixed-methods (a combination of qualitative and quantitative), 8 were reviews (e.g., review of literature, policies, and guidelines), and 5 were a combination of a review and theory (Figure~\ref{fig:papers-methods}). CHI and CSCW are heavily focused on qualitative studies (as expected since they are the premier venues for human-centered studies); however, they lack papers derived from theoretical reviews. AIES and FAccT cover a broader range of research methods with a preference for theoretical papers.

Papers did not have a consistent way of reporting demographics. The typical demographics were gender, job title, age, and prior experience in a particular topic of interest for the research (Table~\ref{tab:paper-demographics}). Out of the papers that reported the location of research or participants, the primary focus was on North America (n=47), and many did not report demographics (n=28) (Table~\ref{tab:paper-continent}).

\begin{table}
\centering
\caption{Count of HCER-AI papers per demographic value (n=103, includes papers with some information about the participants). Prior experience is in a particular topic of interest for the research. Reported demographics with at least 3 papers in the respective demographic.}
\label{tab:paper-demographics}
\begin{tabular}{@{}lr@{}}
\toprule
\textbf{Demographic} & \textbf{Count of HCER-AI papers} \\ \midrule
    Gender & 59 (57\%) \\
    Age & 48 (47\%) \\
    Job title & 43 (42\%) \\
    Prior experience & 21 (20\%) \\
    Education & 16 (16\%) \\
    Ethnicity & 10 (10\%) \\
    Industry type & 6 (0.6\%) \\
    Organization type & 3 (0.3\%) \\
    Income & 3 (0.3\%) \\ \bottomrule
\end{tabular}
\end{table}

\begin{table}
\centering
\caption{Count of HCER-AI papers per continent (n=164). Not applicable includes theory and review papers.}
\label{tab:paper-continent}
\begin{tabular}{@{}lr@{}}
\toprule
\textbf{Continent} & \textbf{Count of HCER-AI papers} \\ \midrule
    North America & 47 (29\%) \\
    Asia & 18 (11\%) \\
    Europe & 15 (9\%) \\
    Africa & 3 (2\%) \\
    Oceania & 1 (0\%) \\
    Not applicable & 69 (42\%) \\
    Not reported & 28 (17\%) \\ \bottomrule
\end{tabular}
\end{table}

\subsubsection*{\textbf{Recommendations for Future Research}}
Future researchers should consider reporting participant demographics to help with replicability and to understand how generalizable the results are. As discussed, there is a lack of consistency when reporting demographics. The community's own findings emphasize that building an AI that respects its users is highly dependent on the target population (see Section~\ref{sec:fairness} for details). Therefore, we see a need to provide a framework for reporting demographics for future researchers if they want to make their research comparable and replicable, which echoes \citet{sambasivan2021everyone}'s finding of: ``\emph{the need for incentivizing data excellence in which academic papers should evolve to offer data documentation, provenance, and ethics as mandatory disclosure.}''~\cite{sambasivan2021everyone}

Furthermore, considering conferences that often suffer from an excessive focus on Western societies~\cite{linxen2021how, van_berkel2023methodology, yfantidou2023beyond, septiandri2023weird}, publishing this information will help with research transparency.\footnote{For example, based on the data from CHI papers published between 2016 to 2020, ``\emph{73\% of CHI study findings are based on Western participant samples, representing less than 12\% of the world's population.}''\cite{linxen2021how}} It may also create a push for extending the research to other populations. Especially with the current trend and growth in HCER-AI research, more people will be influenced by the research done in this community. Thus, additional care is required to make the research inclusive and ensure that everyone is not lumped into the most powerful group, for instance, by contextualizing the findings. To achieve this, one approach could be by encouraging one of the several proposed methods for reporting datasets, models, and participant demographics (e.g., Data Cards~\cite{pushkarna2022data}, Model Cards~\cite{crisan2022interactive}, Datasheets for Datasets~\cite{gebru2021datasheets}, FactSheets~\cite{richards2019factsheets}, and Data Statements~\cite{bender2018data}). The outcomes of these methods could serve as input for a visualization tool to support easy and quick probing of transparency and inclusivity.

We also want to highlight the invaluable contribution of a group of researchers that shed light on less studied populations, such as India in this case. Five papers in our set had a sample from India with shared authors. Therefore, the community is interested in accepting papers from less studied countries in the Global South. We encourage more researchers to study marginalized populations.

\section{Results: Research Themes in HCER-AI}
\label{sec:themes}
\begin{figure}
  \centering
  \includegraphics[width=\linewidth]{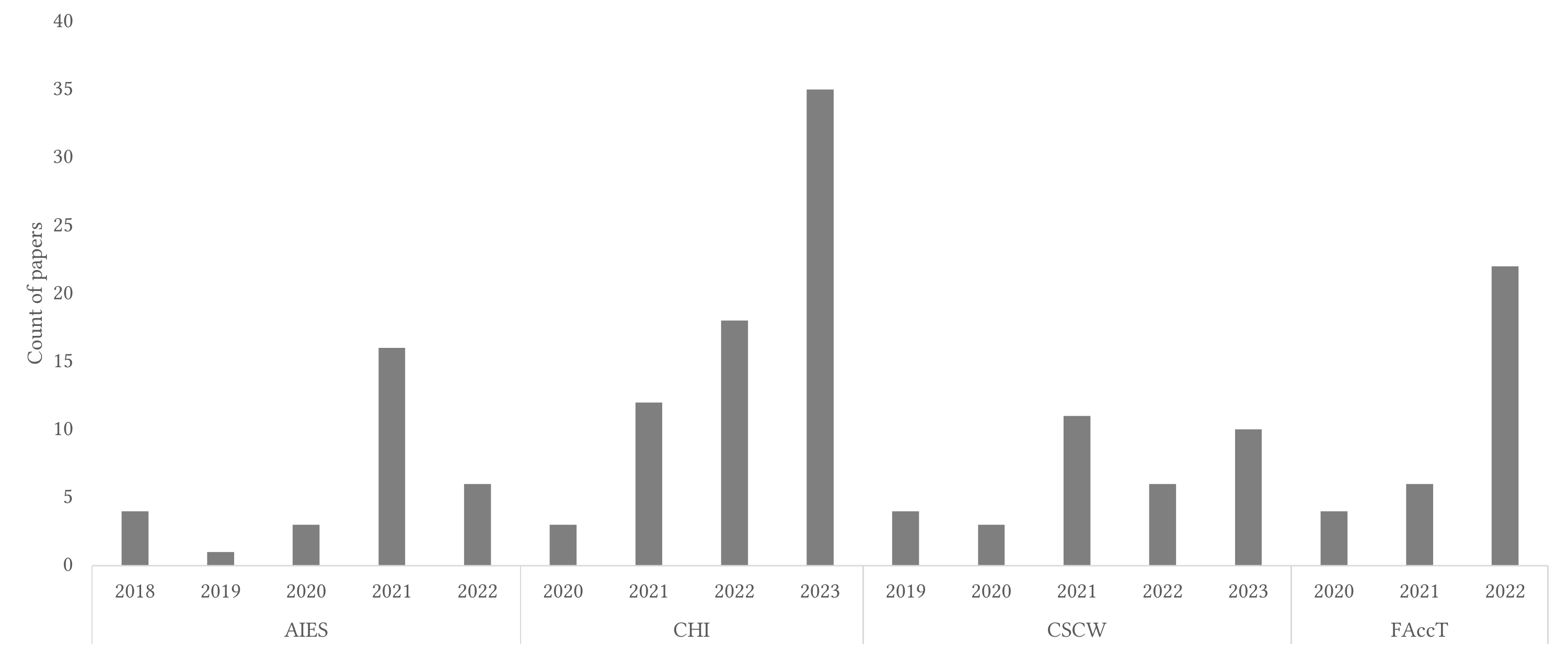}
  \caption{Count of HCER-AI papers per year. There is an exponential growth in HCER-AI research within the past few years.}
  \label{fig:trends}
\end{figure}

\begin{figure*}
\centering
\includegraphics[width=\linewidth]{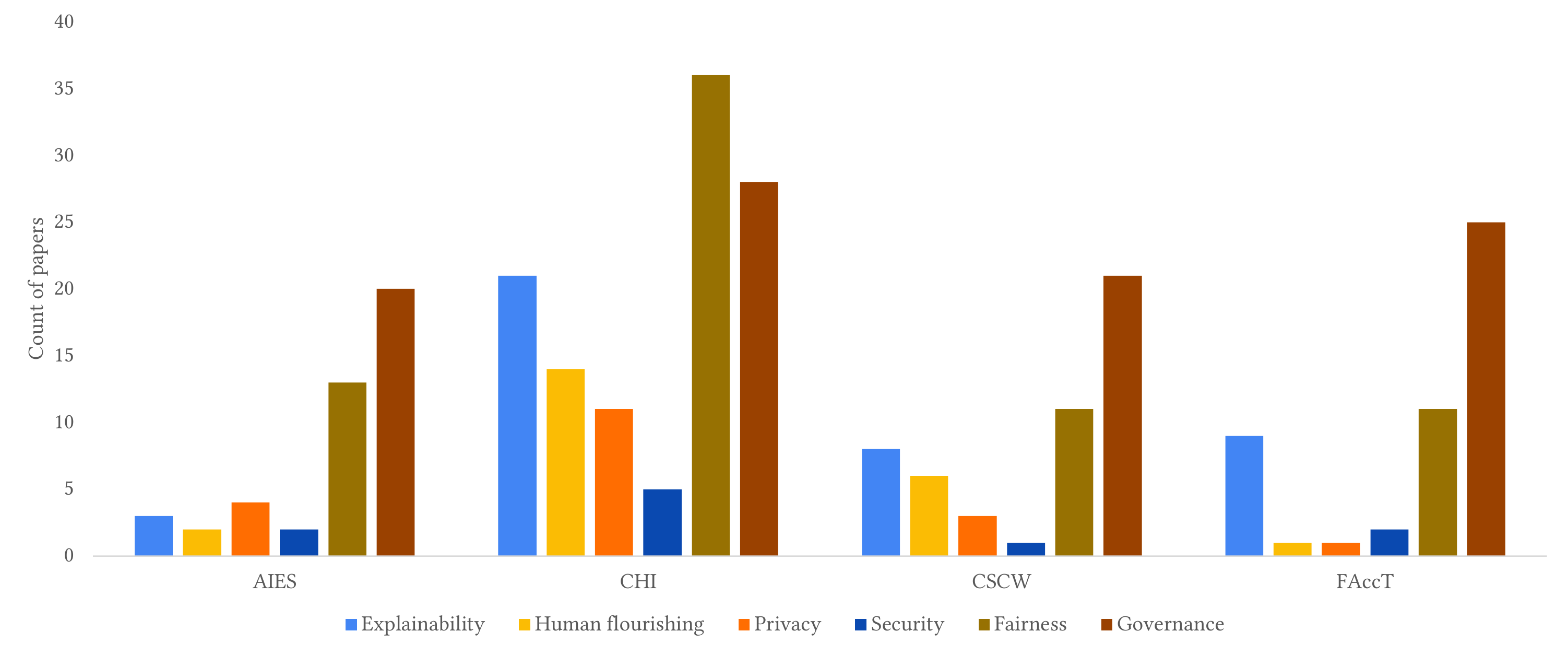}
\caption{Count of HCER-AI papers per research theme. The top three most popular research themes among the reviewed papers are governance, fairness, and explainability. Themes are not mutually exclusive.}
\label{fig:themes}
\end{figure*}

There has been an exponential growth in HCER-AI research in the past few years (Figure~\ref{fig:trends}). Based on our thematic analysis of 164 papers in this area, these efforts fall into six main research themes (Figure~\ref{fig:themes}): governance (n=94), fairness (n=71), explainability (n=41), human flourishing (n=23), privacy (n=19), and security (n=10). All conferences had at least one paper in each research theme focused on governance and fairness. While AIES and FAccT are smaller, with a narrow focus on AI and its ethical considerations, CHI and CSCW have a slightly more diverse research portfolio, possibly because they are larger conferences and attract a broader audience (human-computer interaction broadly defined).

\begin{figure}
\centering
\includegraphics[width=.8\linewidth]{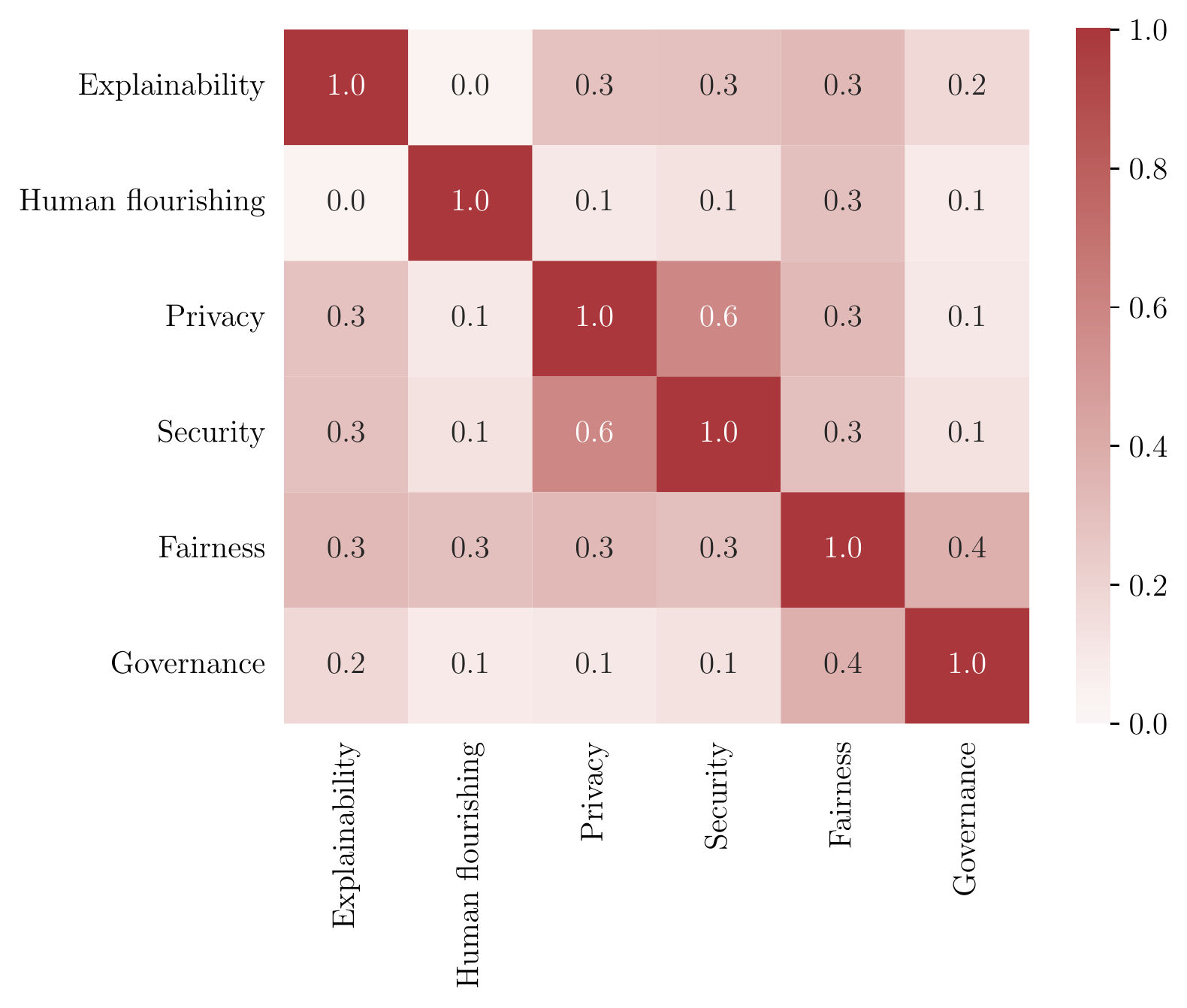}
\caption{Similarities between the research themes (n=164). The closer the number gets to 1, the more likely the pair appears together in a paper. For example, (privacy and security) and (fairness and governance) are often researched together.}
\label{fig:similarities}
\end{figure}

We created a pairwise similarity matrix for the pairs of research themes. Based on Figure~\ref{fig:similarities}, we observed that privacy and security, as well as fairness and governance, are the most common pairs (the closer the number gets to 1, the more likely the pair appears together in a paper). On the other hand, many pairs, such as governance with security, privacy, and human flourishing, did not appear together in our reviewed papers. For example, papers on governance often had a more general view of AI, except for some overlapping research with fairness and explainability. Across the four conferences, security had the least number of published papers. One possible explanation for this is the well-established nature of security as a field, particularly within computer science. Prominent conferences covering human factors, such as USENIX Security and IEEE S\&P, may attract papers at the intersection of security and responsible AI. Future research may explore similar areas in privacy and security conferences with human subjects.

In the following we discuss each theme, with the number of resources in each theme indicated next to the theme title (\texttt{\#N}).

\subsection{Governance \texttt{(\#94)}}
All conferences actively publish papers related to AI governance. Aside from general guidelines and critical views on AI governance, papers here fall into two sub-themes: guidelines about how AI should work, be governed, and regulated~\cite{muller2022forgetting, langer2022look, maas2018regulating, erdélyi2020ai, whittlestone2019role, schiff2020what, henriksen2021situated, costanza_chock2022who, ha2022south, weidinger2022taxonomy, stapleton2022imagining, knowles2021sanction, abdalla2021grey, pushkarna2022data, smith2022real, crisan2022interactive,sloane2022german, ashurst2022disentangling}; and tools for auditing AI~\cite{henriksen2021situated, krafft2021action, kroll2021outlining, ramesh2022how, black2022algorithmic, costanza_chock2022who, knowles2021sanction}.

\subsubsection{AI Guidelines} The three main stakeholders in the AI ecosystem are individuals (e.g., users, developers, and AI experts), organizations (e.g., ACM, companies, and independent institutes), and national and international bodies (e.g., UNICEF and regulators)~\cite{deshpande2022responsible, whittlestone2019role, wang2023designing}.\footnote{\MM{Despite the increasing interest in more collectivist views ~\cite{mcnaney2018enabling, vitos2017supporting, vigil_hayes2017indigenous,peters2013bridging, reinecke2013doodle}, these considerations were not present in our AI-centric search results.}} \marios{Several organizations and communities have already started to respond to this urgent need by sharing responsible AI guidelines~\cite{varanasi2023it, wang2023designing}. These guidelines are typically used not only for addressing AI's design challenges but also for education, cross-functional communication, and developing internal resources, as illustrated in a study with Google's People + AI Guidebook~\cite{yildirim2023investigating}.} From a theoretical standpoint, AI can raise several tensions between \marios{ethical} values. For instance, improving service quality may require additional data collection from users~\cite{whittlestone2019role}. These trade-offs are further confirmed by people who work for or co-founded a startup where AI entrepreneurs face a dilemma between academic integrity and potentially overrated marketing campaigns~\cite{winecoff2022artificial}. Addressing these trade-offs requires understanding AI's ethical challenges~\cite{schiff2020what, eicher2018jill}, developing principles for precautionary policy making~\cite{ha2022south, terzis2020onward}, educating students about AI at schools~\cite{lin2021engaging}, and building practical recommendations for the safe and trustworthy deployment of AI~\cite{maas2018regulating, eicher2018jill, baughan2023mixed, burgess2023healthcare}. Previous actions and strikes by employees in other industries represent one direction that may support promoting ethical values in the AI industry, but they require more commitment from employees than just open letters and petitions~\cite{boag2022tech}. As existing legal liability systems fall short in assigning responsibility when potentially harmful conduct occurs from using AI, the need to establish AI schemes that ensure responsible development of AI is further highlighted~\cite{erdélyi2020ai, bietti2020ethics}. Considering individual experiences, empirically understanding needs, and breaking traditional abstract views of society to enable research communities to be self-reflective and mindful of the previously mentioned trade-off in AI are directions that we, as researchers, need to take to move toward ethics in AI~\cite{siapka2022towards, washington2020whose, stark2021ethics, rismani2023plane, shahid2023decolonizing}.

Funding sources often drive academic research. AI research is no exception; hiring graduate students and researchers requires faculty and universities to attract funding to recruit individuals for research. The AI industry's indirect influence on the governance of AI may come from these fundings that shape academic research~\cite{abdalla2021grey, young2022confronting}. In a sample of 149 faculties from four prestigious North American universities, about half of those with known funding sources received money from large technology companies like Google and Microsoft. This type of influence resonates with the tobacco industry's influence when academic research was exploited to shift the public's negative opinion toward the benefits and positives of tobacco~\cite{abdalla2021grey}. The arbitrary reporting of limitations in AI research exacerbates such concerns. The machine learning research community lacks a single, agreed-upon definition of limitations and does not have a standardized process for disclosing and discussing limitations, in addition to often \marios{being non-inclusive~\cite{chi2021reconfiguring}} and lacking model\MM{-work} documentation~\cite{crisan2022interactive} \MM{or data-work documentation~\cite{sambasivan2021everyone, muller2022forgetting}}. On a positive note, recommendations to fill these gaps exist. REAL ML, for instance, provides a set of guided activities to help machine learning researchers recognize, explore, and articulate the limitations of their research~\cite{smith2022real}. Alternatively, frameworks have been proposed to disentangle different components of ethical research in machine learning to allow AI researchers, practitioners, and regulators to systematically analyze existing cultural understandings, histories, and social practices of ethical AI~\cite{sloane2022german, ashurst2022disentangling}, \marios{guiding the creation of human-AI teams~\cite{flathmann2021modeling}}.

In addition to the body of research on more organized and private AI companies and research institutes, research on open-source communities shows that the nature of open-source and unrestricted use of code propagates within the AI open-source community as well~\cite{widder2022limits}. Working on open-source AI projects creates a sense of neutrality and inevitability toward the technology and its consequences, while there can be serious ramifications to these projects, such as use cases for deep fakes. Similar approaches such as documentation~\cite{miceli2022documenting} and nudging developers into responsible development are suggested to alleviate the harm~\cite{widder2022limits}. However, the community's adoption of these guidelines and frameworks is still unknown.

\subsubsection{Tools for AI Auditing and Research} This category of papers aims to aid experts, activists, and independent bodies in improving the quality and impact of auditing AI~\cite{costanza_chock2022who, lam2023model} \marios{(especially in the age of large language models~\cite{zhou2023synthetic})}, to understand accountability as a responsibility put on the shoulders of AI engineers~\cite{henriksen2021situated}, \marios{to discuss the right of contestability~\cite{lyons2021conceptualising, yurrita2022towards}, and to engage with theories from moral philosophy~\cite{nashed2021ethically}}. At the initial stages of a project, design cards~\cite{elsayed_ali2023responsible} can help bring knowledge in areas such as creative inspiration, human insights, problem definition, and team building~\cite{hsieh2023what}. \marios{Other tools include computational notebooks~\cite{ayobi2023computational} to help practitioners explore machine learning models, workflows that encourage designers to explore model behavior and failure patterns early in the design process~\cite{moore2023failurenotes}, data probes to help practitioners surface the well-being and positionalities that shape their work strategies~\cite{lee2023fostering}, and exhibitions to teach critical thinking~\cite{lee2023fostering}.} Another idea for creating a foundation for audit is to apply the tracing and debugging mechanism from software engineering~\cite{kroll2021outlining, balayn2023faulty}. Version control systems keep track of changes in code repositories, and similar ideas could be applied in auditing AI, which requires structure, logs, and wide adoption~\cite{kroll2021outlining}. Besides technical documentation about how things work and their respective risks~\cite{pushkarna2022data, weidinger2022taxonomy}, many users (average users without technical knowledge and interest in \textit{how} systems work) are interested in \textit{what} the system does, how it can affect them, and whether it is fundamentally reasonable to replace the traditional mechanism with AI~\cite{knowles2021sanction, stapleton2022imagining}. 

\marios{Only by comprehending the foundations of algorithmic decision-making can we establish legitimate control over algorithmic processes~\cite{burrell2019when}. Recently, scholars have begun to discuss the right to contestability, which entails the ability to contest algorithmic decisions~\cite{lyons2021conceptualising, alfrink2023contestable, yurrita2022towards}. For example, a study conducted through a series of participatory design workshops found that by designing for contestability, users can actively shape and exert influence over algorithmic decision-making~\cite{vaccaro2021contestability}. Another study showed that, in high-stakes AI, physicians' trust in AI was less dependent on their general acceptance of AI but more on their contestable experiences with AI~\cite{verma2023rethinking}. Another large body of work aims to engage AI practitioners with ethical theories. For example, by expanding and defending the Ethical Gravity Thesis, scholars proposed a framework that allows one to situate ethical problems at the appropriate level of abstraction, which can be used to target appropriate interventions~\cite{kasirzadeh2021ethical}. Another study provided a mathematical framework to model how much learning is required for an intelligent agent to behave morally with negligible error~\cite{shaw2018towards}, while another developed a computational model for building moral autonomous vehicles by learning and generalizing from human moral judgments~\cite{kim2018computational}.}

\subsubsection*{\textbf{Recommendations for Future Research}}
Future research could study the specifics of governing other research themes in depth. Most of the current governance research is general except for fairness; therefore, studying the governance of AI with respect to privacy, security, human flourishing, and explainability is needed (Figure~\ref{fig:similarities}). Furthermore, providing usable documentation and communicating AI's abilities and implications to the public to create trust in AI is necessary and needs research beyond simple explanations that may cause over- or under-trust issues~\cite{liao2022designing}. One suggestion is to involve users and communities in helping situate the requirements within the target individuals and groups~\cite{ramesh2022how, sambasivan2021re}. The rise of ``user-driven audits'' on social media such as Twitter (i.e., users hypothesizing, collecting data, amplifying, contextualizing, and escalating AI issues and possible harms) can provide a window into people's expectations and provide a flow of integration of AI by the communities~\cite{wang2019designing}.

The industry's impact on academic integrity is a known issue. In privacy, for example, similar influences have been observed, discussed, and documented~\cite{waldman2021industry}. As pointed out in the research methods of HCER-AI (\S\ref{sec:research-methods}), transparency in data collection, reporting data characteristics, and funding resources with an on-demand visualization tool to provide an overview of these aspects could help. \MM{While participatory design approaches have often been used for user-originated or worker-originated critiques of technologies, it is not yet clear whether and how they can be applied to complex algorithmic systems~\cite{delgado2022uncommon, neuhauser2011participatory, zytko2022participatory}.} Designers of design cards also need to further extend the scope of these tools to evaluation and later stages of AI projects beyond the current design cards that focus on ideation and initial stages.

\subsection{Fairness \texttt{(\#71)}}
\label{sec:fairness}
Scholars and practitioners argue that fairness is contextual, cultural, and individually dependent, among many other dimensions; one size does not fit all~\cite{docherty2022re, kapania2022because, wang2022informing, cheng2021soliciting, lee2021who, sambasivan2021re, kasirzadeh2021use, richardson2021towards, ashktorab2023fairness, lewicki2023out, sharma2021fair, aka2021measuring, cruz_cortés2020invitation, yang2022enhancing, deng2023understanding}. Therefore, when designing experiments, contextual considerations should be considered to reduce errors in research design and results~\cite{kapania2022because, niforatos2020would, ashktorab2023fairness}. For example, conclusions about individuals' perceptions or expectations of fairness may not be replicable when the country, age, or gender group changes~\cite{kapania2022because, wang2022informing, simons2021machine}. Some biases are inherent in practitioners'~\cite{ashktorab2023fairness} or \marios{data annotators'~\cite{kapania2023hunt}} backgrounds, and some are also present in the inevitable changes to datasets and models during AI development~\cite{muller2022forgetting, muller2021designing}. For example, during the data annotation \marios{and curation} process, when raw data is prepared for the final model, disagreement between annotators may get lost, resulting in the final ground truth being based on the agreement biased towards a certain annotator or group~\cite{muller2022forgetting, muller2021designing} \marios{and questioning its credibility~\cite{chen2023is}}. \marios{This is why scholars proposed new AI data curation approaches grounded in feminist epistemology and informed by critical theories of race and feminist principles~\cite{leavy2021ethical}.} While the field of algorithmic fairness has primarily explored the notion of fairness as \emph{treating individuals alike} or arguing for \emph{social inclusion}~\cite{huang2022social}, there has been a recent debate on the concept of \emph{vertical equity}---appropriately accounting for relevant differences across individuals---which has also been a central fairness component in many public policy settings~\cite{black2022algorithmic}. The empirical notion of fairness from non-experts also shows that people may accept an AI making inferences about portraits when the use case is advertising but do not find it permissible for AI to make inferences about an individual based on their gender, age, or race, demonstrating the complexity of fairness~\cite{engelmann2022what}.

Terminology (e.g., artificial intelligence, algorithm, and robot) also impacts individuals' perceptions of fair AI and how they trust the system's decisions. Therefore, when researchers conduct research with human participants and use various terms to refer to an automated decision-making system, participants may have a mixed understanding and expectation of trust and fairness from the system when the terms addressing the studied system change~\cite{langer2022look}. Another example that can impact participants' perceptions of an AI's fairness is people's lived experiences. When a marginalized group has low trust in a human decision-maker for historical reasons (e.g., not trusting a medical doctor's decision), people have similar expectations of fairness from an AI \MM{(e.g., \cite{lee2021who, nakao2022involving, sambasivan2021re})}. 

Conversely, if an individual has a positive historical experience with the human decision-maker (e.g., trusting a medical doctor's decision), the individual may place higher trust in the AI's decisions as well~\cite{lee2021who}. While in other situations, people often consider AI more capable than humans but less morally trustworthy~\cite{tolmeijer2022capable}. Overall, human agents are often held responsible in decision-making~\cite{tolmeijer2022capable, lima2021human}. To alleviate the gap between designers and users of AI, designers could bring users and diverse viewpoints into the design stages of AI to create a shared understanding of fairness between different stakeholders~\cite{yildirim2022how, subramonyam2022solving, park2022designing, raz2021face, to2021reducing, choi2023creator, burrell2019when}. To further consider biases in the deployment stages, humans can be integrated into the loop with AI~\cite{cheng2022how, reitmaier2022opportunities, raz2021face}, resulting in human-algorithm collaboration~\cite{donahue2022human, zheng2023competent}. For example, in deciding whether a child was maltreated, when AI decisions were combined with a human decision-maker, the number of racially biased decisions was reduced compared to when AI was the only decision-maker. Hence, the human in the loop may improve decisions by taking an overarching assessment and filling AI's limitations in decision-making~\cite{cheng2022how}.

\subsubsection*{\textbf{Recommendations for Future Research}}
Future research could study fairness from multiple perspectives (e.g., culture, gender, ethnicity, age) and compare expectations or understandings in various groups \MM{\cite{lee2021who, nakao2022involving, sambasivan2021re}}. Multiple studies show the value of diversity in AI development and research teams to uncover potential biases and gaps, a practice that academics and industry need to adopt~\cite{septiandri2023weird, ashktorab2023fairness}. Furthermore, the studied toolkits provide metrics to evaluate fairness, primarily for AI developers (or experts). Researching how to present these metrics to a non-expert user who is impacted by the system could be a direction to pursue: how to present fairness to the public with minimal knowledge about fairness (e.g., is \emph{statistical parity} the most understandable way of presenting fairness to a senior adult or a child about the fairness of an AI?). These tools can be helpful for non-expert users to understand AI decisions and make informed decisions, as typically, interactions between people and machines range between two extremes: humans either tend to under-rely on an algorithm by ignoring its recommendations (algorithmic aversion) or over-rely on it by blindly accepting any recommendation (automation bias) \cite{de_arteaga2020case, he2023knowing}. Additionally, some scholars have argued that an AI can somehow decide ``objectively,''~\cite{houser2019can}. In contrast, others have argued that humans insert their biases into the data they wrangle and the models they build~\cite{muller2022forgetting}. The former ``objectivity'' argument could lead to expectations of better treatment by an AI rather than a person. The latter ``bias'' argument could lead to expectations of equally bad or worse treatment by an AI. While toolkits for audit are focused on fairness, we believe there is space to explore practices and build toolkits for audit and accountability in other aspects like privacy and security. Another aspect to consider is that accountability could provide explainability over decisions but make people feel surveilled~\cite{ehsan2021expanding}; therefore, finding a balance between accountability, privacy, and fairness is a challenge that requires further exploration.

\subsection{Explainability \texttt{(\#41)} }
\label{sec:understandability}
If an AI operates as an opaque box,\footnote{Where a decision is generated without explanation or interpretation, preventing users from understanding why and how the decision was made} users may have trouble trusting the outputs. Consequently, a recent movement in AI seeks to make these decisions comprehensible to garner users' trust~\cite{liao2020questioning, bertrand2022how, kim2023bubbleu, yuan2023contextualizing}, and explores ways to measure their impact~\cite{cabitza2023ai}. In human resource management, decisions made by AI can be further elucidated to gain employees' trust, helping them understand the rationale behind a decision with clear evaluation criteria (which may instill a sense of fairness)~\cite{park2021human, park2022designing}. However, explanations may not always absolve the AI of blame~\cite{lima2023blaming}. Often, trust issues depend on the personal relevance of the explanations and what is at stake~\cite{yuan2023contextualizing}, \marios{or on the lack of tools assisting AI practitioners in generating meaningful explanations~\cite{liao2023designerly}}. Paradoxically, explainability can sometimes conflict with privacy and security~\cite{hall2022supporting, park2022designing, park2021human}. We further discuss this trade-off in Sections~\ref{sec:privacy} and \ref{sec:security}.

However, explanations need to be human-centered, usable, and contextual~\cite{liao2020questioning, bansal2021does, ehsan2021expanding, long2022family, zhang2022debiased, raz2021face, lee2021participatory, toreini2020relationship} and may even need to be augmented with more than just simple text~\cite{yang2023harnessing}. Otherwise, they could create a false sense of control or be used as a scapegoat for responsibility~\cite{lima2022conflict}. For example, when explanations were employed in a human-AI team, they did not help improve the accuracy of the final decision~\cite{bansal2021does}. However, when combined with an opaque box model that elucidated \textit{what} the system does, users' confusion was reduced~\cite{bell2022it}. In either case, not all explanations are beneficial. To enhance explanations, AI developers could profit from posing \emph{how} and \emph{why not} questions to consider users' expectations and needs~\cite{liao2020questioning}, and could also utilize participatory design approaches~\cite{lee2021participatory}. Similar to asking the ``how'' questions, an alternative framework, Sensible AI~\cite{kaur2022sensible}---for interpretability was proposed, grounded in Weick's sensemaking theory~\cite{weick1995sensemaking}. From organizational studies, sensemaking describes the individual, environmental, social, and organizational context affecting human understanding. Similarly, Sensible AI translates these properties into the human-machine context. Another way to improve explanations is to add graphics and control sliders to help users understand an AI's decision, making them feel more in control~\cite{viswanathan2022situational}. Explanations can also help improve AI literacy for public and educational purposes, potentially leading to broader adoption and understanding of AI implications and harms (e.g., teaching kids about feature selection in machine learning using visual explanations)\cite{long2022family}. \marios{Another study examined the role of cognition in understanding explanations~\cite{buçinca2021trust}, proposing three cognitive forcing interventions to compel people to engage more thoughtfully with the AI-generated explanations.}

\subsubsection*{\textbf{Recommendations for Future Research}}
Future research is still needed to improve the explainability of AI's inner workings, decisions, and outputs, as there is some criticism about the value of explanations like SHAP for humans~\cite{kumar2020problems}. Toolkits for communicating datasets also need to track changes over time, from raw data to the final dataset used in the model, and changes that occur post-deployment, to mitigate the forgettance of data work---as \citet{muller2022forgetting} put it: ``\emph{Forgettance in data science is when each action tends to push previous actions into the infrastructure, where the action itself and its consequence are easily forgotten.}''\cite{muller2022forgetting} One way of documenting datasets is through a standardized process, such as using datasheets~\cite{gebru2021datasheets}, nutrition labels~\cite{chmielinski2022dataset}, or data statements~\cite{bender2018data}. Furthermore, taking a cue from computer hardware and industry safety standards, datasheets for datasets should include dataset provenance, key characteristics, relevant regulations, test results, and also more subjective yet noteworthy information (e.g., the potential bias of a dataset~\cite{gebru2021datasheets}).

\subsection{Human Flourishing \texttt{(\#23)}}
Sustainable growth and the well-being of humans (also known as human flourishing) using or being affected by AI were discussed in several studies~\cite{wang2022informing, docherty2022re, wang2022whose, steiger2021psychological, beede2020human, morrison2021social, lee2021participatory}. For example, if an employer wants to increase productivity using AI from a sustainable perspective, employees' expectations should be considered alongside implementing positive changes, such as monetary compensation~\cite{park2022designing}. AI can also facilitate real-time communication for users of augmentative and alternative communication~\cite{valencia2023less}. AI's impact on children has been deemed crucial due to its potential for long-term effects. Instances of using robots for education at early stages of child development can shape children's perceptions of robots. Moreover, entertainment systems that collect children's data may affect their entire life if service providers do not respect their privacy and keep their data indefinitely~\cite{wang2022informing}.\footnote{\MM{Adults are also susceptible to persuasion by, for example, ``cute'' -- but lethal -- robots~\cite{kemper2022k9, vaisman2021your, young2021danger}.}} \marios{In HCI literature, safety drivers working with Autonomous Vehicles (AV) often comprise an under-explored population impacted by AI. A study found that, as front-line workers, safety drivers are forced to shoulder risks accumulated from the upstream AV industry while confronting restricted self-development prospects when working for AV development~\cite{chu2023work}.} Similarly, socio-environmental factors must also be considered for the successful implementation of AI in real-world settings (e.g., a blind person navigating social interactions or the country of post-deployment) \cite{beede2020human, morrison2021social}. For example, offering explanations and options for employers to customize an AI-based scheduling system resulted in a smoother transition to AI and less resistance from users~\cite{lee2021participatory}.

A few papers focused on the labor behind AI~\cite{wang2022whose, steiger2021psychological, klinova2021ai}. AI creates new markets and opportunities which bring wealth to those who will be more in demand (e.g., AI developers) and shrinks the wealth of those who will be less in demand (e.g., customer service agents replaced by an AI chatbot) \cite{klinova2021ai}. However, new AI markets may not always be accommodating and lucrative for all. As part of the data creation process, annotators often work long hours with low wages, especially when cost-effective methods of annotating large datasets are in demand, leading AI companies to hire annotators from third-party annotation companies in the Global South for poor wages. In these companies, some annotators start working with the expectation that annotation work will open up opportunities for high-paying jobs in the future and serve as a stepping stone to becoming an engineer. However, this is often a fallacy, and these annotators rarely progress in their careers~\cite{wang2022whose}. Similarly, content moderators are sometimes exposed to inappropriate or disturbing content, causing severe stress and trauma. Some of these individuals were not fully aware of the risks associated with the job when they signed up for it due to a lack of transparency in the job description~\cite{steiger2021psychological}.

Either of these jobs (annotation or content moderation) is often perceived as a ``dirty'' job and receives minimal attention from companies~\cite {steiger2021psychological, wang2022whose, sambasivan2021everyone}, \marios{or as ``patchwork,'' that is, the human labor that occurs in the space between what AI purports to do and what it actually accomplishes~\cite{fox2023patchwork}.} Nevertheless, their contributions to the AI economy are invaluable\footnote{``\emph{Without the work and labor poured into the data annotation process, ML efforts are no more than sandcastles.}''~\cite{wang2022whose}} and foundational (by generating datasets and keeping AI safe from inappropriate content that others do not have to deal with). The following four recommendations can serve as a starting point to help these individuals manage their jobs' challenges~\cite{gray2019ghost, moreschi2020brazilian,steiger2021psychological}: (1) providing a detailed job description outlining its associated risks could give candidates an understanding of whether the job is appropriate for them, (2) limiting the number of contents they view or annotate per day or week, (3) creating a supportive community to help with stress relief, and (4) companies should consider offering employee benefits, including physical and mental health benefits, to these "ghost" workers.

\subsubsection*{\textbf{Recommendations for Future Research}}
Future research could study the long-term effects of AI on people, the people who directly use AI (e.g., a person using a fitness tracker with an AI coach\MM{~\cite{garcia2019expanding}}), those who are indirectly affected by AI (e.g., a person receiving an AI-assisted loan decision from a credit institute\MM{~\cite{hertzberg2010information, kizilaslan2017can}}, or the people who may lose their jobs or fear that they may lose their job because of AI~\cite{bbc2023AI}), and the people involved in AI's development and deployment (e.g., annotators and developers). \MM{A longer view of the impacts of AI algorithms could include neighborhoods affected by over-policing through AI technologies~\cite{gee2021reducing}.} For instance, as AI in workplaces becomes a technological intervention, which may have different impacts on different groups within that workplace~\cite{park2021human, park2022designing}, multiple workplace scenarios can be analyzed through value-sensitive design~\cite{muller2022extending}. \citet{docherty2022re} argue that the tech industry's focus on user engagement and time spent on the platform is a reductionist view of human well-being. When measuring well-being, an individual's specific needs and characteristics should be considered. Building on this idea, future researchers could study the well-being of people impacted by AI beyond limiting engagement time and think about human flourishing (i.e., ``\emph{the ability to live a good life}''~\cite{lab2022human}) from a positive computing viewpoint~\cite{calvo2014positive}. Furthermore, although not discussed in the reviewed papers, we see a need to study AI's environmental effects on natural resources from a nature-centered perspective. For instance, a tool could be designed to remind users about the energy consumption of their AI tools or to provide information about the energy consumption of AI models (including all stages, such as design, development, and deployment) with an easy-to-understand label, like the Energy Star.

\subsection{Privacy \texttt{(\#19)}}
\label{sec:privacy}
Several papers have touched on the privacy ramifications of AI~\cite{mlynar2022ai, lee2021who, zhang2022debiased, ehsan2021expanding, park2022designing, park2021human, yildirim2022how, subramonyam2022solving, wang2022informing, hall2022supporting}. For example, the use of human resource management tools in companies, which bear privacy implications, can make employees uncomfortable. These systems may be as simple as controlling a gate and storing that information for performance measures. However, collecting more data to improve the model's accuracy can induce a sense of surveillance~\cite{greiffenhagen2023work, constantinides2022good}. Nevertheless, from the viewpoint of those deploying the system, such data collection may be seen as a means of providing a fair assessment of the employee's performance~\cite{park2022designing, park2021human}. Such systems may create conceptual mismatches between normative (or cultural) and legal expectations regarding the use of personal data~\cite{nielsen2021measuring}. Designers and engineers acknowledge this tension between privacy and collecting more data or sharing datasets between different groups or teams to explore new opportunities. \marios{In fact, there is evidence that while people support the sharing of their data to improve technology, they also express concerns over commercial use, associated metadata, and the lack of transparency about the impact of their data~\cite{kamikubo2023contributing}.} Nonetheless, when a company is a large organization (especially when client data is involved), sharing data is difficult because data ownership can be vague and challenging~\cite{yildirim2022how, subramonyam2022solving}. In contrast, smaller companies grapple with several tensions, for example, the trade-off between privacy and ubiquity, resource management and performance optimization, or access and monopolization~\cite{hopkins2021machine}.

A few papers discuss the trade-off between privacy and explainability~\cite{zhang2022debiased, ehsan2021expanding, hall2022supporting}. Providing additional information about the model for explainability may compromise privacy or vice versa. For instance, in a study with people's WiFi data, participants were initially private and sensitive about sharing their WiFi data. However, showing visualizations about their data usage made them feel more comfortable sharing their data because it fostered a sense of trust~\cite{hall2022supporting}. On the other hand, privacy can become a barrier to providing explanations in image classification systems. For example, when images are obfuscated for privacy reasons, providing explanations for the classification may reveal the identities of people in the images~\cite{zhang2022debiased}. \marios{Other scholars have focused on how privacy legislation is discussed among different sets of relationships (e.g., between companies and investors~\cite{wong2023privacy}) through the lens of law and ethics~\cite{benthall2021artificial}. For example, a study found that startups with data-sharing partnerships with high-technology firms or prior experience with privacy regulations are more prone to adopting ethical AI principles~\cite{bessen2022cost}. Furthermore, they are more inclined to take costly actions, such as eliminating training data or rejecting business opportunities, to adhere to their ethical AI policies.}

\subsubsection*{\textbf{Recommendations for Future Research}}
Future research may explore the trade-off between privacy (and/or security) and explainability, as there is a thin line between what \emph{can} be monitored and what \emph{should} be monitored~\cite{constantinides2022good}. One perspective is to look at this trade-off from the lens of historian and philosopher Yuval Noah Harari, who posits that the digital platforms, whether powered by AI or not, need to adhere to three basic rules to protect humans from ``\emph{digital dictatorships}''~\cite{harari2021lessons}. These rules are (1) reasonable use cases for data collection, meaning that any data being collected should be used to assist people rather than manipulate, control, or harm them (e.g., providing Indigenous data sovereignty over culturally-sensitive data~\cite{kukutai2020indigenous, marley2019indigenous, tsosie2019tribal, walter2019indigenous}), (2) surveillance should be bidirectional, implying that if an entity (e.g., an organization or a government) increases surveillance of individuals, accountability on the entity's side should correspondingly increase, and (3) the elimination of data monopolies, which arise from the concentration of data in a single entity. Additionally, the General Data Protection Regulation (GDPR)~\cite{parliament2018general} in Europe argues for limiting the storage time only to legitimate purposes and doing regular reviews to delete unneeded data. Human-centered approaches and design methods can be employed to explore solutions for these problems through committees that oversee data collection and, ideally, ensure that data is in the hands of the individuals.

\subsection{Security \texttt{(\#10)}}
\label{sec:security}
AI security aims to ensure safety and reduce harm to individuals~\cite{wang2022informing, toreini2020relationship}. It involves enhancing a system's ability to resist external threats by testing its resilience against vulnerabilities and cyber-attacks, all while safeguarding the integrity and confidentiality of personal data~\cite{fjeld2020principled}. However, security may conflict with explainability. For instance, in the case of human resource management systems, disclosing information about models for explainability purposes could pose a security risk. Such openness could harm the organization's reputation and provide attack opportunities if all model details are publicly available~\cite{park2022designing, park2021human}. Furthermore, using AI in security-sensitive decision-making can be questionable. If AI decisions yield many false positives, reliance on them can create unwanted stress, as the operator in charge of making the final decision might feel compelled to act on them. This can be particularly stressful for junior staff who may feel obliged to report every incident~\cite{ehsan2021expanding}.

\subsubsection*{\textbf{Recommendations for Future Research}}
Future research may benefit from studying the security aspects of AI from a human-centered perspective. While there is a specific security and privacy track in CHI that publishes papers about the human factor of security and privacy in various technologies such as smartphones, apps, and browsers, and investigates the role of software developers in secure development, we believe the community can benefit from a dedicated focus on what it means to provide human-centered security in AI. This should consider the trade-off between explainability and security, which echoes the classic trade-off between usability and security~\cite{cranor2005security, tahaei2019survey}. Creating usable security systems is often challenging. In response to this challenge, a dedicated venue, the Symposium on Usable Privacy and Security (SOUPS), was established in 2005. Implementing more complicated security (or privacy) mechanisms can burden users, deter them from using the system, clash with their mental models, and lead to insecure actions. Similarly, providing more explanations for AI outcomes may cause users to overtrust the system. Therefore, finding the right balance between the two calls for further research.

\section{Results: Patents of HCER-AI}
\label{sec:patents-results}
The current state of patents in HCER-AI is encapsulated by the following themes: explainability (n=13), fairness (n=11), governance (n=6), human flourishing (n=4), privacy (n=4), and security (n=2). The themes in patents and research papers somewhat follow each other, except for governance being the major theme in research, which is expected because research venues cover theoretical, review, and discourse papers, while patents are more on the industry application level (Table~\ref{tab:papers-patents-counts}). Also, given that the first academic paper on the subject was published in 2018, it is unsurprising to see a small set of patents in this area. HCI research takes 10.5 years to appear in a patent application, and this lag is increasing over time~\cite{cao2023breaking}. The exponential rise in research papers in HCER-AI presents an opportunity for early-adopter industries to work on related patents.

\begin{table}[h]
\caption{Count of HCER-AI papers (n=164) and patents (n=28) in each theme.}
\label{tab:papers-patents-counts}
\begin{tabular}{@{}lllllll@{}}
\toprule
 & Governance & Fairness & Explainability & Human flourishing & Privacy & Security \\ \midrule
Research papers & 94 (57\%) & 71 (43\%) & 41 (25\%) & 23 (14\%) & 19 (12\%) & 10 (6\%) \\
Patents & 6 (21\%) & 11 (39\%) & 13 (46\%) & 4 (14\%) & 4 (14\%) & 2 (7\%) \\ \bottomrule
\end{tabular}
\end{table}

\section{Mapping the HCER-AI Landscape With the NIST AI Risk Management Framework} We also mapped our constructed HCER-AI research themes onto the U.S. National Institute of Standards and Technology (NIST) Artificial Intelligence Risk Management Framework, published in 2023. We chose NIST because it is a recent framework from a renowned organization for developing frameworks and standards. Alternatives include the Principled Artificial Intelligence from the Berkman Klein Center~\cite{fjeld2020principled}, which aligns with the NIST framework. Such mapping provides insights into the different perspectives on the landscape from the viewpoints of a standardization body versus academic research. This approach helps highlight areas needing further attention, potentially directing more funding and focus toward those areas to balance the research portfolio.

\begin{itemize}
    \item \textbf{Governance} (n=94): maps with NIST's ``Govern'' function: ``\textit{cross-cutting function that is infused throughout AI risk management and enables the other functions of the process},'' ``Accountable and Transparent'' risk: ``\textit{accountability presupposes transparency. Transparency reflects the extent to which information about an AI system and its outputs is available to individuals interacting with such a system -- regardless of whether they are even aware that they are doing so},'' and ``Valid and Reliable'' risk: ``\textit{Validity and reliability for deployed AI systems are often assessed by ongoing testing or monitoring that confirms a system is performing as intended}.''
    \item \textbf{Fairness} (n=71): maps with NIST's ``Fair -- with Harmful Bias Managed'' risk: ``\textit{includes concerns for equality and equity by addressing issues such as harmful bias and discrimination}.''
    \item \textbf{Explainability} (n=41): maps with NIST's ``Explainable and Interpretable'' risk: ``\textit{refers to a representation of the mechanisms underlying AI systems' operation, whereas interpretability refers to the meaning of AI systems' output in the context of their designed functional purposes}.''
    \item \textbf{Human flourishing} (n=23): AI should support ``\textit{happiness and life satisfaction, meaning and purpose, character and virtue, and close social relationships}.''~\cite{willen2022rethinking} This theme also maps loosely with the NIST's ``Safe'' risk: ``\textit{AI systems should not under defined conditions, lead to a state in which human life, health, property, or the environment is endangered}.''
    \item \textbf{Privacy} (n=19): maps with NIST's ``Privacy-Enhanced'' risk: ``\textit{refers generally to the norms and practices that help to safeguard human autonomy, identity, and dignity}.''
    \item \textbf{Security} (n=10): maps with NIST's ``Secure and Resilient'' risk: ``\textit{AI systems, as well as the ecosystems in which they are deployed, may be resilient if they can withstand unexpected adverse events or changes in their environment or use}.''
\end{itemize}

\section{Posteriori Analysis Using ChatGPT 4.0}
\label{sec:results-chatgpt}
Motivated by recent work on using AI for qualitative analysis~\cite{byun2023dispensing, abram2020methods, tahaei2020so}, we explored the possibilities of using ChatGPT in a similar analysis without manual analysis based only on the abstracts (because of input limitations of AI models). We used all the abstracts of the 164 papers as input for ChatGPT to evaluate its ability to produce a similar report to ours. The primary tasks were finding research themes, methods, and recommendations (our RQs). Noting that, we used ChatGPT 4.0, like any other text clustering and classification tool, to probe for topics we might have missed during our research. This process was a complete posthoc procedure after human authors wrote and finalized all the paper sections except this one. All of our prompts and all the content, besides the main headlines and topics, produced by ChatGPT are intentionally left in Appendix~\ref{app:chatgpt-report} to avoid confusion between our analysis and findings vs. AI-generated analysis and findings.

We first asked for \textit{research themes} from the abstract following the thematic analysis proposed by \citet{braun2006using} (our approach). ChatGPT came up with the following four research themes: ethical implications of AI, practical applications of AI, understanding and documenting data, and user engagement. We further asked for the \textit{research methods} used in the abstracts with a caveat that not all papers mention their methods in their abstracts. The answer was more nuanced than the research themes: interviews, participatory design, co-design workshops, prototyping, observation, critical discourse analysis, iterative co-design activities, characterization of collective actions, and case studies. We followed the exploration with \textit{research recommendations} and future research directions based on the thematic analysis. The following recommendations were generated: user-centric design, greater transparency in AI, AI in marginalized communities, collective action in tech industry, real-world AI evaluations, AI in disaster risk management, effective AI communication, AI ethics integration in education, work-integrated learning for AI, and AI governance practices. We did the same analysis for research themes of patents; the resulting themes were: AI-driven decision making and predictions, transparency, interpretability, and explainability in AI, and machine learning model performance \& enhancement.

The research themes produced by ChatGPT (ethical implications of AI, practical applications of AI, understanding and documenting data, and user engagement) could help get a general understanding of the dataset but were not enough to get an in-depth grasp of the data. Our research themes (governance, fairness, explainability, human flourishing, privacy, and security) are also a construct of our positionality (see Section~\ref{sec:positionality}). Despite the differences in both lists, AI governance and AI's ethical implications are visible. We also recognize that the classification generated by AI does not constitute a compelling taxonomy, with some themes containing others without being of the same level of importance. Possibly, because the AI-generated themes may be meant for general consumption and do not directly correlate with specific research areas of our analysis. Overall, we speculate that AI-assisted research tools like ChatGPT may, in the future, act as a quick secondary point to help researchers with unseen perspectives---considering that it took us less than two hours to produce these results compared to months of work for the rest of the paper. Future research may explore how researchers can better leverage AI to deliver valuable literature reviews by conducting more experiments, comparisons with human-generated analysis, and prompt engineering.

\section{Conclusion}
\label{sec:conclusion}
We collected 228 records related to HCER-AI from four conference proceedings: AIES, CHI, CSCW, and FAccT, as well as 67 patent applications. We then selected and thematically analyzed 164 research papers and 28 patents from these records. We found that the HCER-AI landscape emphasizes governance, fairness, and explainability. The research community needs to put effort into HCER-AI beyond fairness and create usable tools for non-expert users to audit AI and think about privacy and security from a human-centered viewpoint. While these topics are part of the broader conversation in responsible AI, there is still a need for future research to bring them into practice.

\begin{acks}
We thank Ali Septiandri, Edyta Bogucka, Ke Zhou, Lucy Havens, Sanja Scepanovic, and Vera Liao, for their constructive feedback that helped improve the paper.
\end{acks}

\bibliographystyle{ACM-Reference-Format}
\bibliography{main}


\begin{thebibliography}{277}


\ifx \showCODEN    \undefined \def \showCODEN     #1{\unskip}     \fi
\ifx \showDOI      \undefined \def \showDOI       #1{#1}\fi
\ifx \showISBNx    \undefined \def \showISBNx     #1{\unskip}     \fi
\ifx \showISBNxiii \undefined \def \showISBNxiii  #1{\unskip}     \fi
\ifx \showISSN     \undefined \def \showISSN      #1{\unskip}     \fi
\ifx \showLCCN     \undefined \def \showLCCN      #1{\unskip}     \fi
\ifx \shownote     \undefined \def \shownote      #1{#1}          \fi
\ifx \showarticletitle \undefined \def \showarticletitle #1{#1}   \fi
\ifx \showURL      \undefined \def \showURL       {\relax}        \fi
\providecommand\bibfield[2]{#2}
\providecommand\bibinfo[2]{#2}
\providecommand\natexlab[1]{#1}
\providecommand\showeprint[2][]{arXiv:#2}

\bibitem[Abdalla and Abdalla(2021)]%
        {abdalla2021grey}
\bibfield{author}{\bibinfo{person}{Mohamed Abdalla} {and}
  \bibinfo{person}{Moustafa Abdalla}.} \bibinfo{year}{2021}\natexlab{}.
\newblock \showarticletitle{{The Grey Hoodie Project: Big Tobacco, Big Tech,
  and the Threat on Academic Integrity}}. In
  \bibinfo{booktitle}{\emph{Proceedings of the 2021 {AAAI}/{ACM} {Conference}
  on {AI}, {Ethics}, and {Society}}} \emph{(\bibinfo{series}{{AIES} '21})}.
  \bibinfo{publisher}{ACM}.
\newblock
\urldef\tempurl%
\url{https://doi.org/10.1145/3461702.3462563}
\showDOI{\tempurl}


\bibitem[Abdul et~al\mbox{.}(2018)]%
        {abdul2018trends}
\bibfield{author}{\bibinfo{person}{Ashraf Abdul}, \bibinfo{person}{Jo
  Vermeulen}, \bibinfo{person}{Danding Wang}, \bibinfo{person}{Brian~Y. Lim},
  {and} \bibinfo{person}{Mohan Kankanhalli}.} \bibinfo{year}{2018}\natexlab{}.
\newblock \showarticletitle{{Trends and Trajectories for Explainable,
  Accountable and Intelligible Systems: An HCI Research Agenda}}. In
  \bibinfo{booktitle}{\emph{Proceedings of the 2018 CHI Conference on Human
  Factors in Computing Systems}} \emph{(\bibinfo{series}{CHI '18})}.
  \bibinfo{publisher}{ACM}.
\newblock
\urldef\tempurl%
\url{https://doi.org/10.1145/3173574.3174156}
\showDOI{\tempurl}


\bibitem[Abram et~al\mbox{.}(2020)]%
        {abram2020methods}
\bibfield{author}{\bibinfo{person}{Marissa~D. Abram}, \bibinfo{person}{Karen~T.
  Mancini}, {and} \bibinfo{person}{R.~David Parker}.}
  \bibinfo{year}{2020}\natexlab{}.
\newblock \showarticletitle{Methods to Integrate Natural Language Processing
  Into Qualitative Research}.
\newblock \bibinfo{journal}{\emph{International Journal of Qualitative
  Methods}} (\bibinfo{year}{2020}).
\newblock
\urldef\tempurl%
\url{https://doi.org/10.1177/1609406920984608}
\showDOI{\tempurl}


\bibitem[Aka et~al\mbox{.}(2021)]%
        {aka2021measuring}
\bibfield{author}{\bibinfo{person}{Osman Aka}, \bibinfo{person}{Ken Burke},
  \bibinfo{person}{Alex Bauerle}, \bibinfo{person}{Christina Greer}, {and}
  \bibinfo{person}{Margaret Mitchell}.} \bibinfo{year}{2021}\natexlab{}.
\newblock \showarticletitle{{Measuring Model Biases in the Absence of Ground
  Truth}}. In \bibinfo{booktitle}{\emph{Proceedings of the 2021 {AAAI}/{ACM}
  {Conference} on {AI}, {Ethics}, and {Society}}}
  \emph{(\bibinfo{series}{{AIES} '21})}. \bibinfo{publisher}{ACM}.
\newblock
\urldef\tempurl%
\url{https://doi.org/10.1145/3461702.3462557}
\showDOI{\tempurl}


\bibitem[Alfrink et~al\mbox{.}(2023)]%
        {alfrink2023contestable}
\bibfield{author}{\bibinfo{person}{Kars Alfrink}, \bibinfo{person}{Ianus
  Keller}, \bibinfo{person}{Neelke Doorn}, {and} \bibinfo{person}{Gerd
  Kortuem}.} \bibinfo{year}{2023}\natexlab{}.
\newblock \showarticletitle{{Contestable Camera Cars: A Speculative Design
  Exploration of Public AI That Is Open and Responsive to Dispute}}. In
  \bibinfo{booktitle}{\emph{Proceedings of the 2023 {CHI} {Conference} on
  {Human} {Factors} in {Computing} {Systems}}} \emph{(\bibinfo{series}{{CHI}
  '23})}. \bibinfo{publisher}{ACM}.
\newblock
\urldef\tempurl%
\url{https://doi.org/10.1145/3544548.3580984}
\showDOI{\tempurl}


\bibitem[Angelov et~al\mbox{.}(2021)]%
        {angelov2021explainable}
\bibfield{author}{\bibinfo{person}{Plamen~P. Angelov},
  \bibinfo{person}{Eduardo~A. Soares}, \bibinfo{person}{Richard Jiang},
  \bibinfo{person}{Nicholas~I. Arnold}, {and} \bibinfo{person}{Peter~M.
  Atkinson}.} \bibinfo{year}{2021}\natexlab{}.
\newblock \showarticletitle{{Explainable artificial intelligence: an analytical
  review}}.
\newblock \bibinfo{journal}{\emph{WIREs Data Mining and Knowledge Discovery}}
  (\bibinfo{year}{2021}).
\newblock
\urldef\tempurl%
\url{https://doi.org/10.1002/widm.1424}
\showDOI{\tempurl}


\bibitem[Aragon et~al\mbox{.}(2022)]%
        {aragon2022human}
\bibfield{author}{\bibinfo{person}{Cecilia Aragon}, \bibinfo{person}{Shion
  Guha}, \bibinfo{person}{Marina Kogan}, \bibinfo{person}{Michael Muller},
  {and} \bibinfo{person}{Gina Neff}.} \bibinfo{year}{2022}\natexlab{}.
\newblock \bibinfo{booktitle}{\emph{{Human-centered data science: An
  introduction}}}.
\newblock \bibinfo{publisher}{MIT Press}.
\newblock
\urldef\tempurl%
\url{https://mitpress.mit.edu/9780262543217/human-centered-data-science/}
\showURL{%
\tempurl}


\bibitem[Arthur et~al\mbox{.}(2008)]%
        {arthur2008racial}
\bibfield{author}{\bibinfo{person}{Melanie Arthur}, \bibinfo{person}{Jerris~R
  Hedges}, \bibinfo{person}{Craig~D Newgard}, \bibinfo{person}{Brain~S Diggs},
  {and} \bibinfo{person}{Richard~J Mullins}.} \bibinfo{year}{2008}\natexlab{}.
\newblock \showarticletitle{{Racial disparities in mortality among adults
  hospitalized after injury}}.
\newblock \bibinfo{journal}{\emph{Medical care}} (\bibinfo{year}{2008}).
\newblock
\urldef\tempurl%
\url{https://doi.org/10.1097/MLR.0b013e31815b9d8e}
\showDOI{\tempurl}


\bibitem[Ashktorab et~al\mbox{.}(2023)]%
        {ashktorab2023fairness}
\bibfield{author}{\bibinfo{person}{Zahra Ashktorab}, \bibinfo{person}{Benjamin
  Hoover}, \bibinfo{person}{Mayank Agarwal}, \bibinfo{person}{Casey Dugan},
  \bibinfo{person}{Werner Geyer}, \bibinfo{person}{Hao~Bang Yang}, {and}
  \bibinfo{person}{Mikhail Yurochkin}.} \bibinfo{year}{2023}\natexlab{}.
\newblock \showarticletitle{{Fairness Evaluation in Text Classification:
  Machine Learning Practitioner Perspectives of Individual and Group
  Fairness}}. In \bibinfo{booktitle}{\emph{Proceedings of the 2023 {CHI}
  {Conference} on {Human} {Factors} in {Computing} {Systems}}}
  \emph{(\bibinfo{series}{{CHI} '23})}. \bibinfo{publisher}{ACM}.
\newblock
\urldef\tempurl%
\url{https://doi.org/10.1145/3544548.3581227}
\showDOI{\tempurl}


\bibitem[Ashurst et~al\mbox{.}(2022)]%
        {ashurst2022disentangling}
\bibfield{author}{\bibinfo{person}{Carolyn Ashurst}, \bibinfo{person}{Solon
  Barocas}, \bibinfo{person}{Rosie Campbell}, {and} \bibinfo{person}{Deborah
  Raji}.} \bibinfo{year}{2022}\natexlab{}.
\newblock \showarticletitle{{Disentangling the Components of Ethical Research
  in Machine Learning}}. In \bibinfo{booktitle}{\emph{2022 {ACM} {Conference}
  on {Fairness}, {Accountability}, and {Transparency}}}
  \emph{(\bibinfo{series}{{FAccT} '22})}. \bibinfo{publisher}{ACM}.
\newblock
\urldef\tempurl%
\url{https://doi.org/10.1145/3531146.3533781}
\showDOI{\tempurl}


\bibitem[{Associated Press}(2021)]%
        {press2021judge}
\bibfield{author}{\bibinfo{person}{{Associated Press}}.}
  \bibinfo{year}{2021}\natexlab{}.
\newblock \bibinfo{booktitle}{\emph{{Judge approves \$650m settlement of
  privacy lawsuit against Facebook}}}.
\newblock {Guardian News \& Media Limited}.
\newblock
\urldef\tempurl%
\url{https://www.theguardian.com/technology/2021/feb/27/facebook-illinois-privacy-lawsuit-settlement}
\showURL{%
Retrieved January 2023 from \tempurl}


\bibitem[Ayobi et~al\mbox{.}(2023)]%
        {ayobi2023computational}
\bibfield{author}{\bibinfo{person}{Amid Ayobi}, \bibinfo{person}{Jacob Hughes},
  \bibinfo{person}{Christopher~J Duckworth}, \bibinfo{person}{Jakub~J Dylag},
  \bibinfo{person}{Sam James}, \bibinfo{person}{Paul Marshall},
  \bibinfo{person}{Matthew Guy}, \bibinfo{person}{Anitha Kumaran},
  \bibinfo{person}{Adriane Chapman}, \bibinfo{person}{Michael Boniface}, {and}
  \bibinfo{person}{Aisling~Ann O'Kane}.} \bibinfo{year}{2023}\natexlab{}.
\newblock \showarticletitle{{Computational Notebooks as Co-Design Tools:
  Engaging Young Adults Living with Diabetes, Family Carers, and Clinicians
  with Machine Learning Models}}. In \bibinfo{booktitle}{\emph{Proceedings of
  the 2023 {CHI} {Conference} on {Human} {Factors} in {Computing} {Systems}}}
  \emph{(\bibinfo{series}{{CHI} '23})}. \bibinfo{publisher}{ACM}.
\newblock
\urldef\tempurl%
\url{https://doi.org/10.1145/3544548.3581424}
\showDOI{\tempurl}


\bibitem[Baeza-Yates(2018)]%
        {baeza_yates2018bias}
\bibfield{author}{\bibinfo{person}{Ricardo Baeza-Yates}.}
  \bibinfo{year}{2018}\natexlab{}.
\newblock \showarticletitle{{Bias on the Web}}.
\newblock \bibinfo{journal}{\emph{Commun. ACM}} (\bibinfo{date}{May}
  \bibinfo{year}{2018}).
\newblock
\showISSN{0001-0782}
\urldef\tempurl%
\url{https://doi.org/10.1145/3209581}
\showDOI{\tempurl}


\bibitem[Balayn et~al\mbox{.}(2023)]%
        {balayn2023faulty}
\bibfield{author}{\bibinfo{person}{Agathe Balayn}, \bibinfo{person}{Natasa
  Rikalo}, \bibinfo{person}{Jie Yang}, {and} \bibinfo{person}{Alessandro
  Bozzon}.} \bibinfo{year}{2023}\natexlab{}.
\newblock \showarticletitle{Faulty or Ready? Handling Failures in Deep-Learning
  Computer Vision Models until Deployment: A Study of Practices, Challenges,
  and Needs}. In \bibinfo{booktitle}{\emph{Proceedings of the 2023 CHI
  Conference on Human Factors in Computing Systems}}
  \emph{(\bibinfo{series}{CHI '23})}. \bibinfo{publisher}{ACM}.
\newblock
\urldef\tempurl%
\url{https://doi.org/10.1145/3544548.3581555}
\showDOI{\tempurl}


\bibitem[Banovic et~al\mbox{.}(2023)]%
        {banovic2023being}
\bibfield{author}{\bibinfo{person}{Nikola Banovic}, \bibinfo{person}{Zhuoran
  Yang}, \bibinfo{person}{Aditya Ramesh}, {and} \bibinfo{person}{Alice Liu}.}
  \bibinfo{year}{2023}\natexlab{}.
\newblock \showarticletitle{{Being Trustworthy is Not Enough: How Untrustworthy
  Artificial Intelligence (AI) Can Deceive the End-Users and Gain Their
  Trust}}.
\newblock \bibinfo{journal}{\emph{Proc. ACM Hum.-Comput. Interact.}}
  (\bibinfo{date}{April} \bibinfo{year}{2023}).
\newblock
\urldef\tempurl%
\url{https://doi.org/10.1145/3579460}
\showDOI{\tempurl}


\bibitem[Bansal et~al\mbox{.}(2021)]%
        {bansal2021does}
\bibfield{author}{\bibinfo{person}{Gagan Bansal}, \bibinfo{person}{Tongshuang
  Wu}, \bibinfo{person}{Joyce Zhou}, \bibinfo{person}{Raymond Fok},
  \bibinfo{person}{Besmira Nushi}, \bibinfo{person}{Ece Kamar},
  \bibinfo{person}{Marco~Tulio Ribeiro}, {and} \bibinfo{person}{Daniel Weld}.}
  \bibinfo{year}{2021}\natexlab{}.
\newblock \showarticletitle{{Does the Whole Exceed Its Parts? The Effect of AI
  Explanations on Complementary Team Performance}}. In
  \bibinfo{booktitle}{\emph{Proceedings of the 2021 {CHI} {Conference} on
  {Human} {Factors} in {Computing} {Systems}}} \emph{(\bibinfo{series}{{CHI}
  '21})}. \bibinfo{publisher}{ACM}.
\newblock
\urldef\tempurl%
\url{https://doi.org/10.1145/3411764.3445717}
\showDOI{\tempurl}


\bibitem[{Barredo Arrieta} et~al\mbox{.}(2020)]%
        {arrieta2020explainable}
\bibfield{author}{\bibinfo{person}{Alejandro {Barredo Arrieta}},
  \bibinfo{person}{Natalia Díaz-Rodríguez}, \bibinfo{person}{Javier {Del
  Ser}}, \bibinfo{person}{Adrien Bennetot}, \bibinfo{person}{Siham Tabik},
  \bibinfo{person}{Alberto Barbado}, \bibinfo{person}{Salvador Garcia},
  \bibinfo{person}{Sergio Gil-Lopez}, \bibinfo{person}{Daniel Molina},
  \bibinfo{person}{Richard Benjamins}, \bibinfo{person}{Raja Chatila}, {and}
  \bibinfo{person}{Francisco Herrera}.} \bibinfo{year}{2020}\natexlab{}.
\newblock \showarticletitle{{Explainable Artificial Intelligence (XAI):
  Concepts, taxonomies, opportunities and challenges toward responsible AI}}.
\newblock \bibinfo{journal}{\emph{Information Fusion}} (\bibinfo{year}{2020}).
\newblock
\urldef\tempurl%
\url{https://doi.org/10.1016/j.inffus.2019.12.012}
\showDOI{\tempurl}


\bibitem[Baughan et~al\mbox{.}(2023)]%
        {baughan2023mixed}
\bibfield{author}{\bibinfo{person}{Amanda Baughan}, \bibinfo{person}{Xuezhi
  Wang}, \bibinfo{person}{Ariel Liu}, \bibinfo{person}{Allison Mercurio},
  \bibinfo{person}{Jilin Chen}, {and} \bibinfo{person}{Xiao Ma}.}
  \bibinfo{year}{2023}\natexlab{}.
\newblock \showarticletitle{{A Mixed-Methods Approach to Understanding User
  Trust after Voice Assistant Failures}}. In
  \bibinfo{booktitle}{\emph{Proceedings of the 2023 {CHI} {Conference} on
  {Human} {Factors} in {Computing} {Systems}}} \emph{(\bibinfo{series}{{CHI}
  '23})}. \bibinfo{publisher}{ACM}.
\newblock
\urldef\tempurl%
\url{https://doi.org/10.1145/3544548.3581152}
\showDOI{\tempurl}


\bibitem[Bawa et~al\mbox{.}(2020)]%
        {bawa2020do}
\bibfield{author}{\bibinfo{person}{Anshul Bawa}, \bibinfo{person}{Pranav
  Khadpe}, \bibinfo{person}{Pratik Joshi}, \bibinfo{person}{Kalika Bali}, {and}
  \bibinfo{person}{Monojit Choudhury}.} \bibinfo{year}{2020}\natexlab{}.
\newblock \showarticletitle{{Do Multilingual Users Prefer Chat-Bots That
  Code-Mix? Let's Nudge and Find Out!}}
\newblock \bibinfo{journal}{\emph{Proc. ACM Hum.-Comput. Interact.}}
  (\bibinfo{date}{May} \bibinfo{year}{2020}).
\newblock
\urldef\tempurl%
\url{https://doi.org/10.1145/3392846}
\showDOI{\tempurl}


\bibitem[Beede et~al\mbox{.}(2020)]%
        {beede2020human}
\bibfield{author}{\bibinfo{person}{Emma Beede}, \bibinfo{person}{Elizabeth
  Baylor}, \bibinfo{person}{Fred Hersch}, \bibinfo{person}{Anna Iurchenko},
  \bibinfo{person}{Lauren Wilcox}, \bibinfo{person}{Paisan Ruamviboonsuk},
  {and} \bibinfo{person}{Laura~M. Vardoulakis}.}
  \bibinfo{year}{2020}\natexlab{}.
\newblock \showarticletitle{{A Human-Centered Evaluation of a Deep Learning
  System Deployed in Clinics for the Detection of Diabetic Retinopathy}}. In
  \bibinfo{booktitle}{\emph{Proceedings of the 2020 {CHI} {Conference} on
  {Human} {Factors} in {Computing} {Systems}}} \emph{(\bibinfo{series}{{CHI}
  '20})}. \bibinfo{publisher}{ACM}.
\newblock
\urldef\tempurl%
\url{https://doi.org/10.1145/3313831.3376718}
\showDOI{\tempurl}


\bibitem[Bell et~al\mbox{.}(2022)]%
        {bell2022it}
\bibfield{author}{\bibinfo{person}{Andrew Bell}, \bibinfo{person}{Ian
  Solano-Kamaiko}, \bibinfo{person}{Oded Nov}, {and} \bibinfo{person}{Julia
  Stoyanovich}.} \bibinfo{year}{2022}\natexlab{}.
\newblock \showarticletitle{{It’s Just Not That Simple: An Empirical Study of
  the Accuracy-Explainability Trade-off in Machine Learning for Public
  Policy}}. In \bibinfo{booktitle}{\emph{2022 {ACM} {Conference} on {Fairness},
  {Accountability}, and {Transparency}}} \emph{(\bibinfo{series}{{FAccT}
  '22})}. \bibinfo{publisher}{ACM}.
\newblock
\urldef\tempurl%
\url{https://doi.org/10.1145/3531146.3533090}
\showDOI{\tempurl}


\bibitem[Bender and Friedman(2018)]%
        {bender2018data}
\bibfield{author}{\bibinfo{person}{Emily~M. Bender} {and}
  \bibinfo{person}{Batya Friedman}.} \bibinfo{year}{2018}\natexlab{}.
\newblock \showarticletitle{{Data Statements for Natural Language Processing:
  Toward Mitigating System Bias and Enabling Better Science}}.
\newblock \bibinfo{journal}{\emph{Transactions of the Association for
  Computational Linguistics}} (\bibinfo{year}{2018}).
\newblock
\urldef\tempurl%
\url{https://doi.org/10.1162/tacl_a_00041}
\showDOI{\tempurl}


\bibitem[Benthall and Goldenfein(2021)]%
        {benthall2021artificial}
\bibfield{author}{\bibinfo{person}{Sebastian Benthall} {and}
  \bibinfo{person}{Jake Goldenfein}.} \bibinfo{year}{2021}\natexlab{}.
\newblock \showarticletitle{{Artificial Intelligence and the Purpose of Social
  Systems}}. In \bibinfo{booktitle}{\emph{Proceedings of the 2021 {AAAI}/{ACM}
  {Conference} on {AI}, {Ethics}, and {Society}}}
  \emph{(\bibinfo{series}{{AIES} '21})}. \bibinfo{publisher}{ACM}.
\newblock
\urldef\tempurl%
\url{https://doi.org/10.1145/3461702.3462526}
\showDOI{\tempurl}


\bibitem[Bergram et~al\mbox{.}(2022)]%
        {bergram2022digital}
\bibfield{author}{\bibinfo{person}{Kristoffer Bergram}, \bibinfo{person}{Marija
  Djokovic}, \bibinfo{person}{Val\'{e}ry Bezen\c{c}on}, {and}
  \bibinfo{person}{Adrian Holzer}.} \bibinfo{year}{2022}\natexlab{}.
\newblock \showarticletitle{{The Digital Landscape of Nudging: A Systematic
  Literature Review of Empirical Research on Digital Nudges}}. In
  \bibinfo{booktitle}{\emph{Proceedings of the 2022 CHI Conference on Human
  Factors in Computing Systems}} \emph{(\bibinfo{series}{CHI '22})}.
  \bibinfo{publisher}{ACM}.
\newblock
\urldef\tempurl%
\url{https://doi.org/10.1145/3491102.3517638}
\showDOI{\tempurl}


\bibitem[Bertrand et~al\mbox{.}(2022)]%
        {bertrand2022how}
\bibfield{author}{\bibinfo{person}{Astrid Bertrand}, \bibinfo{person}{Rafik
  Belloum}, \bibinfo{person}{James~R. Eagan}, {and} \bibinfo{person}{Winston
  Maxwell}.} \bibinfo{year}{2022}\natexlab{}.
\newblock \showarticletitle{{How Cognitive Biases Affect XAI-Assisted
  Decision-Making: A Systematic Review}}. In
  \bibinfo{booktitle}{\emph{Proceedings of the 2022 {AAAI}/{ACM} {Conference}
  on {AI}, {Ethics}, and {Society}}} \emph{(\bibinfo{series}{{AIES} '22})}.
  \bibinfo{publisher}{ACM}.
\newblock
\urldef\tempurl%
\url{https://doi.org/10.1145/3514094.3534164}
\showDOI{\tempurl}


\bibitem[Bessen et~al\mbox{.}(2022)]%
        {bessen2022cost}
\bibfield{author}{\bibinfo{person}{James Bessen},
  \bibinfo{person}{Stephen~Michael Impink}, {and} \bibinfo{person}{Robert
  Seamans}.} \bibinfo{year}{2022}\natexlab{}.
\newblock \showarticletitle{{The Cost of Ethical AI Development for AI
  Startups}}. In \bibinfo{booktitle}{\emph{Proceedings of the 2022 {AAAI}/{ACM}
  {Conference} on {AI}, {Ethics}, and {Society}}}
  \emph{(\bibinfo{series}{{AIES} '22})}. \bibinfo{publisher}{ACM}.
\newblock
\urldef\tempurl%
\url{https://doi.org/10.1145/3514094.3534195}
\showDOI{\tempurl}


\bibitem[Bietti(2020)]%
        {bietti2020ethics}
\bibfield{author}{\bibinfo{person}{Elettra Bietti}.}
  \bibinfo{year}{2020}\natexlab{}.
\newblock \showarticletitle{{From Ethics Washing to Ethics Bashing: A View on
  Tech Ethics from within Moral Philosophy}}. In
  \bibinfo{booktitle}{\emph{Proceedings of the 2020 {Conference} on {Fairness},
  {Accountability}, and {Transparency}}} \emph{(\bibinfo{series}{{FAT}* '20})}.
  \bibinfo{publisher}{ACM}.
\newblock
\urldef\tempurl%
\url{https://doi.org/10.1145/3351095.3372860}
\showDOI{\tempurl}


\bibitem[Bingley et~al\mbox{.}(2023)]%
        {bingley2023where}
\bibfield{author}{\bibinfo{person}{William~J. Bingley},
  \bibinfo{person}{Caitlin Curtis}, \bibinfo{person}{Steven Lockey},
  \bibinfo{person}{Alina Bialkowski}, \bibinfo{person}{Nicole Gillespie},
  \bibinfo{person}{S.~Alexander Haslam}, \bibinfo{person}{Ryan~K.L. Ko},
  \bibinfo{person}{Niklas Steffens}, \bibinfo{person}{Janet Wiles}, {and}
  \bibinfo{person}{Peter Worthy}.} \bibinfo{year}{2023}\natexlab{}.
\newblock \showarticletitle{{Where is the human in human-centered AI? Insights
  from developer priorities and user experiences}}.
\newblock \bibinfo{journal}{\emph{Computers in Human Behavior}}
  (\bibinfo{year}{2023}).
\newblock
\showISSN{0747-5632}
\urldef\tempurl%
\url{https://doi.org/10.1016/j.chb.2022.107617}
\showDOI{\tempurl}


\bibitem[Black et~al\mbox{.}(2022)]%
        {black2022algorithmic}
\bibfield{author}{\bibinfo{person}{Emily Black}, \bibinfo{person}{Hadi Elzayn},
  \bibinfo{person}{Alexandra Chouldechova}, \bibinfo{person}{Jacob Goldin},
  {and} \bibinfo{person}{Daniel Ho}.} \bibinfo{year}{2022}\natexlab{}.
\newblock \showarticletitle{{Algorithmic Fairness and Vertical Equity: Income
  Fairness with IRS Tax Audit Models}}. In \bibinfo{booktitle}{\emph{2022 {ACM}
  {Conference} on {Fairness}, {Accountability}, and {Transparency}}}
  \emph{(\bibinfo{series}{{FAccT} '22})}. \bibinfo{publisher}{ACM}.
\newblock
\urldef\tempurl%
\url{https://doi.org/10.1145/3531146.3533204}
\showDOI{\tempurl}


\bibitem[Boag et~al\mbox{.}(2022)]%
        {boag2022tech}
\bibfield{author}{\bibinfo{person}{William Boag}, \bibinfo{person}{Harini
  Suresh}, \bibinfo{person}{Bianca Lepe}, {and} \bibinfo{person}{Catherine
  D'Ignazio}.} \bibinfo{year}{2022}\natexlab{}.
\newblock \showarticletitle{{Tech Worker Organizing for Power and
  Accountability}}. In \bibinfo{booktitle}{\emph{2022 {ACM} {Conference} on
  {Fairness}, {Accountability}, and {Transparency}}}
  \emph{(\bibinfo{series}{{FAccT} '22})}. \bibinfo{publisher}{ACM}.
\newblock
\urldef\tempurl%
\url{https://doi.org/10.1145/3531146.3533111}
\showDOI{\tempurl}


\bibitem[Boyarskaya et~al\mbox{.}(2020)]%
        {boyarskaya2020overcoming}
\bibfield{author}{\bibinfo{person}{Margarita Boyarskaya},
  \bibinfo{person}{Alexandra Olteanu}, {and} \bibinfo{person}{Kate Crawford}.}
  \bibinfo{year}{2020}\natexlab{}.
\newblock \showarticletitle{{Overcoming Failures of Imagination in AI Infused
  System Development and Deployment}}. In \bibinfo{booktitle}{\emph{In the
  Navigating the Broader Impacts of AI Research Workshop at NeurIPS 2020}}.
\newblock
\urldef\tempurl%
\url{https://www.microsoft.com/en-us/research/publication/overcoming-failures-of-imagination-in-ai-infused-system-development-and-deployment/}
\showURL{%
\tempurl}


\bibitem[Braun and Clarke(2006)]%
        {braun2006using}
\bibfield{author}{\bibinfo{person}{Virginia Braun} {and}
  \bibinfo{person}{Victoria Clarke}.} \bibinfo{year}{2006}\natexlab{}.
\newblock \showarticletitle{{Using thematic analysis in psychology}}.
\newblock \bibinfo{journal}{\emph{Qualitative Research in Psychology}}
  (\bibinfo{year}{2006}).
\newblock
\urldef\tempurl%
\url{https://doi.org/10.1191/1478088706qp063oa}
\showDOI{\tempurl}


\bibitem[Buolamwini and Gebru(2018)]%
        {buolamwini2018gendera}
\bibfield{author}{\bibinfo{person}{Joy Buolamwini} {and}
  \bibinfo{person}{Timnit Gebru}.} \bibinfo{year}{2018}\natexlab{}.
\newblock \showarticletitle{{Gender Shades: Intersectional Accuracy Disparities
  in Commercial Gender Classification}}. In
  \bibinfo{booktitle}{\emph{Proceedings of the 1st {Conference} on {Fairness},
  {Accountability} and {Transparency}}} \emph{(\bibinfo{series}{Proceedings of
  {Machine} {Learning} {Research}})},
  \bibfield{editor}{\bibinfo{person}{Sorelle~A. Friedler} {and}
  \bibinfo{person}{Christo Wilson}} (Eds.). \bibinfo{publisher}{PMLR}.
\newblock
\urldef\tempurl%
\url{https://proceedings.mlr.press/v81/buolamwini18a.html}
\showURL{%
\tempurl}


\bibitem[Burgess et~al\mbox{.}(2023)]%
        {burgess2023healthcare}
\bibfield{author}{\bibinfo{person}{Eleanor~R. Burgess}, \bibinfo{person}{Ivana
  Jankovic}, \bibinfo{person}{Melissa Austin}, \bibinfo{person}{Nancy Cai},
  \bibinfo{person}{Adela Kapuundefinedcińska}, \bibinfo{person}{Suzanne
  Currie}, \bibinfo{person}{J.~Marc Overhage}, \bibinfo{person}{Erika~S Poole},
  {and} \bibinfo{person}{Jofish Kaye}.} \bibinfo{year}{2023}\natexlab{}.
\newblock \showarticletitle{{Healthcare AI Treatment Decision Support: Design
  Principles to Enhance Clinician Adoption and Trust}}. In
  \bibinfo{booktitle}{\emph{Proceedings of the 2023 {CHI} {Conference} on
  {Human} {Factors} in {Computing} {Systems}}} \emph{(\bibinfo{series}{{CHI}
  '23})}. \bibinfo{publisher}{ACM}.
\newblock
\urldef\tempurl%
\url{https://doi.org/10.1145/3544548.3581251}
\showDOI{\tempurl}


\bibitem[Burrell et~al\mbox{.}(2019)]%
        {burrell2019when}
\bibfield{author}{\bibinfo{person}{Jenna Burrell}, \bibinfo{person}{Zoe Kahn},
  \bibinfo{person}{Anne Jonas}, {and} \bibinfo{person}{Daniel Griffin}.}
  \bibinfo{year}{2019}\natexlab{}.
\newblock \showarticletitle{{When Users Control the Algorithms: Values
  Expressed in Practices on Twitter}}.
\newblock \bibinfo{journal}{\emph{Proc. ACM Hum.-Comput. Interact.}}
  (\bibinfo{date}{Nov.} \bibinfo{year}{2019}).
\newblock
\urldef\tempurl%
\url{https://doi.org/10.1145/3359240}
\showDOI{\tempurl}


\bibitem[Buçinca et~al\mbox{.}(2021)]%
        {buçinca2021trust}
\bibfield{author}{\bibinfo{person}{Zana Buçinca},
  \bibinfo{person}{Maja~Barbara Malaya}, {and} \bibinfo{person}{Krzysztof~Z.
  Gajos}.} \bibinfo{year}{2021}\natexlab{}.
\newblock \showarticletitle{{To Trust or to Think: Cognitive Forcing Functions
  Can Reduce Overreliance on AI in AI-Assisted Decision-Making}}.
\newblock \bibinfo{journal}{\emph{Proc. ACM Hum.-Comput. Interact.}}
  (\bibinfo{date}{April} \bibinfo{year}{2021}).
\newblock
\urldef\tempurl%
\url{https://doi.org/10.1145/3449287}
\showDOI{\tempurl}


\bibitem[Byun et~al\mbox{.}(2023)]%
        {byun2023dispensing}
\bibfield{author}{\bibinfo{person}{Courtni Byun}, \bibinfo{person}{Piper
  Vasicek}, {and} \bibinfo{person}{Kevin Seppi}.}
  \bibinfo{year}{2023}\natexlab{}.
\newblock \showarticletitle{Dispensing with Humans in Human-Computer
  Interaction Research}. In \bibinfo{booktitle}{\emph{Extended Abstracts of the
  2023 CHI Conference on Human Factors in Computing Systems}}
  \emph{(\bibinfo{series}{CHI EA '23})}. \bibinfo{publisher}{ACM}, Article
  \bibinfo{articleno}{413}, \bibinfo{numpages}{26}~pages.
\newblock
\urldef\tempurl%
\url{https://doi.org/10.1145/3544549.3582749}
\showDOI{\tempurl}


\bibitem[Cabitza et~al\mbox{.}(2023)]%
        {cabitza2023ai}
\bibfield{author}{\bibinfo{person}{Federico Cabitza}, \bibinfo{person}{Andrea
  Campagner}, \bibinfo{person}{Riccardo Angius}, \bibinfo{person}{Chiara
  Natali}, {and} \bibinfo{person}{Carlo Reverberi}.}
  \bibinfo{year}{2023}\natexlab{}.
\newblock \showarticletitle{{AI Shall Have No Dominion: On How to Measure
  Technology Dominance in AI-Supported Human Decision-Making}}. In
  \bibinfo{booktitle}{\emph{Proceedings of the 2023 {CHI} {Conference} on
  {Human} {Factors} in {Computing} {Systems}}} \emph{(\bibinfo{series}{{CHI}
  '23})}. \bibinfo{publisher}{ACM}.
\newblock
\urldef\tempurl%
\url{https://doi.org/10.1145/3544548.3581095}
\showDOI{\tempurl}


\bibitem[Cabrera et~al\mbox{.}(2021)]%
        {cabrera2021discovering}
\bibfield{author}{\bibinfo{person}{Ángel~Alexander Cabrera},
  \bibinfo{person}{Abraham~J. Druck}, \bibinfo{person}{Jason~I. Hong}, {and}
  \bibinfo{person}{Adam Perer}.} \bibinfo{year}{2021}\natexlab{}.
\newblock \showarticletitle{{Discovering and Validating AI Errors With
  Crowdsourced Failure Reports}}.
\newblock \bibinfo{journal}{\emph{Proc. ACM Hum.-Comput. Interact.}}
  (\bibinfo{date}{Oct.} \bibinfo{year}{2021}).
\newblock
\urldef\tempurl%
\url{https://doi.org/10.1145/3479569}
\showDOI{\tempurl}


\bibitem[Calvo and Peters(2014)]%
        {calvo2014positive}
\bibfield{author}{\bibinfo{person}{Rafael~A Calvo} {and}
  \bibinfo{person}{Dorian Peters}.} \bibinfo{year}{2014}\natexlab{}.
\newblock \bibinfo{booktitle}{\emph{{Positive Computing: Technology for
  Wellbeing and Human Potential}}}.
\newblock \bibinfo{publisher}{MIT press}.
\newblock
\urldef\tempurl%
\url{https://mitpress.mit.edu/9780262533706/positive-computing/}
\showURL{%
\tempurl}


\bibitem[Cao et~al\mbox{.}(2023)]%
        {cao2023breaking}
\bibfield{author}{\bibinfo{person}{Hancheng Cao}, \bibinfo{person}{Yujie Lu},
  \bibinfo{person}{Yuting Deng}, \bibinfo{person}{Daniel Mcfarland}, {and}
  \bibinfo{person}{Michael~S. Bernstein}.} \bibinfo{year}{2023}\natexlab{}.
\newblock \showarticletitle{Breaking Out of the Ivory Tower: A Large-Scale
  Analysis of Patent Citations to HCI Research}. In
  \bibinfo{booktitle}{\emph{Proceedings of the 2023 CHI Conference on Human
  Factors in Computing Systems}} \emph{(\bibinfo{series}{CHI '23})}.
  \bibinfo{publisher}{ACM}.
\newblock
\urldef\tempurl%
\url{https://doi.org/10.1145/3544548.3581108}
\showDOI{\tempurl}


\bibitem[Capel and Brereton(2023)]%
        {capel2023what}
\bibfield{author}{\bibinfo{person}{Tara Capel} {and} \bibinfo{person}{Margot
  Brereton}.} \bibinfo{year}{2023}\natexlab{}.
\newblock \showarticletitle{{What is Human-Centered about Human-Centered AI? A
  Map of the Research Landscape}}. In \bibinfo{booktitle}{\emph{Proceedings of
  the 2023 {CHI} {Conference} on {Human} {Factors} in {Computing} {Systems}}}
  \emph{(\bibinfo{series}{{CHI} '23})}. \bibinfo{publisher}{ACM}, Article
  \bibinfo{articleno}{359}.
\newblock
\urldef\tempurl%
\url{https://doi.org/10.1145/3544548.3580959}
\showDOI{\tempurl}


\bibitem[Caton and Haas(2020)]%
        {caton2020fairness}
\bibfield{author}{\bibinfo{person}{Simon Caton} {and}
  \bibinfo{person}{Christian Haas}.} \bibinfo{year}{2020}\natexlab{}.
\newblock \bibinfo{title}{{Fairness in Machine Learning: A Survey}}.
\newblock
\newblock
\urldef\tempurl%
\url{https://doi.org/10.48550/ARXIV.2010.04053}
\showDOI{\tempurl}


\bibitem[Chancellor(2023)]%
        {chancellor2023practices}
\bibfield{author}{\bibinfo{person}{Stevie Chancellor}.}
  \bibinfo{year}{2023}\natexlab{}.
\newblock \showarticletitle{{Toward Practices for Human-Centered Machine
  Learning}}.
\newblock \bibinfo{journal}{\emph{Commun. ACM}} (\bibinfo{date}{Feb.}
  \bibinfo{year}{2023}).
\newblock
\showISSN{0001-0782}
\urldef\tempurl%
\url{https://doi.org/10.1145/3530987}
\showDOI{\tempurl}


\bibitem[Chancellor et~al\mbox{.}(2019)]%
        {chancellor2019who}
\bibfield{author}{\bibinfo{person}{Stevie Chancellor}, \bibinfo{person}{Eric
  P.~S. Baumer}, {and} \bibinfo{person}{Munmun De~Choudhury}.}
  \bibinfo{year}{2019}\natexlab{}.
\newblock \showarticletitle{{Who is the "Human" in Human-Centered Machine
  Learning: The Case of Predicting Mental Health from Social Media}}.
\newblock \bibinfo{journal}{\emph{Proc. ACM Hum.-Comput. Interact.}}
  (\bibinfo{date}{Nov.} \bibinfo{year}{2019}).
\newblock
\urldef\tempurl%
\url{https://doi.org/10.1145/3359249}
\showDOI{\tempurl}


\bibitem[Chen and Sundar(2023)]%
        {chen2023is}
\bibfield{author}{\bibinfo{person}{Cheng Chen} {and} \bibinfo{person}{S.~Shyam
  Sundar}.} \bibinfo{year}{2023}\natexlab{}.
\newblock \showarticletitle{{Is This AI Trained on Credible Data? The Effects
  of Labeling Quality and Performance Bias on User Trust}}. In
  \bibinfo{booktitle}{\emph{Proceedings of the 2023 {CHI} {Conference} on
  {Human} {Factors} in {Computing} {Systems}}} \emph{(\bibinfo{series}{{CHI}
  '23})}. \bibinfo{publisher}{ACM}.
\newblock
\urldef\tempurl%
\url{https://doi.org/10.1145/3544548.3580805}
\showDOI{\tempurl}


\bibitem[Cheng et~al\mbox{.}(2022)]%
        {cheng2022how}
\bibfield{author}{\bibinfo{person}{Hao-Fei Cheng}, \bibinfo{person}{Logan
  Stapleton}, \bibinfo{person}{Anna Kawakami}, \bibinfo{person}{Venkatesh
  Sivaraman}, \bibinfo{person}{Yanghuidi Cheng}, \bibinfo{person}{Diana Qing},
  \bibinfo{person}{Adam Perer}, \bibinfo{person}{Kenneth Holstein},
  \bibinfo{person}{Zhiwei~Steven Wu}, {and} \bibinfo{person}{Haiyi Zhu}.}
  \bibinfo{year}{2022}\natexlab{}.
\newblock \showarticletitle{{How Child Welfare Workers Reduce Racial
  Disparities in Algorithmic Decisions}}. In
  \bibinfo{booktitle}{\emph{Proceedings of the 2022 {CHI} {Conference} on
  {Human} {Factors} in {Computing} {Systems}}} \emph{(\bibinfo{series}{{CHI}
  '22})}. \bibinfo{publisher}{ACM}.
\newblock
\urldef\tempurl%
\url{https://doi.org/10.1145/3491102.3501831}
\showDOI{\tempurl}


\bibitem[Cheng et~al\mbox{.}(2021)]%
        {cheng2021soliciting}
\bibfield{author}{\bibinfo{person}{Hao-Fei Cheng}, \bibinfo{person}{Logan
  Stapleton}, \bibinfo{person}{Ruiqi Wang}, \bibinfo{person}{Paige Bullock},
  \bibinfo{person}{Alexandra Chouldechova}, \bibinfo{person}{Zhiwei
  Steven~Steven Wu}, {and} \bibinfo{person}{Haiyi Zhu}.}
  \bibinfo{year}{2021}\natexlab{}.
\newblock \showarticletitle{{Soliciting Stakeholders’ Fairness Notions in
  Child Maltreatment Predictive Systems}}. In
  \bibinfo{booktitle}{\emph{Proceedings of the 2021 {CHI} {Conference} on
  {Human} {Factors} in {Computing} {Systems}}} \emph{(\bibinfo{series}{{CHI}
  '21})}. \bibinfo{publisher}{ACM}.
\newblock
\urldef\tempurl%
\url{https://doi.org/10.1145/3411764.3445308}
\showDOI{\tempurl}


\bibitem[Chi et~al\mbox{.}(2021)]%
        {chi2021reconfiguring}
\bibfield{author}{\bibinfo{person}{Nicole Chi}, \bibinfo{person}{Emma Lurie},
  {and} \bibinfo{person}{Deirdre~K. Mulligan}.}
  \bibinfo{year}{2021}\natexlab{}.
\newblock \showarticletitle{{Reconfiguring Diversity and Inclusion for AI
  Ethics}}. In \bibinfo{booktitle}{\emph{Proceedings of the 2021 {AAAI}/{ACM}
  {Conference} on {AI}, {Ethics}, and {Society}}}
  \emph{(\bibinfo{series}{{AIES} '21})}. \bibinfo{publisher}{ACM}.
\newblock
\urldef\tempurl%
\url{https://doi.org/10.1145/3461702.3462622}
\showDOI{\tempurl}


\bibitem[Chmielinski et~al\mbox{.}(2022)]%
        {chmielinski2022dataset}
\bibfield{author}{\bibinfo{person}{Kasia~S. Chmielinski},
  \bibinfo{person}{Sarah Newman}, \bibinfo{person}{Matt Taylor},
  \bibinfo{person}{Josh Joseph}, \bibinfo{person}{Kemi Thomas},
  \bibinfo{person}{Jessica Yurkofsky}, {and} \bibinfo{person}{Yue~Chelsea
  Qiu}.} \bibinfo{year}{2022}\natexlab{}.
\newblock \bibinfo{title}{{The Dataset Nutrition Label (2nd Gen): Leveraging
  Context to Mitigate Harms in Artificial Intelligence}}.
\newblock
\newblock
\urldef\tempurl%
\url{https://doi.org/10.48550/ARXIV.2201.03954}
\showDOI{\tempurl}


\bibitem[Choi et~al\mbox{.}(2023)]%
        {choi2023creator}
\bibfield{author}{\bibinfo{person}{Yoonseo Choi}, \bibinfo{person}{Eun~Jeong
  Kang}, \bibinfo{person}{Min~Kyung Lee}, {and} \bibinfo{person}{Juho Kim}.}
  \bibinfo{year}{2023}\natexlab{}.
\newblock \showarticletitle{{Creator-Friendly Algorithms: Behaviors,
  Challenges, and Design Opportunities in Algorithmic Platforms}}. In
  \bibinfo{booktitle}{\emph{Proceedings of the 2023 {CHI} {Conference} on
  {Human} {Factors} in {Computing} {Systems}}} \emph{(\bibinfo{series}{{CHI}
  '23})}. \bibinfo{publisher}{ACM}.
\newblock
\urldef\tempurl%
\url{https://doi.org/10.1145/3544548.3581386}
\showDOI{\tempurl}


\bibitem[Chu et~al\mbox{.}(2023)]%
        {chu2023work}
\bibfield{author}{\bibinfo{person}{Mengdi Chu}, \bibinfo{person}{Keyu Zong},
  \bibinfo{person}{Xin Shu}, \bibinfo{person}{Jiangtao Gong},
  \bibinfo{person}{Zhicong Lu}, \bibinfo{person}{Kaimin Guo},
  \bibinfo{person}{Xinyi Dai}, {and} \bibinfo{person}{Guyue Zhou}.}
  \bibinfo{year}{2023}\natexlab{}.
\newblock \showarticletitle{{Work with AI and Work for AI: Autonomous Vehicle
  Safety Drivers’ Lived Experiences}}. In
  \bibinfo{booktitle}{\emph{Proceedings of the 2023 {CHI} {Conference} on
  {Human} {Factors} in {Computing} {Systems}}} \emph{(\bibinfo{series}{{CHI}
  '23})}. \bibinfo{publisher}{ACM}.
\newblock
\urldef\tempurl%
\url{https://doi.org/10.1145/3544548.3581564}
\showDOI{\tempurl}


\bibitem[Constantinides and Quercia(2022)]%
        {constantinides2022good}
\bibfield{author}{\bibinfo{person}{Marios Constantinides} {and}
  \bibinfo{person}{Daniele Quercia}.} \bibinfo{year}{2022}\natexlab{}.
\newblock \showarticletitle{{Good Intentions, Bad Inventions: How Employees
  Judge Pervasive Technologies in the Workplace}}.
\newblock \bibinfo{journal}{\emph{IEEE Pervasive Computing}}
  (\bibinfo{year}{2022}).
\newblock
\urldef\tempurl%
\url{https://doi.org/10.1109/MPRV.2022.3217408}
\showDOI{\tempurl}


\bibitem[Costanza-Chock et~al\mbox{.}(2022)]%
        {costanza_chock2022who}
\bibfield{author}{\bibinfo{person}{Sasha Costanza-Chock},
  \bibinfo{person}{Inioluwa~Deborah Raji}, {and} \bibinfo{person}{Joy
  Buolamwini}.} \bibinfo{year}{2022}\natexlab{}.
\newblock \showarticletitle{{Who Audits the Auditors? Recommendations from a
  Field Scan of the Algorithmic Auditing Ecosystem}}. In
  \bibinfo{booktitle}{\emph{2022 {ACM} {Conference} on {Fairness},
  {Accountability}, and {Transparency}}} \emph{(\bibinfo{series}{{FAccT}
  '22})}. \bibinfo{publisher}{ACM}.
\newblock
\urldef\tempurl%
\url{https://doi.org/10.1145/3531146.3533213}
\showDOI{\tempurl}


\bibitem[Cranor and Garfinkel(2005)]%
        {cranor2005security}
\bibfield{author}{\bibinfo{person}{Lorrie~Faith Cranor} {and}
  \bibinfo{person}{Simson Garfinkel}.} \bibinfo{year}{2005}\natexlab{}.
\newblock \bibinfo{booktitle}{\emph{{Security and Usability: Designing Secure
  Systems That People Can Use}}}.
\newblock \bibinfo{publisher}{O'Reilly Media, Inc.}
\newblock
\urldef\tempurl%
\url{https://www.oreilly.com/library/view/security-and-usability/0596008279/}
\showURL{%
\tempurl}


\bibitem[Crisan et~al\mbox{.}(2022)]%
        {crisan2022interactive}
\bibfield{author}{\bibinfo{person}{Anamaria Crisan}, \bibinfo{person}{Margaret
  Drouhard}, \bibinfo{person}{Jesse Vig}, {and} \bibinfo{person}{Nazneen
  Rajani}.} \bibinfo{year}{2022}\natexlab{}.
\newblock \showarticletitle{{Interactive Model Cards: A Human-Centered Approach
  to Model Documentation}}. In \bibinfo{booktitle}{\emph{2022 {ACM}
  {Conference} on {Fairness}, {Accountability}, and {Transparency}}}
  \emph{(\bibinfo{series}{{FAccT} '22})}. \bibinfo{publisher}{ACM}.
\newblock
\urldef\tempurl%
\url{https://doi.org/10.1145/3531146.3533108}
\showDOI{\tempurl}


\bibitem[Cruz~Cortés and Ghosh(2020)]%
        {cruz_cortés2020invitation}
\bibfield{author}{\bibinfo{person}{Efrén Cruz~Cortés} {and}
  \bibinfo{person}{Debashis Ghosh}.} \bibinfo{year}{2020}\natexlab{}.
\newblock \showarticletitle{{An Invitation to System-Wide Algorithmic
  Fairness}}. In \bibinfo{booktitle}{\emph{Proceedings of the {AAAI}/{ACM}
  {Conference} on {AI}, {Ethics}, and {Society}}}
  \emph{(\bibinfo{series}{{AIES} '20})}. \bibinfo{publisher}{ACM}.
\newblock
\urldef\tempurl%
\url{https://doi.org/10.1145/3375627.3375860}
\showDOI{\tempurl}


\bibitem[De-Arteaga et~al\mbox{.}(2020)]%
        {de_arteaga2020case}
\bibfield{author}{\bibinfo{person}{Maria De-Arteaga}, \bibinfo{person}{Riccardo
  Fogliato}, {and} \bibinfo{person}{Alexandra Chouldechova}.}
  \bibinfo{year}{2020}\natexlab{}.
\newblock \showarticletitle{{A Case for Humans-in-the-Loop: Decisions in the
  Presence of Erroneous Algorithmic Scores}}. In
  \bibinfo{booktitle}{\emph{Proceedings of the 2020 CHI Conference on Human
  Factors in Computing Systems}} \emph{(\bibinfo{series}{CHI '20})}.
  \bibinfo{publisher}{ACM}.
\newblock
\urldef\tempurl%
\url{https://doi.org/10.1145/3313831.3376638}
\showDOI{\tempurl}


\bibitem[Delgado et~al\mbox{.}(2022)]%
        {delgado2022uncommon}
\bibfield{author}{\bibinfo{person}{Fernando Delgado}, \bibinfo{person}{Solon
  Barocas}, {and} \bibinfo{person}{Karen Levy}.}
  \bibinfo{year}{2022}\natexlab{}.
\newblock \showarticletitle{{An Uncommon Task: Participatory Design in Legal
  AI}}.
\newblock \bibinfo{journal}{\emph{Proc. ACM Hum.-Comput. Interact.}}, Article
  \bibinfo{articleno}{51} (\bibinfo{date}{April} \bibinfo{year}{2022}).
\newblock
\urldef\tempurl%
\url{https://doi.org/10.1145/3512898}
\showDOI{\tempurl}


\bibitem[Deng et~al\mbox{.}(2023)]%
        {deng2023understanding}
\bibfield{author}{\bibinfo{person}{Wesley~Hanwen Deng}, \bibinfo{person}{Boyuan
  Guo}, \bibinfo{person}{Alicia Devrio}, \bibinfo{person}{Hong Shen},
  \bibinfo{person}{Motahhare Eslami}, {and} \bibinfo{person}{Kenneth
  Holstein}.} \bibinfo{year}{2023}\natexlab{}.
\newblock \showarticletitle{{Understanding Practices, Challenges, and
  Opportunities for User-Engaged Algorithm Auditing in Industry Practice}}. In
  \bibinfo{booktitle}{\emph{Proceedings of the 2023 {CHI} {Conference} on
  {Human} {Factors} in {Computing} {Systems}}} \emph{(\bibinfo{series}{{CHI}
  '23})}. \bibinfo{publisher}{ACM}.
\newblock
\urldef\tempurl%
\url{https://doi.org/10.1145/3544548.3581026}
\showDOI{\tempurl}


\bibitem[Deshpande and Sharp(2022)]%
        {deshpande2022responsible}
\bibfield{author}{\bibinfo{person}{Advait Deshpande} {and}
  \bibinfo{person}{Helen Sharp}.} \bibinfo{year}{2022}\natexlab{}.
\newblock \showarticletitle{{Responsible AI Systems: Who Are the
  Stakeholders?}}. In \bibinfo{booktitle}{\emph{Proceedings of the 2022
  {AAAI}/{ACM} {Conference} on {AI}, {Ethics}, and {Society}}}
  \emph{(\bibinfo{series}{{AIES} '22})}. \bibinfo{publisher}{ACM}.
\newblock
\urldef\tempurl%
\url{https://doi.org/10.1145/3514094.3534187}
\showDOI{\tempurl}


\bibitem[Dieber and Kirrane(2020)]%
        {dieber2020why}
\bibfield{author}{\bibinfo{person}{Jürgen Dieber} {and}
  \bibinfo{person}{Sabrina Kirrane}.} \bibinfo{year}{2020}\natexlab{}.
\newblock \bibinfo{title}{{Why model why? Assessing the strengths and
  limitations of LIME}}.
\newblock
\newblock
\urldef\tempurl%
\url{https://doi.org/10.48550/ARXIV.2012.00093}
\showDOI{\tempurl}


\bibitem[Docherty and Biega(2022)]%
        {docherty2022re}
\bibfield{author}{\bibinfo{person}{Niall Docherty} {and}
  \bibinfo{person}{Asia~J. Biega}.} \bibinfo{year}{2022}\natexlab{}.
\newblock \showarticletitle{{(Re)Politicizing Digital Well-Being: Beyond User
  Engagements}}. In \bibinfo{booktitle}{\emph{Proceedings of the 2022 {CHI}
  {Conference} on {Human} {Factors} in {Computing} {Systems}}}
  \emph{(\bibinfo{series}{{CHI} '22})}. \bibinfo{publisher}{ACM}.
\newblock
\urldef\tempurl%
\url{https://doi.org/10.1145/3491102.3501857}
\showDOI{\tempurl}


\bibitem[Donahue et~al\mbox{.}(2022)]%
        {donahue2022human}
\bibfield{author}{\bibinfo{person}{Kate Donahue}, \bibinfo{person}{Alexandra
  Chouldechova}, {and} \bibinfo{person}{Krishnaram Kenthapadi}.}
  \bibinfo{year}{2022}\natexlab{}.
\newblock \showarticletitle{{Human-Algorithm Collaboration: Achieving
  Complementarity and Avoiding Unfairness}}. In \bibinfo{booktitle}{\emph{2022
  {ACM} {Conference} on {Fairness}, {Accountability}, and {Transparency}}}
  \emph{(\bibinfo{series}{{FAccT} '22})}. \bibinfo{publisher}{ACM}.
\newblock
\urldef\tempurl%
\url{https://doi.org/10.1145/3531146.3533221}
\showDOI{\tempurl}


\bibitem[Ehsan et~al\mbox{.}(2021a)]%
        {ehsan2021expanding}
\bibfield{author}{\bibinfo{person}{Upol Ehsan}, \bibinfo{person}{Q.~Vera Liao},
  \bibinfo{person}{Michael Muller}, \bibinfo{person}{Mark~O. Riedl}, {and}
  \bibinfo{person}{Justin~D. Weisz}.} \bibinfo{year}{2021}\natexlab{a}.
\newblock \showarticletitle{{Expanding Explainability: Towards Social
  Transparency in AI Systems}}. In \bibinfo{booktitle}{\emph{Proceedings of the
  2021 {CHI} {Conference} on {Human} {Factors} in {Computing} {Systems}}}
  \emph{(\bibinfo{series}{{CHI} '21})}. \bibinfo{publisher}{ACM}.
\newblock
\urldef\tempurl%
\url{https://doi.org/10.1145/3411764.3445188}
\showDOI{\tempurl}


\bibitem[Ehsan et~al\mbox{.}(2023)]%
        {ehsan2023charting}
\bibfield{author}{\bibinfo{person}{Upol Ehsan}, \bibinfo{person}{Koustuv Saha},
  \bibinfo{person}{Munmun De~Choudhury}, {and} \bibinfo{person}{Mark~O.
  Riedl}.} \bibinfo{year}{2023}\natexlab{}.
\newblock \showarticletitle{{Charting the Sociotechnical Gap in Explainable AI:
  A Framework to Address the Gap in XAI}}.
\newblock \bibinfo{journal}{\emph{Proc. ACM Hum.-Comput. Interact.}}
  (\bibinfo{date}{April} \bibinfo{year}{2023}).
\newblock
\urldef\tempurl%
\url{https://doi.org/10.1145/3579467}
\showDOI{\tempurl}


\bibitem[Ehsan et~al\mbox{.}(2021b)]%
        {ehsan2021operationalizing}
\bibfield{author}{\bibinfo{person}{Upol Ehsan}, \bibinfo{person}{Philipp
  Wintersberger}, \bibinfo{person}{Q.~Vera Liao}, \bibinfo{person}{Martina
  Mara}, \bibinfo{person}{Marc Streit}, \bibinfo{person}{Sandra Wachter},
  \bibinfo{person}{Andreas Riener}, {and} \bibinfo{person}{Mark~O. Riedl}.}
  \bibinfo{year}{2021}\natexlab{b}.
\newblock \showarticletitle{{Operationalizing Human-Centered Perspectives in
  Explainable AI}}. In \bibinfo{booktitle}{\emph{Extended Abstracts of the 2021
  CHI Conference on Human Factors in Computing Systems}}
  \emph{(\bibinfo{series}{CHI EA '21})}. \bibinfo{publisher}{ACM}.
\newblock
\urldef\tempurl%
\url{https://doi.org/10.1145/3411763.3441342}
\showDOI{\tempurl}


\bibitem[Eicher et~al\mbox{.}(2018)]%
        {eicher2018jill}
\bibfield{author}{\bibinfo{person}{Bobbie Eicher}, \bibinfo{person}{Lalith
  Polepeddi}, {and} \bibinfo{person}{Ashok Goel}.}
  \bibinfo{year}{2018}\natexlab{}.
\newblock \showarticletitle{{Jill Watson Doesn't Care If You're Pregnant:
  Grounding AI Ethics in Empirical Studies}}. In
  \bibinfo{booktitle}{\emph{Proceedings of the 2018 {AAAI}/{ACM} {Conference}
  on {AI}, {Ethics}, and {Society}}} \emph{(\bibinfo{series}{{AIES} '18})}.
  \bibinfo{publisher}{ACM}.
\newblock
\urldef\tempurl%
\url{https://doi.org/10.1145/3278721.3278760}
\showDOI{\tempurl}


\bibitem[Elsayed-Ali et~al\mbox{.}(2023)]%
        {elsayed_ali2023responsible}
\bibfield{author}{\bibinfo{person}{Salma Elsayed-Ali}, \bibinfo{person}{Sara~E
  Berger}, \bibinfo{person}{Vagner Figueredo~De Santana}, {and}
  \bibinfo{person}{Juana~Catalina Becerra~Sandoval}.}
  \bibinfo{year}{2023}\natexlab{}.
\newblock \showarticletitle{{Responsible \&amp; Inclusive Cards: An Online Card
  Tool to Promote Critical Reflection in Technology Industry Work Practices}}.
  In \bibinfo{booktitle}{\emph{Proceedings of the 2023 {CHI} {Conference} on
  {Human} {Factors} in {Computing} {Systems}}} \emph{(\bibinfo{series}{{CHI}
  '23})}. \bibinfo{publisher}{ACM}.
\newblock
\urldef\tempurl%
\url{https://doi.org/10.1145/3544548.3580771}
\showDOI{\tempurl}


\bibitem[Engelmann et~al\mbox{.}(2022)]%
        {engelmann2022what}
\bibfield{author}{\bibinfo{person}{Severin Engelmann}, \bibinfo{person}{Chiara
  Ullstein}, \bibinfo{person}{Orestis Papakyriakopoulos}, {and}
  \bibinfo{person}{Jens Grossklags}.} \bibinfo{year}{2022}\natexlab{}.
\newblock \showarticletitle{{What People Think AI Should Infer From Faces}}. In
  \bibinfo{booktitle}{\emph{2022 {ACM} {Conference} on {Fairness},
  {Accountability}, and {Transparency}}} \emph{(\bibinfo{series}{{FAccT}
  '22})}. \bibinfo{publisher}{ACM}.
\newblock
\urldef\tempurl%
\url{https://doi.org/10.1145/3531146.3533080}
\showDOI{\tempurl}


\bibitem[Erdélyi and Erdélyi(2020)]%
        {erdélyi2020ai}
\bibfield{author}{\bibinfo{person}{Olivia~J. Erdélyi} {and}
  \bibinfo{person}{Gábor Erdélyi}.} \bibinfo{year}{2020}\natexlab{}.
\newblock \showarticletitle{{The AI Liability Puzzle and a Fund-Based
  Work-Around}}. In \bibinfo{booktitle}{\emph{Proceedings of the {AAAI}/{ACM}
  {Conference} on {AI}, {Ethics}, and {Society}}}
  \emph{(\bibinfo{series}{{AIES} '20})}. \bibinfo{publisher}{ACM}.
\newblock
\urldef\tempurl%
\url{https://doi.org/10.1145/3375627.3375806}
\showDOI{\tempurl}


\bibitem[Feinberg(2017)]%
        {feinberg2017design}
\bibfield{author}{\bibinfo{person}{Melanie Feinberg}.}
  \bibinfo{year}{2017}\natexlab{}.
\newblock \showarticletitle{{A Design Perspective on Data}}. In
  \bibinfo{booktitle}{\emph{Proceedings of the 2017 CHI Conference on Human
  Factors in Computing Systems}} \emph{(\bibinfo{series}{CHI '17})}.
  \bibinfo{publisher}{ACM}.
\newblock
\urldef\tempurl%
\url{https://doi.org/10.1145/3025453.3025837}
\showDOI{\tempurl}


\bibitem[Feuston and Brubaker(2021)]%
        {feuston2021putting}
\bibfield{author}{\bibinfo{person}{Jessica~L. Feuston} {and}
  \bibinfo{person}{Jed~R. Brubaker}.} \bibinfo{year}{2021}\natexlab{}.
\newblock \showarticletitle{{Putting Tools in Their Place: The Role of Time and
  Perspective in Human-AI Collaboration for Qualitative Analysis}}.
\newblock \bibinfo{journal}{\emph{Proc. ACM Hum.-Comput. Interact.}}
  (\bibinfo{date}{Oct.} \bibinfo{year}{2021}).
\newblock
\urldef\tempurl%
\url{https://doi.org/10.1145/3479856}
\showDOI{\tempurl}


\bibitem[Fjeld et~al\mbox{.}(2020)]%
        {fjeld2020principled}
\bibfield{author}{\bibinfo{person}{Jessica Fjeld}, \bibinfo{person}{Nele
  Achten}, \bibinfo{person}{Hannah Hilligoss}, \bibinfo{person}{Adam Nagy},
  {and} \bibinfo{person}{Madhulika Srikumar}.} \bibinfo{year}{2020}\natexlab{}.
\newblock \showarticletitle{{Principled artificial intelligence: Mapping
  consensus in ethical and rights-based approaches to principles for AI}}.
\newblock \bibinfo{journal}{\emph{Berkman Klein Center Research Publication}}
  (\bibinfo{year}{2020}).
\newblock
\urldef\tempurl%
\url{https://doi.org/10.2139/ssrn.3518482}
\showDOI{\tempurl}


\bibitem[Flathmann et~al\mbox{.}(2021)]%
        {flathmann2021modeling}
\bibfield{author}{\bibinfo{person}{Christopher Flathmann},
  \bibinfo{person}{Beau~G. Schelble}, \bibinfo{person}{Rui Zhang}, {and}
  \bibinfo{person}{Nathan~J. McNeese}.} \bibinfo{year}{2021}\natexlab{}.
\newblock \showarticletitle{{Modeling and Guiding the Creation of Ethical
  Human-AI Teams}}. In \bibinfo{booktitle}{\emph{Proceedings of the 2021
  {AAAI}/{ACM} {Conference} on {AI}, {Ethics}, and {Society}}}
  \emph{(\bibinfo{series}{{AIES} '21})}. \bibinfo{publisher}{ACM}.
\newblock
\urldef\tempurl%
\url{https://doi.org/10.1145/3461702.3462573}
\showDOI{\tempurl}


\bibitem[Fox et~al\mbox{.}(2023)]%
        {fox2023patchwork}
\bibfield{author}{\bibinfo{person}{Sarah~E. Fox}, \bibinfo{person}{Samantha
  Shorey}, \bibinfo{person}{Esther~Y. Kang}, \bibinfo{person}{Dominique
  Montiel~Valle}, {and} \bibinfo{person}{Estefania Rodriguez}.}
  \bibinfo{year}{2023}\natexlab{}.
\newblock \showarticletitle{{Patchwork: The Hidden, Human Labor of AI
  Integration within Essential Work}}.
\newblock \bibinfo{journal}{\emph{Proc. ACM Hum.-Comput. Interact.}}
  (\bibinfo{date}{April} \bibinfo{year}{2023}).
\newblock
\urldef\tempurl%
\url{https://doi.org/10.1145/3579514}
\showDOI{\tempurl}


\bibitem[Frluckaj et~al\mbox{.}(2022)]%
        {frluckaj2022gender}
\bibfield{author}{\bibinfo{person}{Hana Frluckaj}, \bibinfo{person}{Laura
  Dabbish}, \bibinfo{person}{David~Gray Widder},
  \bibinfo{person}{Huilian~Sophie Qiu}, {and} \bibinfo{person}{James~D.
  Herbsleb}.} \bibinfo{year}{2022}\natexlab{}.
\newblock \showarticletitle{{Gender and Participation in Open Source Software
  Development}}.
\newblock \bibinfo{journal}{\emph{Proc. ACM Hum.-Comput. Interact.}}
  (\bibinfo{date}{Nov.} \bibinfo{year}{2022}).
\newblock
\urldef\tempurl%
\url{https://doi.org/10.1145/3555190}
\showDOI{\tempurl}


\bibitem[Garcia and Cifor(2019)]%
        {garcia2019expanding}
\bibfield{author}{\bibinfo{person}{Patricia Garcia} {and}
  \bibinfo{person}{Marika Cifor}.} \bibinfo{year}{2019}\natexlab{}.
\newblock \showarticletitle{{Expanding Our Reflexive Toolbox: Collaborative
  Possibilities for Examining Socio-Technical Systems Using Duoethnography}}.
\newblock \bibinfo{journal}{\emph{Proc. ACM Hum.-Comput. Interact.}}, Article
  \bibinfo{articleno}{190} (\bibinfo{date}{Nov.} \bibinfo{year}{2019}).
\newblock
\urldef\tempurl%
\url{https://doi.org/10.1145/3359292}
\showDOI{\tempurl}


\bibitem[Gebru et~al\mbox{.}(2021)]%
        {gebru2021datasheets}
\bibfield{author}{\bibinfo{person}{Timnit Gebru}, \bibinfo{person}{Jamie
  Morgenstern}, \bibinfo{person}{Briana Vecchione},
  \bibinfo{person}{Jennifer~Wortman Vaughan}, \bibinfo{person}{Hanna Wallach},
  \bibinfo{person}{Hal~Daum\'{e} III}, {and} \bibinfo{person}{Kate Crawford}.}
  \bibinfo{year}{2021}\natexlab{}.
\newblock \showarticletitle{{Datasheets for Datasets}}.
\newblock \bibinfo{journal}{\emph{Commun. ACM}} (\bibinfo{date}{Nov.}
  \bibinfo{year}{2021}).
\newblock
\urldef\tempurl%
\url{https://doi.org/10.1145/3458723}
\showDOI{\tempurl}


\bibitem[Gee(2021)]%
        {gee2021reducing}
\bibfield{author}{\bibinfo{person}{Harvey Gee}.}
  \bibinfo{year}{2021}\natexlab{}.
\newblock \bibinfo{title}{{Reducing Gun Violence with ShotSpotter Gunshot
  Detection Technology and Community-Based Plans: What Works?}}
\newblock
\newblock
\urldef\tempurl%
\url{https://scholarsbank.uoregon.edu/xmlui/handle/1794/27170}
\showURL{%
\tempurl}


\bibitem[{Geoffrey A. Fowler}(2021)]%
        {fowler2021there}
\bibfield{author}{\bibinfo{person}{{Geoffrey A. Fowler}}.}
  \bibinfo{year}{2021}\natexlab{}.
\newblock \bibinfo{booktitle}{\emph{{There's no escape from Facebook, even if
  you don't use it}}}.
\newblock {The Washington Post}.
\newblock
\urldef\tempurl%
\url{https://www.washingtonpost.com/technology/2021/08/29/facebook-privacy-monopoly/}
\showURL{%
Retrieved January 2023 from \tempurl}


\bibitem[{Google}(2022)]%
        {e2022responsible}
\bibfield{author}{\bibinfo{person}{{Google}}.} \bibinfo{year}{2022}\natexlab{}.
\newblock \bibinfo{booktitle}{\emph{{Responsible AI practices}}}.
\newblock
\urldef\tempurl%
\url{https://ai.google/responsibilities/responsible-ai-practices/}
\showURL{%
Retrieved February 2023 from \tempurl}


\bibitem[Gray and Suri(2019)]%
        {gray2019ghost}
\bibfield{author}{\bibinfo{person}{Mary~L Gray} {and}
  \bibinfo{person}{Siddharth Suri}.} \bibinfo{year}{2019}\natexlab{}.
\newblock \bibinfo{booktitle}{\emph{{Ghost Work: How to Stop Silicon Valley
  from Building a New Global Underclass}}}.
\newblock \bibinfo{publisher}{Eamon Dolan Books}.
\newblock
\urldef\tempurl%
\url{https://ghostwork.info/}
\showURL{%
\tempurl}


\bibitem[Greiffenhagen et~al\mbox{.}(2023)]%
        {greiffenhagen2023work}
\bibfield{author}{\bibinfo{person}{Christian Greiffenhagen},
  \bibinfo{person}{Xinzhi Xu}, {and} \bibinfo{person}{Stuart Reeves}.}
  \bibinfo{year}{2023}\natexlab{}.
\newblock \showarticletitle{{The Work to Make Facial Recognition Work}}.
\newblock \bibinfo{journal}{\emph{Proc. ACM Hum.-Comput. Interact.}}
  (\bibinfo{date}{April} \bibinfo{year}{2023}).
\newblock
\urldef\tempurl%
\url{https://doi.org/10.1145/3579531}
\showDOI{\tempurl}


\bibitem[Gu et~al\mbox{.}(2021)]%
        {gu2021lessons}
\bibfield{author}{\bibinfo{person}{Hongyan Gu}, \bibinfo{person}{Jingbin
  Huang}, \bibinfo{person}{Lauren Hung}, {and} \bibinfo{person}{Xiang~'Anthony'
  Chen}.} \bibinfo{year}{2021}\natexlab{}.
\newblock \showarticletitle{{Lessons Learned from Designing an AI-Enabled
  Diagnosis Tool for Pathologists}}.
\newblock \bibinfo{journal}{\emph{Proc. ACM Hum.-Comput. Interact.}}
  (\bibinfo{date}{April} \bibinfo{year}{2021}).
\newblock
\urldef\tempurl%
\url{https://doi.org/10.1145/3449084}
\showDOI{\tempurl}


\bibitem[Ha(2022)]%
        {ha2022south}
\bibfield{author}{\bibinfo{person}{You~Jeen Ha}.}
  \bibinfo{year}{2022}\natexlab{}.
\newblock \showarticletitle{{South Korean Public Value Coproduction
  Towards‘AI for Humanity’: A Synergy of Sociocultural Norms and
  Multistakeholder Deliberation in Bridging the Design and Implementation of
  National AI Ethics Guidelines}}. In \bibinfo{booktitle}{\emph{2022 {ACM}
  {Conference} on {Fairness}, {Accountability}, and {Transparency}}}
  \emph{(\bibinfo{series}{{FAccT} '22})}. \bibinfo{publisher}{ACM}.
\newblock
\urldef\tempurl%
\url{https://doi.org/10.1145/3531146.3533091}
\showDOI{\tempurl}


\bibitem[Hall et~al\mbox{.}(2022)]%
        {hall2022supporting}
\bibfield{author}{\bibinfo{person}{Kaely Hall}, \bibinfo{person}{Dong~Whi Yoo},
  \bibinfo{person}{Wenrui Zhang}, \bibinfo{person}{Mehrab Bin~Morshed},
  \bibinfo{person}{Vedant Das~Swain}, \bibinfo{person}{Gregory~D. Abowd},
  \bibinfo{person}{Munmun De~Choudhury}, \bibinfo{person}{Alex Endert},
  \bibinfo{person}{John Stasko}, {and} \bibinfo{person}{Jennifer~G Kim}.}
  \bibinfo{year}{2022}\natexlab{}.
\newblock \showarticletitle{{Supporting the Contact Tracing Process with WiFi
  Location Data: Opportunities and Challenges}}. In
  \bibinfo{booktitle}{\emph{Proceedings of the 2022 {CHI} {Conference} on
  {Human} {Factors} in {Computing} {Systems}}} \emph{(\bibinfo{series}{{CHI}
  '22})}. \bibinfo{publisher}{ACM}.
\newblock
\urldef\tempurl%
\url{https://doi.org/10.1145/3491102.3517703}
\showDOI{\tempurl}


\bibitem[Havens et~al\mbox{.}(2020)]%
        {havens2020situated}
\bibfield{author}{\bibinfo{person}{Lucy Havens}, \bibinfo{person}{Melissa
  Terras}, \bibinfo{person}{Benjamin Bach}, {and} \bibinfo{person}{Beatrice
  Alex}.} \bibinfo{year}{2020}\natexlab{}.
\newblock \showarticletitle{{Situated Data, Situated Systems: A Methodology to
  Engage with Power Relations in Natural Language Processing Research}}. In
  \bibinfo{booktitle}{\emph{Proceedings of the Second Workshop on Gender Bias
  in Natural Language Processing}}. \bibinfo{publisher}{Association for
  Computational Linguistics}.
\newblock
\urldef\tempurl%
\url{https://aclanthology.org/2020.gebnlp-1.10}
\showURL{%
\tempurl}


\bibitem[He et~al\mbox{.}(2023)]%
        {he2023knowing}
\bibfield{author}{\bibinfo{person}{Gaole He}, \bibinfo{person}{Lucie Kuiper},
  {and} \bibinfo{person}{Ujwal Gadiraju}.} \bibinfo{year}{2023}\natexlab{}.
\newblock \showarticletitle{Knowing About Knowing: An Illusion of Human
  Competence Can Hinder Appropriate Reliance on AI Systems}. In
  \bibinfo{booktitle}{\emph{Proceedings of the 2023 CHI Conference on Human
  Factors in Computing Systems}} \emph{(\bibinfo{series}{CHI '23})}.
  \bibinfo{publisher}{ACM}.
\newblock
\urldef\tempurl%
\url{https://doi.org/10.1145/3544548.3581025}
\showDOI{\tempurl}


\bibitem[{Health Equity \& Policy Lab}(2022)]%
        {lab2022human}
\bibfield{author}{\bibinfo{person}{{Health Equity \& Policy Lab}}.}
  \bibinfo{year}{2022}\natexlab{}.
\newblock \bibinfo{booktitle}{\emph{{Human Flourishing}}}.
\newblock {University of Pennsylvania}.
\newblock
\urldef\tempurl%
\url{https://www.healthequityandpolicylab.com/human-flourishing}
\showURL{%
Retrieved January 2023 from \tempurl}


\bibitem[Heger et~al\mbox{.}(2022)]%
        {heger2022understanding}
\bibfield{author}{\bibinfo{person}{Amy~K. Heger}, \bibinfo{person}{Liz~B.
  Marquis}, \bibinfo{person}{Mihaela Vorvoreanu}, \bibinfo{person}{Hanna
  Wallach}, {and} \bibinfo{person}{Jennifer Wortman~Vaughan}.}
  \bibinfo{year}{2022}\natexlab{}.
\newblock \showarticletitle{{Understanding Machine Learning Practitioners' Data
  Documentation Perceptions, Needs, Challenges, and Desiderata}}.
\newblock \bibinfo{journal}{\emph{Proc. ACM Hum.-Comput. Interact.}}
  (\bibinfo{date}{Nov.} \bibinfo{year}{2022}).
\newblock
\urldef\tempurl%
\url{https://doi.org/10.1145/3555760}
\showDOI{\tempurl}


\bibitem[Henriksen et~al\mbox{.}(2021)]%
        {henriksen2021situated}
\bibfield{author}{\bibinfo{person}{Anne Henriksen}, \bibinfo{person}{Simon
  Enni}, {and} \bibinfo{person}{Anja Bechmann}.}
  \bibinfo{year}{2021}\natexlab{}.
\newblock \showarticletitle{{Situated Accountability: Ethical Principles,
  Certification Standards, and Explanation Methods in Applied AI}}. In
  \bibinfo{booktitle}{\emph{Proceedings of the 2021 {AAAI}/{ACM} {Conference}
  on {AI}, {Ethics}, and {Society}}} \emph{(\bibinfo{series}{{AIES} '21})}.
  \bibinfo{publisher}{ACM}.
\newblock
\urldef\tempurl%
\url{https://doi.org/10.1145/3461702.3462564}
\showDOI{\tempurl}


\bibitem[Hertzberg et~al\mbox{.}(2010)]%
        {hertzberg2010information}
\bibfield{author}{\bibinfo{person}{Andrew Hertzberg},
  \bibinfo{person}{Jose~Maria Liberti}, {and} \bibinfo{person}{Daniel
  Paravisini}.} \bibinfo{year}{2010}\natexlab{}.
\newblock \showarticletitle{{Information and incentives inside the firm:
  Evidence from loan officer rotation}}.
\newblock \bibinfo{journal}{\emph{The Journal of Finance}}
  (\bibinfo{year}{2010}).
\newblock
\urldef\tempurl%
\url{https://doi.org/10.1111/j.1540-6261.2010.01553.x}
\showDOI{\tempurl}


\bibitem[Holl\"{a}nder et~al\mbox{.}(2021)]%
        {holl_ander2021taxonomy}
\bibfield{author}{\bibinfo{person}{Kai Holl\"{a}nder}, \bibinfo{person}{Mark
  Colley}, \bibinfo{person}{Enrico Rukzio}, {and} \bibinfo{person}{Andreas
  Butz}.} \bibinfo{year}{2021}\natexlab{}.
\newblock \showarticletitle{{A Taxonomy of Vulnerable Road Users for HCI Based
  On A Systematic Literature Review}}. In \bibinfo{booktitle}{\emph{Proceedings
  of the 2021 CHI Conference on Human Factors in Computing Systems}}
  \emph{(\bibinfo{series}{CHI '21})}. \bibinfo{publisher}{ACM}.
\newblock
\urldef\tempurl%
\url{https://doi.org/10.1145/3411764.3445480}
\showDOI{\tempurl}


\bibitem[Holstein et~al\mbox{.}(2023)]%
        {holstein2023supporting}
\bibfield{author}{\bibinfo{person}{Kenneth Holstein}, \bibinfo{person}{Maria
  De-Arteaga}, \bibinfo{person}{Lakshmi Tumati}, {and}
  \bibinfo{person}{Yanghuidi Cheng}.} \bibinfo{year}{2023}\natexlab{}.
\newblock \showarticletitle{{Toward Supporting Perceptual Complementarity in
  Human-AI Collaboration via Reflection on Unobservables}}.
\newblock \bibinfo{journal}{\emph{Proc. ACM Hum.-Comput. Interact.}}
  (\bibinfo{date}{April} \bibinfo{year}{2023}).
\newblock
\urldef\tempurl%
\url{https://doi.org/10.1145/3579628}
\showDOI{\tempurl}


\bibitem[Hopkins and Booth(2021)]%
        {hopkins2021machine}
\bibfield{author}{\bibinfo{person}{Aspen Hopkins} {and} \bibinfo{person}{Serena
  Booth}.} \bibinfo{year}{2021}\natexlab{}.
\newblock \showarticletitle{{Machine Learning Practices Outside Big Tech: How
  Resource Constraints Challenge Responsible Development}}. In
  \bibinfo{booktitle}{\emph{Proceedings of the 2021 {AAAI}/{ACM} {Conference}
  on {AI}, {Ethics}, and {Society}}} \emph{(\bibinfo{series}{{AIES} '21})}.
  \bibinfo{publisher}{ACM}.
\newblock
\urldef\tempurl%
\url{https://doi.org/10.1145/3461702.3462527}
\showDOI{\tempurl}


\bibitem[Houser(2019)]%
        {houser2019can}
\bibfield{author}{\bibinfo{person}{Kimberly~A Houser}.}
  \bibinfo{year}{2019}\natexlab{}.
\newblock \showarticletitle{Can AI solve the diversity problem in the tech
  industry: Mitigating noise and bias in employment decision-making}.
\newblock \bibinfo{journal}{\emph{Stan. Tech. L. Rev.}} (\bibinfo{year}{2019}).
\newblock
\urldef\tempurl%
\url{https://ssrn.com/abstract=3344751}
\showURL{%
\tempurl}


\bibitem[Hsieh et~al\mbox{.}(2023)]%
        {hsieh2023what}
\bibfield{author}{\bibinfo{person}{Gary Hsieh}, \bibinfo{person}{Brett~A.
  Halperin}, \bibinfo{person}{Evan Schmitz}, \bibinfo{person}{Yen~Nee Chew},
  {and} \bibinfo{person}{Yuan-Chi Tseng}.} \bibinfo{year}{2023}\natexlab{}.
\newblock \showarticletitle{{What is in the Cards: Exploring Uses, Patterns,
  and Trends in Design Cards}}. In \bibinfo{booktitle}{\emph{Proceedings of the
  2023 {CHI} {Conference} on {Human} {Factors} in {Computing} {Systems}}}
  \emph{(\bibinfo{series}{{CHI} '23})}. \bibinfo{publisher}{ACM}.
\newblock
\urldef\tempurl%
\url{https://doi.org/10.1145/3544548.3580712}
\showDOI{\tempurl}


\bibitem[Huang and Liem(2022)]%
        {huang2022social}
\bibfield{author}{\bibinfo{person}{Han-Yin Huang} {and}
  \bibinfo{person}{Cynthia C.~S. Liem}.} \bibinfo{year}{2022}\natexlab{}.
\newblock \showarticletitle{{Social Inclusion in Curated Contexts: Insights
  from Museum Practices}}. In \bibinfo{booktitle}{\emph{2022 {ACM} {Conference}
  on {Fairness}, {Accountability}, and {Transparency}}}
  \emph{(\bibinfo{series}{{FAccT} '22})}. \bibinfo{publisher}{ACM}.
\newblock
\urldef\tempurl%
\url{https://doi.org/10.1145/3531146.3533095}
\showDOI{\tempurl}


\bibitem[Jakesch et~al\mbox{.}(2022)]%
        {jakesch2022how}
\bibfield{author}{\bibinfo{person}{Maurice Jakesch}, \bibinfo{person}{Zana
  Buçinca}, \bibinfo{person}{Saleema Amershi}, {and}
  \bibinfo{person}{Alexandra Olteanu}.} \bibinfo{year}{2022}\natexlab{}.
\newblock \showarticletitle{{How Different Groups Prioritize Ethical Values for
  Responsible AI}}. In \bibinfo{booktitle}{\emph{2022 {ACM} {Conference} on
  {Fairness}, {Accountability}, and {Transparency}}}
  \emph{(\bibinfo{series}{{FAccT} '22})}. \bibinfo{publisher}{ACM}.
\newblock
\urldef\tempurl%
\url{https://doi.org/10.1145/3531146.3533097}
\showDOI{\tempurl}


\bibitem[Jia et~al\mbox{.}(2022)]%
        {jia2022understanding}
\bibfield{author}{\bibinfo{person}{Chenyan Jia}, \bibinfo{person}{Alexander
  Boltz}, \bibinfo{person}{Angie Zhang}, \bibinfo{person}{Anqing Chen}, {and}
  \bibinfo{person}{Min~Kyung Lee}.} \bibinfo{year}{2022}\natexlab{}.
\newblock \showarticletitle{{Understanding Effects of Algorithmic vs. Community
  Label on Perceived Accuracy of Hyper-Partisan Misinformation}}.
\newblock \bibinfo{journal}{\emph{Proc. ACM Hum.-Comput. Interact.}}
  (\bibinfo{date}{Nov.} \bibinfo{year}{2022}).
\newblock
\urldef\tempurl%
\url{https://doi.org/10.1145/3555096}
\showDOI{\tempurl}


\bibitem[Jones et~al\mbox{.}(2020)]%
        {jones2020covid}
\bibfield{author}{\bibinfo{person}{Bernadette Jones},
  \bibinfo{person}{Paula~Toko King}, \bibinfo{person}{Gabrielle Baker}, {and}
  \bibinfo{person}{Tristram Ingham}.} \bibinfo{year}{2020}\natexlab{}.
\newblock \showarticletitle{{COVID-19, intersectionality, and health equity for
  indigenous peoples with lived experience of disability}}.
\newblock \bibinfo{journal}{\emph{American Indian Culture and Research
  Journal}} (\bibinfo{year}{2020}).
\newblock
\urldef\tempurl%
\url{https://doi.org/10.17953/aicrj.44.2.jones}
\showDOI{\tempurl}


\bibitem[{Josie Cox}(2023)]%
        {bbc2023AI}
\bibfield{author}{\bibinfo{person}{{Josie Cox}}.}
  \bibinfo{year}{2023}\natexlab{}.
\newblock \bibinfo{booktitle}{\emph{{AI anxiety: The workers who fear losing
  their jobs to artificial intelligence}}}.
\newblock
\urldef\tempurl%
\url{https://www.bbc.com/worklife/article/20230418-ai-anxiety-artificial-intelligence-replace-jobs}
\showURL{%
Retrieved June 2023 from \tempurl}


\bibitem[Kamikubo et~al\mbox{.}(2023)]%
        {kamikubo2023contributing}
\bibfield{author}{\bibinfo{person}{Rie Kamikubo}, \bibinfo{person}{Kyungjun
  Lee}, {and} \bibinfo{person}{Hernisa Kacorri}.}
  \bibinfo{year}{2023}\natexlab{}.
\newblock \showarticletitle{{Contributing to Accessibility Datasets:
  Reflections on Sharing Study Data by Blind People}}. In
  \bibinfo{booktitle}{\emph{Proceedings of the 2023 {CHI} {Conference} on
  {Human} {Factors} in {Computing} {Systems}}} \emph{(\bibinfo{series}{{CHI}
  '23})}. \bibinfo{publisher}{ACM}.
\newblock
\urldef\tempurl%
\url{https://doi.org/10.1145/3544548.3581337}
\showDOI{\tempurl}


\bibitem[Kang et~al\mbox{.}(2022)]%
        {kang2022how}
\bibfield{author}{\bibinfo{person}{Jimoon Kang}, \bibinfo{person}{June~Seop
  Yoon}, {and} \bibinfo{person}{Byungjoo Lee}.}
  \bibinfo{year}{2022}\natexlab{}.
\newblock \showarticletitle{{How AI-Based Training Affected the Performance of
  Professional Go Players}}. In \bibinfo{booktitle}{\emph{Proceedings of the
  2022 {CHI} {Conference} on {Human} {Factors} in {Computing} {Systems}}}
  \emph{(\bibinfo{series}{{CHI} '22})}. \bibinfo{publisher}{ACM}.
\newblock
\urldef\tempurl%
\url{https://doi.org/10.1145/3491102.3517540}
\showDOI{\tempurl}


\bibitem[Kapania et~al\mbox{.}(2022)]%
        {kapania2022because}
\bibfield{author}{\bibinfo{person}{Shivani Kapania}, \bibinfo{person}{Oliver
  Siy}, \bibinfo{person}{Gabe Clapper}, \bibinfo{person}{Azhagu~Meena SP},
  {and} \bibinfo{person}{Nithya Sambasivan}.} \bibinfo{year}{2022}\natexlab{}.
\newblock \showarticletitle{{”Because AI is 100\% Right and Safe”: User
  Attitudes and Sources of AI Authority in India}}. In
  \bibinfo{booktitle}{\emph{Proceedings of the 2022 {CHI} {Conference} on
  {Human} {Factors} in {Computing} {Systems}}} \emph{(\bibinfo{series}{{CHI}
  '22})}. \bibinfo{publisher}{ACM}.
\newblock
\urldef\tempurl%
\url{https://doi.org/10.1145/3491102.3517533}
\showDOI{\tempurl}


\bibitem[Kapania et~al\mbox{.}(2023)]%
        {kapania2023hunt}
\bibfield{author}{\bibinfo{person}{Shivani Kapania}, \bibinfo{person}{Alex~S
  Taylor}, {and} \bibinfo{person}{Ding Wang}.} \bibinfo{year}{2023}\natexlab{}.
\newblock \showarticletitle{{A Hunt for the Snark: Annotator Diversity in Data
  Practices}}. In \bibinfo{booktitle}{\emph{Proceedings of the 2023 {CHI}
  {Conference} on {Human} {Factors} in {Computing} {Systems}}}
  \emph{(\bibinfo{series}{{CHI} '23})}. \bibinfo{publisher}{ACM}.
\newblock
\urldef\tempurl%
\url{https://doi.org/10.1145/3544548.3580645}
\showDOI{\tempurl}


\bibitem[Kasirzadeh and Klein(2021)]%
        {kasirzadeh2021ethical}
\bibfield{author}{\bibinfo{person}{Atoosa Kasirzadeh} {and}
  \bibinfo{person}{Colin Klein}.} \bibinfo{year}{2021}\natexlab{}.
\newblock \showarticletitle{{The Ethical Gravity Thesis: Marrian Levels and the
  Persistence of Bias in Automated Decision-Making Systems}}. In
  \bibinfo{booktitle}{\emph{Proceedings of the 2021 {AAAI}/{ACM} {Conference}
  on {AI}, {Ethics}, and {Society}}} \emph{(\bibinfo{series}{{AIES} '21})}.
  \bibinfo{publisher}{ACM}.
\newblock
\urldef\tempurl%
\url{https://doi.org/10.1145/3461702.3462606}
\showDOI{\tempurl}


\bibitem[Kasirzadeh and Smart(2021)]%
        {kasirzadeh2021use}
\bibfield{author}{\bibinfo{person}{Atoosa Kasirzadeh} {and}
  \bibinfo{person}{Andrew Smart}.} \bibinfo{year}{2021}\natexlab{}.
\newblock \showarticletitle{{The Use and Misuse of Counterfactuals in Ethical
  Machine Learning}}. In \bibinfo{booktitle}{\emph{Proceedings of the 2021
  {ACM} {Conference} on {Fairness}, {Accountability}, and {Transparency}}}
  \emph{(\bibinfo{series}{{FAccT} '21})}. \bibinfo{publisher}{ACM}.
\newblock
\urldef\tempurl%
\url{https://doi.org/10.1145/3442188.3445886}
\showDOI{\tempurl}


\bibitem[Kaur et~al\mbox{.}(2022)]%
        {kaur2022sensible}
\bibfield{author}{\bibinfo{person}{Harmanpreet Kaur}, \bibinfo{person}{Eytan
  Adar}, \bibinfo{person}{Eric Gilbert}, {and} \bibinfo{person}{Cliff Lampe}.}
  \bibinfo{year}{2022}\natexlab{}.
\newblock \showarticletitle{{Sensible AI: Re-Imagining Interpretability and
  Explainability Using Sensemaking Theory}}. In \bibinfo{booktitle}{\emph{2022
  {ACM} {Conference} on {Fairness}, {Accountability}, and {Transparency}}}
  \emph{(\bibinfo{series}{{FAccT} '22})}. \bibinfo{publisher}{ACM}.
\newblock
\urldef\tempurl%
\url{https://doi.org/10.1145/3531146.3533135}
\showDOI{\tempurl}


\bibitem[Kemper and Kolain(2022)]%
        {kemper2022k9}
\bibfield{author}{\bibinfo{person}{Carolin Kemper} {and}
  \bibinfo{person}{Michael Kolain}.} \bibinfo{year}{2022}\natexlab{}.
\newblock \showarticletitle{{K9 Police Robots-Strolling Drones, RoboDogs, or
  Lethal Weapons?}}. In \bibinfo{booktitle}{\emph{WeRobot2022 conference}}.
\newblock
\urldef\tempurl%
\url{https://doi.org/10.2139/ssrn.4201692}
\showDOI{\tempurl}


\bibitem[Kim et~al\mbox{.}(2023)]%
        {kim2023bubbleu}
\bibfield{author}{\bibinfo{person}{Minji Kim}, \bibinfo{person}{Kyungjin Lee},
  \bibinfo{person}{Rajesh Balan}, {and} \bibinfo{person}{Youngki Lee}.}
  \bibinfo{year}{2023}\natexlab{}.
\newblock \showarticletitle{{Bubbleu: Exploring Augmented Reality Game Design
  with Uncertain AI-Based Interaction}}. In
  \bibinfo{booktitle}{\emph{Proceedings of the 2023 {CHI} {Conference} on
  {Human} {Factors} in {Computing} {Systems}}} \emph{(\bibinfo{series}{{CHI}
  '23})}. \bibinfo{publisher}{ACM}.
\newblock
\urldef\tempurl%
\url{https://doi.org/10.1145/3544548.3581270}
\showDOI{\tempurl}


\bibitem[Kim et~al\mbox{.}(2018)]%
        {kim2018computational}
\bibfield{author}{\bibinfo{person}{Richard Kim}, \bibinfo{person}{Max
  Kleiman-Weiner}, \bibinfo{person}{Andrés Abeliuk}, \bibinfo{person}{Edmond
  Awad}, \bibinfo{person}{Sohan Dsouza}, \bibinfo{person}{Joshua~B. Tenenbaum},
  {and} \bibinfo{person}{Iyad Rahwan}.} \bibinfo{year}{2018}\natexlab{}.
\newblock \showarticletitle{{A Computational Model of Commonsense Moral
  Decision Making}}. In \bibinfo{booktitle}{\emph{Proceedings of the 2018
  {AAAI}/{ACM} {Conference} on {AI}, {Ethics}, and {Society}}}
  \emph{(\bibinfo{series}{{AIES} '18})}. \bibinfo{publisher}{ACM}.
\newblock
\urldef\tempurl%
\url{https://doi.org/10.1145/3278721.3278770}
\showDOI{\tempurl}


\bibitem[Kissinger et~al\mbox{.}(2021)]%
        {kissinger2021age}
\bibfield{author}{\bibinfo{person}{Henry Kissinger}, \bibinfo{person}{Eric
  Schmidt}, {and} \bibinfo{person}{Daniel~P Huttenlocher}.}
  \bibinfo{year}{2021}\natexlab{}.
\newblock \bibinfo{booktitle}{\emph{{The Age of AI: And Our Human Future}}}.
\newblock \bibinfo{publisher}{John Murray London}.
\newblock


\bibitem[Kizilaslan and Lookman(2017)]%
        {kizilaslan2017can}
\bibfield{author}{\bibinfo{person}{Atay Kizilaslan} {and}
  \bibinfo{person}{Aziz~A Lookman}.} \bibinfo{year}{2017}\natexlab{}.
\newblock \showarticletitle{{Can Economically Intuitive Factors Improve Ability
  of Proprietary Algorithms to Predict Defaults of Peer-to-Peer Loans?}}
\newblock \bibinfo{journal}{\emph{Available at SSRN 2987613}}
  (\bibinfo{year}{2017}).
\newblock
\urldef\tempurl%
\url{https://doi.org/10.2139/ssrn.2987613}
\showDOI{\tempurl}


\bibitem[Klinova and Korinek(2021)]%
        {klinova2021ai}
\bibfield{author}{\bibinfo{person}{Katya Klinova} {and} \bibinfo{person}{Anton
  Korinek}.} \bibinfo{year}{2021}\natexlab{}.
\newblock \showarticletitle{{AI and Shared Prosperity}}. In
  \bibinfo{booktitle}{\emph{Proceedings of the 2021 {AAAI}/{ACM} {Conference}
  on {AI}, {Ethics}, and {Society}}} \emph{(\bibinfo{series}{{AIES} '21})}.
  \bibinfo{publisher}{ACM}.
\newblock
\urldef\tempurl%
\url{https://doi.org/10.1145/3461702.3462619}
\showDOI{\tempurl}


\bibitem[Knowles and Richards(2021)]%
        {knowles2021sanction}
\bibfield{author}{\bibinfo{person}{Bran Knowles} {and} \bibinfo{person}{John~T.
  Richards}.} \bibinfo{year}{2021}\natexlab{}.
\newblock \showarticletitle{{The Sanction of Authority: Promoting Public Trust
  in AI}}. In \bibinfo{booktitle}{\emph{Proceedings of the 2021 {ACM}
  {Conference} on {Fairness}, {Accountability}, and {Transparency}}}
  \emph{(\bibinfo{series}{{FAccT} '21})}. \bibinfo{publisher}{ACM}.
\newblock
\urldef\tempurl%
\url{https://doi.org/10.1145/3442188.3445890}
\showDOI{\tempurl}


\bibitem[Krafft et~al\mbox{.}(2021)]%
        {krafft2021action}
\bibfield{author}{\bibinfo{person}{P.~M. Krafft}, \bibinfo{person}{Meg Young},
  \bibinfo{person}{Michael Katell}, \bibinfo{person}{Jennifer~E. Lee},
  \bibinfo{person}{Shankar Narayan}, \bibinfo{person}{Micah Epstein},
  \bibinfo{person}{Dharma Dailey}, \bibinfo{person}{Bernease Herman},
  \bibinfo{person}{Aaron Tam}, \bibinfo{person}{Vivian Guetler},
  \bibinfo{person}{Corinne Bintz}, \bibinfo{person}{Daniella Raz},
  \bibinfo{person}{Pa~Ousman Jobe}, \bibinfo{person}{Franziska Putz},
  \bibinfo{person}{Brian Robick}, {and} \bibinfo{person}{Bissan Barghouti}.}
  \bibinfo{year}{2021}\natexlab{}.
\newblock \showarticletitle{{An Action-Oriented AI Policy Toolkit for
  Technology Audits by Community Advocates and Activists}}. In
  \bibinfo{booktitle}{\emph{Proceedings of the 2021 {ACM} {Conference} on
  {Fairness}, {Accountability}, and {Transparency}}}
  \emph{(\bibinfo{series}{{FAccT} '21})}. \bibinfo{publisher}{ACM}.
\newblock
\urldef\tempurl%
\url{https://doi.org/10.1145/3442188.3445938}
\showDOI{\tempurl}


\bibitem[Kroll(2021)]%
        {kroll2021outlining}
\bibfield{author}{\bibinfo{person}{Joshua~A. Kroll}.}
  \bibinfo{year}{2021}\natexlab{}.
\newblock \showarticletitle{{Outlining Traceability: A Principle for
  Operationalizing Accountability in Computing Systems}}. In
  \bibinfo{booktitle}{\emph{Proceedings of the 2021 {ACM} {Conference} on
  {Fairness}, {Accountability}, and {Transparency}}}
  \emph{(\bibinfo{series}{{FAccT} '21})}. \bibinfo{publisher}{ACM}.
\newblock
\urldef\tempurl%
\url{https://doi.org/10.1145/3442188.3445937}
\showDOI{\tempurl}


\bibitem[Kukutai et~al\mbox{.}(2020)]%
        {kukutai2020indigenous}
\bibfield{author}{\bibinfo{person}{Tahu Kukutai},
  \bibinfo{person}{Stephanie~Russo Carroll}, {and} \bibinfo{person}{Maggie
  Walter}.} \bibinfo{year}{2020}\natexlab{}.
\newblock \showarticletitle{Indigenous data sovereignty}.
\newblock  (\bibinfo{year}{2020}).
\newblock


\bibitem[Kumar et~al\mbox{.}(2020)]%
        {kumar2020problems}
\bibfield{author}{\bibinfo{person}{I.~Elizabeth Kumar}, \bibinfo{person}{Suresh
  Venkatasubramanian}, \bibinfo{person}{Carlos Scheidegger}, {and}
  \bibinfo{person}{Sorelle~A. Friedler}.} \bibinfo{year}{2020}\natexlab{}.
\newblock \showarticletitle{{Problems with Shapley-Value-Based Explanations as
  Feature Importance Measures}}. In \bibinfo{booktitle}{\emph{Proceedings of
  the 37th International Conference on Machine Learning}}
  \emph{(\bibinfo{series}{ICML'20})}. \bibinfo{publisher}{JMLR.org}.
\newblock
\urldef\tempurl%
\url{https://doi.org/10.48550/arXiv.2002.11097}
\showDOI{\tempurl}


\bibitem[Laker(2022)]%
        {laker2022artificial}
\bibfield{author}{\bibinfo{person}{Benjamin Laker}.}
  \bibinfo{year}{2022}\natexlab{}.
\newblock \bibinfo{booktitle}{\emph{{Artificial Intelligence Can Help Leaders
  Drive Global Economy Forward In 2022}}}.
\newblock Forbes.
\newblock
\urldef\tempurl%
\url{https://www.forbes.com/sites/benjaminlaker/2022/01/19/artificial-intelligence-can-help-leaders-drive-global-economy-forward-in-2022/}
\showURL{%
Retrieved November 2022 from \tempurl}


\bibitem[Lam et~al\mbox{.}(2022)]%
        {lam2022end}
\bibfield{author}{\bibinfo{person}{Michelle~S. Lam},
  \bibinfo{person}{Mitchell~L. Gordon}, \bibinfo{person}{Danaë Metaxa},
  \bibinfo{person}{Jeffrey~T. Hancock}, \bibinfo{person}{James~A. Landay},
  {and} \bibinfo{person}{Michael~S. Bernstein}.}
  \bibinfo{year}{2022}\natexlab{}.
\newblock \showarticletitle{{End-User Audits: A System Empowering Communities
  to Lead Large-Scale Investigations of Harmful Algorithmic Behavior}}.
\newblock \bibinfo{journal}{\emph{Proc. ACM Hum.-Comput. Interact.}}
  (\bibinfo{date}{Nov.} \bibinfo{year}{2022}).
\newblock
\urldef\tempurl%
\url{https://doi.org/10.1145/3555625}
\showDOI{\tempurl}


\bibitem[Lam et~al\mbox{.}(2023)]%
        {lam2023model}
\bibfield{author}{\bibinfo{person}{Michelle~S. Lam}, \bibinfo{person}{Zixian
  Ma}, \bibinfo{person}{Anne Li}, \bibinfo{person}{Izequiel Freitas},
  \bibinfo{person}{Dakuo Wang}, \bibinfo{person}{James~A. Landay}, {and}
  \bibinfo{person}{Michael~S. Bernstein}.} \bibinfo{year}{2023}\natexlab{}.
\newblock \showarticletitle{{Model Sketching: Centering Concepts in Early-Stage
  Machine Learning Model Design}}. In \bibinfo{booktitle}{\emph{Proceedings of
  the 2023 {CHI} {Conference} on {Human} {Factors} in {Computing} {Systems}}}
  \emph{(\bibinfo{series}{{CHI} '23})}. \bibinfo{publisher}{ACM}.
\newblock
\urldef\tempurl%
\url{https://doi.org/10.1145/3544548.3581290}
\showDOI{\tempurl}


\bibitem[Langer et~al\mbox{.}(2022)]%
        {langer2022look}
\bibfield{author}{\bibinfo{person}{Markus Langer}, \bibinfo{person}{Tim
  Hunsicker}, \bibinfo{person}{Tina Feldkamp}, \bibinfo{person}{Cornelius~J.
  König}, {and} \bibinfo{person}{Nina Grgić-Hlača}.}
  \bibinfo{year}{2022}\natexlab{}.
\newblock \showarticletitle{{“Look! It’s a Computer Program! It’s an
  Algorithm! It’s AI!”: Does Terminology Affect Human Perceptions and
  Evaluations of Algorithmic Decision-Making Systems?}}. In
  \bibinfo{booktitle}{\emph{Proceedings of the 2022 {CHI} {Conference} on
  {Human} {Factors} in {Computing} {Systems}}} \emph{(\bibinfo{series}{{CHI}
  '22})}. \bibinfo{publisher}{ACM}.
\newblock
\urldef\tempurl%
\url{https://doi.org/10.1145/3491102.3517527}
\showDOI{\tempurl}


\bibitem[Lazar et~al\mbox{.}(2017a)]%
        {lazar2017chaptera}
\bibfield{author}{\bibinfo{person}{Jonathan Lazar},
  \bibinfo{person}{Jinjuan~Heidi Feng}, {and} \bibinfo{person}{Harry
  Hochheiser}.} \bibinfo{year}{2017}\natexlab{a}.
\newblock \showarticletitle{{Chapter 11 - Analyzing qualitative data}}.
\newblock In \bibinfo{booktitle}{\emph{Research Methods in Human Computer
  Interaction} (\bibinfo{edition}{second edition} ed.)},
  \bibfield{editor}{\bibinfo{person}{Jonathan Lazar},
  \bibinfo{person}{Jinjuan~Heidi Feng}, {and} \bibinfo{person}{Harry
  Hochheiser}} (Eds.). \bibinfo{publisher}{Morgan Kaufmann}.
\newblock
\urldef\tempurl%
\url{https://doi.org/10.1016/B978-0-12-805390-4.00011-X}
\showDOI{\tempurl}


\bibitem[Lazar et~al\mbox{.}(2017b)]%
        {lazar2017chapterb}
\bibfield{author}{\bibinfo{person}{Jonathan Lazar},
  \bibinfo{person}{Jinjuan~Heidi Feng}, {and} \bibinfo{person}{Harry
  Hochheiser}.} \bibinfo{year}{2017}\natexlab{b}.
\newblock \showarticletitle{{Chapter 8 - Interviews and focus groups}}.
\newblock In \bibinfo{booktitle}{\emph{Research Methods in Human Computer
  Interaction} (\bibinfo{edition}{second edition} ed.)},
  \bibfield{editor}{\bibinfo{person}{Jonathan Lazar},
  \bibinfo{person}{Jinjuan~Heidi Feng}, {and} \bibinfo{person}{Harry
  Hochheiser}} (Eds.). \bibinfo{publisher}{Morgan Kaufmann}.
\newblock
\urldef\tempurl%
\url{https://doi.org/10.1016/B978-0-12-805390-4.00008-X}
\showDOI{\tempurl}


\bibitem[Leavy et~al\mbox{.}(2021)]%
        {leavy2021ethical}
\bibfield{author}{\bibinfo{person}{Susan Leavy}, \bibinfo{person}{Eugenia
  Siapera}, {and} \bibinfo{person}{Barry O'Sullivan}.}
  \bibinfo{year}{2021}\natexlab{}.
\newblock \showarticletitle{{Ethical Data Curation for AI: An Approach Based on
  Feminist Epistemology and Critical Theories of Race}}. In
  \bibinfo{booktitle}{\emph{Proceedings of the 2021 {AAAI}/{ACM} {Conference}
  on {AI}, {Ethics}, and {Society}}} \emph{(\bibinfo{series}{{AIES} '21})}.
  \bibinfo{publisher}{ACM}.
\newblock
\urldef\tempurl%
\url{https://doi.org/10.1145/3461702.3462598}
\showDOI{\tempurl}


\bibitem[Leckning et~al\mbox{.}(2021)]%
        {leckning2021patterns}
\bibfield{author}{\bibinfo{person}{Bernard Leckning},
  \bibinfo{person}{Vincent~YF He}, \bibinfo{person}{John~R Condon},
  \bibinfo{person}{Tanja Hirvonen}, \bibinfo{person}{Helen Milroy}, {and}
  \bibinfo{person}{Steven Guthridge}.} \bibinfo{year}{2021}\natexlab{}.
\newblock \showarticletitle{{Patterns of child protection service involvement
  by Aboriginal children associated with a higher risk of self-harm in
  adolescence: A retrospective population cohort study using linked
  administrative data}}.
\newblock \bibinfo{journal}{\emph{Child Abuse \& Neglect}}
  (\bibinfo{year}{2021}).
\newblock
\urldef\tempurl%
\url{https://doi.org/10.1016/j.chiabu.2021.104931}
\showDOI{\tempurl}


\bibitem[Lee(2022)]%
        {lee2022ai}
\bibfield{author}{\bibinfo{person}{Kai-Fu Lee}.}
  \bibinfo{year}{2022}\natexlab{}.
\newblock \bibinfo{booktitle}{\emph{{AI and the human future: Net positive}}}.
\newblock Economics.
\newblock
\urldef\tempurl%
\url{https://impact.economist.com/projects/innovation-matters/blogs/ai-and-the-human-future-net-positive/}
\showURL{%
Retrieved November 2022 from \tempurl}


\bibitem[Lee et~al\mbox{.}(2019)]%
        {lee2019webuildai}
\bibfield{author}{\bibinfo{person}{Min~Kyung Lee}, \bibinfo{person}{Daniel
  Kusbit}, \bibinfo{person}{Anson Kahng}, \bibinfo{person}{Ji~Tae Kim},
  \bibinfo{person}{Xinran Yuan}, \bibinfo{person}{Allissa Chan},
  \bibinfo{person}{Daniel See}, \bibinfo{person}{Ritesh Noothigattu},
  \bibinfo{person}{Siheon Lee}, \bibinfo{person}{Alexandros Psomas}, {and}
  \bibinfo{person}{Ariel~D. Procaccia}.} \bibinfo{year}{2019}\natexlab{}.
\newblock \showarticletitle{{WeBuildAI: Participatory Framework for Algorithmic
  Governance}}.
\newblock \bibinfo{journal}{\emph{Proc. ACM Hum.-Comput. Interact.}}
  (\bibinfo{date}{Nov.} \bibinfo{year}{2019}).
\newblock
\urldef\tempurl%
\url{https://doi.org/10.1145/3359283}
\showDOI{\tempurl}


\bibitem[Lee et~al\mbox{.}(2021)]%
        {lee2021participatory}
\bibfield{author}{\bibinfo{person}{Min~Kyung Lee}, \bibinfo{person}{Ishan
  Nigam}, \bibinfo{person}{Angie Zhang}, \bibinfo{person}{Joel Afriyie},
  \bibinfo{person}{Zhizhen Qin}, {and} \bibinfo{person}{Sicun Gao}.}
  \bibinfo{year}{2021}\natexlab{}.
\newblock \showarticletitle{{Participatory Algorithmic Management: Elicitation
  Methods for Worker Well-Being Models}}. In
  \bibinfo{booktitle}{\emph{Proceedings of the 2021 {AAAI}/{ACM} {Conference}
  on {AI}, {Ethics}, and {Society}}} \emph{(\bibinfo{series}{{AIES} '21})}.
  \bibinfo{publisher}{ACM}.
\newblock
\urldef\tempurl%
\url{https://doi.org/10.1145/3461702.3462628}
\showDOI{\tempurl}


\bibitem[Lee and Rich(2021)]%
        {lee2021who}
\bibfield{author}{\bibinfo{person}{Min~Kyung Lee} {and}
  \bibinfo{person}{Katherine Rich}.} \bibinfo{year}{2021}\natexlab{}.
\newblock \showarticletitle{{Who Is Included in Human Perceptions of AI?: Trust
  and Perceived Fairness around Healthcare AI and Cultural Mistrust}}. In
  \bibinfo{booktitle}{\emph{Proceedings of the 2021 {CHI} {Conference} on
  {Human} {Factors} in {Computing} {Systems}}} \emph{(\bibinfo{series}{{CHI}
  '21})}. \bibinfo{publisher}{ACM}, Article \bibinfo{articleno}{138}.
\newblock
\urldef\tempurl%
\url{https://doi.org/10.1145/3411764.3445570}
\showDOI{\tempurl}


\bibitem[Lee et~al\mbox{.}(2023b)]%
        {lee2023speculating}
\bibfield{author}{\bibinfo{person}{Patrick Yung~Kang Lee},
  \bibinfo{person}{Ning~F. Ma}, \bibinfo{person}{Ig-Jae Kim}, {and}
  \bibinfo{person}{Dongwook Yoon}.} \bibinfo{year}{2023}\natexlab{b}.
\newblock \showarticletitle{{Speculating on Risks of AI Clones to Selfhood and
  Relationships: Doppelganger-Phobia, Identity Fragmentation, and Living
  Memories}}.
\newblock \bibinfo{journal}{\emph{Proc. ACM Hum.-Comput. Interact.}}
  (\bibinfo{date}{April} \bibinfo{year}{2023}).
\newblock
\urldef\tempurl%
\url{https://doi.org/10.1145/3579524}
\showDOI{\tempurl}


\bibitem[Lee et~al\mbox{.}(2023a)]%
        {lee2023fostering}
\bibfield{author}{\bibinfo{person}{Sunok Lee}, \bibinfo{person}{Dasom Choi},
  \bibinfo{person}{Minha Lee}, \bibinfo{person}{Jonghak Choi}, {and}
  \bibinfo{person}{Sangsu Lee}.} \bibinfo{year}{2023}\natexlab{a}.
\newblock \showarticletitle{{Fostering Youth’s Critical Thinking Competency
  About AI through Exhibition}}. In \bibinfo{booktitle}{\emph{Proceedings of
  the 2023 {CHI} {Conference} on {Human} {Factors} in {Computing} {Systems}}}
  \emph{(\bibinfo{series}{{CHI} '23})}. \bibinfo{publisher}{ACM}.
\newblock
\urldef\tempurl%
\url{https://doi.org/10.1145/3544548.3581159}
\showDOI{\tempurl}


\bibitem[{Leonardo Nicoletti and Dina Bass}(2023)]%
        {bloomberg2023humans}
\bibfield{author}{\bibinfo{person}{{Leonardo Nicoletti and Dina Bass}}.}
  \bibinfo{year}{2023}\natexlab{}.
\newblock \bibinfo{booktitle}{\emph{{HUMANS ARE BIASED. GENERATIVE AI IS EVEN
  WORSE}}}.
\newblock
\urldef\tempurl%
\url{https://www.bloomberg.com/graphics/2023-generative-ai-bias/}
\showURL{%
Retrieved June 2023 from \tempurl}


\bibitem[Lewicki et~al\mbox{.}(2023)]%
        {lewicki2023out}
\bibfield{author}{\bibinfo{person}{Kornel Lewicki}, \bibinfo{person}{Michelle
  Seng~Ah Lee}, \bibinfo{person}{Jennifer Cobbe}, {and}
  \bibinfo{person}{Jatinder Singh}.} \bibinfo{year}{2023}\natexlab{}.
\newblock \showarticletitle{{Out of Context: Investigating the Bias and
  Fairness Concerns of “Artificial Intelligence as a Service”}}. In
  \bibinfo{booktitle}{\emph{Proceedings of the 2023 {CHI} {Conference} on
  {Human} {Factors} in {Computing} {Systems}}} \emph{(\bibinfo{series}{{CHI}
  '23})}. \bibinfo{publisher}{ACM}.
\newblock
\urldef\tempurl%
\url{https://doi.org/10.1145/3544548.3581463}
\showDOI{\tempurl}


\bibitem[Li et~al\mbox{.}(2023)]%
        {li2023participation}
\bibfield{author}{\bibinfo{person}{Rena Li}, \bibinfo{person}{Sara Kingsley},
  \bibinfo{person}{Chelsea Fan}, \bibinfo{person}{Proteeti Sinha},
  \bibinfo{person}{Nora Wai}, \bibinfo{person}{Jaimie Lee},
  \bibinfo{person}{Hong Shen}, \bibinfo{person}{Motahhare Eslami}, {and}
  \bibinfo{person}{Jason Hong}.} \bibinfo{year}{2023}\natexlab{}.
\newblock \showarticletitle{{Participation and Division of Labor in User-Driven
  Algorithm Audits: How Do Everyday Users Work Together to Surface Algorithmic
  Harms?}}. In \bibinfo{booktitle}{\emph{Proceedings of the 2023 {CHI}
  {Conference} on {Human} {Factors} in {Computing} {Systems}}}
  \emph{(\bibinfo{series}{{CHI} '23})}. \bibinfo{publisher}{ACM}.
\newblock
\urldef\tempurl%
\url{https://doi.org/10.1145/3544548.3582074}
\showDOI{\tempurl}


\bibitem[Li et~al\mbox{.}(2022)]%
        {li2022fairness}
\bibfield{author}{\bibinfo{person}{Yunqi Li}, \bibinfo{person}{Hanxiong Chen},
  \bibinfo{person}{Shuyuan Xu}, \bibinfo{person}{Yingqiang Ge},
  \bibinfo{person}{Juntao Tan}, \bibinfo{person}{Shuchang Liu}, {and}
  \bibinfo{person}{Yongfeng Zhang}.} \bibinfo{year}{2022}\natexlab{}.
\newblock \bibinfo{title}{{Fairness in Recommendation: A Survey}}.
\newblock
\newblock
\urldef\tempurl%
\url{https://doi.org/10.48550/ARXIV.2205.13619}
\showDOI{\tempurl}


\bibitem[Liao and Sundar(2022)]%
        {liao2022designing}
\bibfield{author}{\bibinfo{person}{Q.Vera Liao} {and} \bibinfo{person}{S.~Shyam
  Sundar}.} \bibinfo{year}{2022}\natexlab{}.
\newblock \showarticletitle{{Designing for Responsible Trust in AI Systems: A
  Communication Perspective}}. In \bibinfo{booktitle}{\emph{2022 {ACM}
  {Conference} on {Fairness}, {Accountability}, and {Transparency}}}
  \emph{(\bibinfo{series}{{FAccT} '22})}. \bibinfo{publisher}{ACM}.
\newblock
\urldef\tempurl%
\url{https://doi.org/10.1145/3531146.3533182}
\showDOI{\tempurl}


\bibitem[Liao et~al\mbox{.}(2020)]%
        {liao2020questioning}
\bibfield{author}{\bibinfo{person}{Q.~Vera Liao}, \bibinfo{person}{Daniel
  Gruen}, {and} \bibinfo{person}{Sarah Miller}.}
  \bibinfo{year}{2020}\natexlab{}.
\newblock \showarticletitle{{Questioning the AI: Informing Design Practices for
  Explainable AI User Experiences}}. In \bibinfo{booktitle}{\emph{Proceedings
  of the 2020 {CHI} {Conference} on {Human} {Factors} in {Computing}
  {Systems}}} \emph{(\bibinfo{series}{{CHI} '20})}. \bibinfo{publisher}{ACM}.
\newblock
\urldef\tempurl%
\url{https://doi.org/10.1145/3313831.3376590}
\showDOI{\tempurl}


\bibitem[Liao et~al\mbox{.}(2023)]%
        {liao2023designerly}
\bibfield{author}{\bibinfo{person}{Q.~Vera Liao}, \bibinfo{person}{Hariharan
  Subramonyam}, \bibinfo{person}{Jennifer Wang}, {and}
  \bibinfo{person}{Jennifer Wortman~Vaughan}.} \bibinfo{year}{2023}\natexlab{}.
\newblock \showarticletitle{{Designerly Understanding: Information Needs for
  Model Transparency to Support Design Ideation for AI-Powered User
  Experience}}. In \bibinfo{booktitle}{\emph{Proceedings of the 2023 {CHI}
  {Conference} on {Human} {Factors} in {Computing} {Systems}}}
  \emph{(\bibinfo{series}{{CHI} '23})}. \bibinfo{publisher}{ACM}.
\newblock
\urldef\tempurl%
\url{https://doi.org/10.1145/3544548.3580652}
\showDOI{\tempurl}


\bibitem[Liao and Varshney(2021)]%
        {liao2021human}
\bibfield{author}{\bibinfo{person}{Q.~Vera Liao} {and} \bibinfo{person}{Kush~R.
  Varshney}.} \bibinfo{year}{2021}\natexlab{}.
\newblock \showarticletitle{{Human-Centered Explainable AI (XAI): From
  Algorithms to User Experiences}}.
\newblock  (\bibinfo{year}{2021}).
\newblock
\urldef\tempurl%
\url{https://doi.org/10.48550/ARXIV.2110.10790}
\showDOI{\tempurl}


\bibitem[Lima et~al\mbox{.}(2021)]%
        {lima2021human}
\bibfield{author}{\bibinfo{person}{Gabriel Lima}, \bibinfo{person}{Nina
  Grgić-Hlača}, {and} \bibinfo{person}{Meeyoung Cha}.}
  \bibinfo{year}{2021}\natexlab{}.
\newblock \showarticletitle{{Human Perceptions on Moral Responsibility of AI: A
  Case Study in AI-Assisted Bail Decision-Making}}. In
  \bibinfo{booktitle}{\emph{Proceedings of the 2021 {CHI} {Conference} on
  {Human} {Factors} in {Computing} {Systems}}} \emph{(\bibinfo{series}{{CHI}
  '21})}. \bibinfo{publisher}{ACM}.
\newblock
\urldef\tempurl%
\url{https://doi.org/10.1145/3411764.3445260}
\showDOI{\tempurl}


\bibitem[Lima et~al\mbox{.}(2023)]%
        {lima2023blaming}
\bibfield{author}{\bibinfo{person}{Gabriel Lima}, \bibinfo{person}{Nina
  Grgić-Hlača}, {and} \bibinfo{person}{Meeyoung Cha}.}
  \bibinfo{year}{2023}\natexlab{}.
\newblock \showarticletitle{{Blaming Humans and Machines: What Shapes
  People’s Reactions to Algorithmic Harm}}. In
  \bibinfo{booktitle}{\emph{Proceedings of the 2023 {CHI} {Conference} on
  {Human} {Factors} in {Computing} {Systems}}} \emph{(\bibinfo{series}{{CHI}
  '23})}. \bibinfo{publisher}{ACM}.
\newblock
\urldef\tempurl%
\url{https://doi.org/10.1145/3544548.3580953}
\showDOI{\tempurl}


\bibitem[Lima et~al\mbox{.}(2022)]%
        {lima2022conflict}
\bibfield{author}{\bibinfo{person}{Gabriel Lima}, \bibinfo{person}{Nina
  Grgić-Hlača}, \bibinfo{person}{Jin~Keun Jeong}, {and}
  \bibinfo{person}{Meeyoung Cha}.} \bibinfo{year}{2022}\natexlab{}.
\newblock \showarticletitle{{The Conflict Between Explainable and Accountable
  Decision-Making Algorithms}}. In \bibinfo{booktitle}{\emph{2022 {ACM}
  {Conference} on {Fairness}, {Accountability}, and {Transparency}}}
  \emph{(\bibinfo{series}{{FAccT} '22})}. \bibinfo{publisher}{ACM}.
\newblock
\urldef\tempurl%
\url{https://doi.org/10.1145/3531146.3534628}
\showDOI{\tempurl}


\bibitem[Lin and Van~Brummelen(2021)]%
        {lin2021engaging}
\bibfield{author}{\bibinfo{person}{Phoebe Lin} {and} \bibinfo{person}{Jessica
  Van~Brummelen}.} \bibinfo{year}{2021}\natexlab{}.
\newblock \showarticletitle{{Engaging Teachers to Co-Design Integrated AI
  Curriculum for K-12 Classrooms}}. In \bibinfo{booktitle}{\emph{Proceedings of
  the 2021 {CHI} {Conference} on {Human} {Factors} in {Computing} {Systems}}}
  \emph{(\bibinfo{series}{{CHI} '21})}. \bibinfo{publisher}{ACM}.
\newblock
\urldef\tempurl%
\url{https://doi.org/10.1145/3411764.3445377}
\showDOI{\tempurl}


\bibitem[Linxen et~al\mbox{.}(2021)]%
        {linxen2021how}
\bibfield{author}{\bibinfo{person}{Sebastian Linxen},
  \bibinfo{person}{Christian Sturm}, \bibinfo{person}{Florian Br\"{u}hlmann},
  \bibinfo{person}{Vincent Cassau}, \bibinfo{person}{Klaus Opwis}, {and}
  \bibinfo{person}{Katharina Reinecke}.} \bibinfo{year}{2021}\natexlab{}.
\newblock \showarticletitle{{How WEIRD is CHI?}}. In
  \bibinfo{booktitle}{\emph{Proceedings of the 2021 CHI Conference on Human
  Factors in Computing Systems}} \emph{(\bibinfo{series}{CHI '21})}.
  \bibinfo{publisher}{ACM}.
\newblock
\urldef\tempurl%
\url{https://doi.org/10.1145/3411764.3445488}
\showDOI{\tempurl}


\bibitem[Long et~al\mbox{.}(2021)]%
        {long2021co}
\bibfield{author}{\bibinfo{person}{Duri Long}, \bibinfo{person}{Takeria Blunt},
  {and} \bibinfo{person}{Brian Magerko}.} \bibinfo{year}{2021}\natexlab{}.
\newblock \showarticletitle{{Co-Designing AI Literacy Exhibits for Informal
  Learning Spaces}}.
\newblock \bibinfo{journal}{\emph{Proc. ACM Hum.-Comput. Interact.}}
  (\bibinfo{date}{Oct.} \bibinfo{year}{2021}).
\newblock
\urldef\tempurl%
\url{https://doi.org/10.1145/3476034}
\showDOI{\tempurl}


\bibitem[Long et~al\mbox{.}(2022)]%
        {long2022family}
\bibfield{author}{\bibinfo{person}{Duri Long}, \bibinfo{person}{Anthony
  Teachey}, {and} \bibinfo{person}{Brian Magerko}.}
  \bibinfo{year}{2022}\natexlab{}.
\newblock \showarticletitle{{Family Learning Talk in AI Literacy Learning
  Activities}}. In \bibinfo{booktitle}{\emph{Proceedings of the 2022 {CHI}
  {Conference} on {Human} {Factors} in {Computing} {Systems}}}
  \emph{(\bibinfo{series}{{CHI} '22})}. \bibinfo{publisher}{ACM}.
\newblock
\urldef\tempurl%
\url{https://doi.org/10.1145/3491102.3502091}
\showDOI{\tempurl}


\bibitem[Lundberg and Lee(2017)]%
        {lundberg2017unified}
\bibfield{author}{\bibinfo{person}{Scott~M. Lundberg} {and}
  \bibinfo{person}{Su-In Lee}.} \bibinfo{year}{2017}\natexlab{}.
\newblock \showarticletitle{{A Unified Approach to Interpreting Model
  Predictions}}. In \bibinfo{booktitle}{\emph{Proceedings of the 31st
  International Conference on Neural Information Processing Systems}}
  \emph{(\bibinfo{series}{NIPS'17})}. \bibinfo{publisher}{Curran Associates
  Inc.}
\newblock
\urldef\tempurl%
\url{https://dl.acm.org/doi/pdf/10.5555/3295222.3295230}
\showURL{%
\tempurl}


\bibitem[Lyn(2020)]%
        {lyn2020risky}
\bibfield{author}{\bibinfo{person}{Alexandra Lyn}.}
  \bibinfo{year}{2020}\natexlab{}.
\newblock \showarticletitle{{Risky Business: Artificial Intelligence and Risk
  Assessments in Sentencing and Bail Procedures in the United States}}.
\newblock \bibinfo{journal}{\emph{Available at SSRN 3831441}}
  (\bibinfo{year}{2020}).
\newblock
\urldef\tempurl%
\url{https://doi.org/10.2139/ssrn.3831441}
\showDOI{\tempurl}


\bibitem[Lyons et~al\mbox{.}(2021)]%
        {lyons2021conceptualising}
\bibfield{author}{\bibinfo{person}{Henrietta Lyons}, \bibinfo{person}{Eduardo
  Velloso}, {and} \bibinfo{person}{Tim Miller}.}
  \bibinfo{year}{2021}\natexlab{}.
\newblock \showarticletitle{{Conceptualising Contestability: Perspectives on
  Contesting Algorithmic Decisions}}.
\newblock \bibinfo{journal}{\emph{Proc. ACM Hum.-Comput. Interact.}}
  (\bibinfo{date}{April} \bibinfo{year}{2021}).
\newblock
\urldef\tempurl%
\url{https://doi.org/10.1145/3449180}
\showDOI{\tempurl}


\bibitem[Maas(2018)]%
        {maas2018regulating}
\bibfield{author}{\bibinfo{person}{Matthijs~M. Maas}.}
  \bibinfo{year}{2018}\natexlab{}.
\newblock \showarticletitle{{Regulating for 'Normal AI Accidents': Operational
  Lessons for the Responsible Governance of Artificial Intelligence
  Deployment}}. In \bibinfo{booktitle}{\emph{Proceedings of the 2018
  {AAAI}/{ACM} {Conference} on {AI}, {Ethics}, and {Society}}}
  \emph{(\bibinfo{series}{{AIES} '18})}. \bibinfo{publisher}{ACM}.
\newblock
\urldef\tempurl%
\url{https://doi.org/10.1145/3278721.3278766}
\showDOI{\tempurl}


\bibitem[Marley(2019)]%
        {marley2019indigenous}
\bibfield{author}{\bibinfo{person}{Tennille~L Marley}.}
  \bibinfo{year}{2019}\natexlab{}.
\newblock \showarticletitle{Indigenous data sovereignty: University
  institutional review board policies and guidelines and research with American
  Indian and Alaska Native communities}.
\newblock \bibinfo{journal}{\emph{American Behavioral Scientist}}
  (\bibinfo{year}{2019}).
\newblock


\bibitem[McDonald and Pan(2020)]%
        {mcdonald2020intersectional}
\bibfield{author}{\bibinfo{person}{Nora McDonald} {and} \bibinfo{person}{Shimei
  Pan}.} \bibinfo{year}{2020}\natexlab{}.
\newblock \showarticletitle{{Intersectional AI: A Study of How Information
  Science Students Think about Ethics and Their Impact}}.
\newblock \bibinfo{journal}{\emph{Proc. ACM Hum.-Comput. Interact.}}
  (\bibinfo{date}{Oct.} \bibinfo{year}{2020}).
\newblock
\urldef\tempurl%
\url{https://doi.org/10.1145/3415218}
\showDOI{\tempurl}


\bibitem[McNaney et~al\mbox{.}(2018)]%
        {mcnaney2018enabling}
\bibfield{author}{\bibinfo{person}{Roisin McNaney}, \bibinfo{person}{John
  Vines}, \bibinfo{person}{Andy Dow}, \bibinfo{person}{Harry Robinson},
  \bibinfo{person}{Heather Robinson}, \bibinfo{person}{Kate McDonald},
  \bibinfo{person}{Leslie Brown}, \bibinfo{person}{Peter Santer},
  \bibinfo{person}{Don Murray}, \bibinfo{person}{Janice Murray},
  \bibinfo{person}{David Green}, {and} \bibinfo{person}{Peter Wright}.}
  \bibinfo{year}{2018}\natexlab{}.
\newblock \showarticletitle{{Enabling the Participation of People with
  Parkinson's and Their Caregivers in Co-Inquiry around Collectivist Health
  Technologies}}. In \bibinfo{booktitle}{\emph{Proceedings of the 2018 CHI
  Conference on Human Factors in Computing Systems}}
  \emph{(\bibinfo{series}{CHI '18})}. \bibinfo{publisher}{ACM}.
\newblock
\urldef\tempurl%
\url{https://doi.org/10.1145/3173574.3174065}
\showDOI{\tempurl}


\bibitem[Mentis et~al\mbox{.}(2016)]%
        {mentis2016crafting}
\bibfield{author}{\bibinfo{person}{Helena~M. Mentis}, \bibinfo{person}{Ahmed
  Rahim}, {and} \bibinfo{person}{Pierre Theodore}.}
  \bibinfo{year}{2016}\natexlab{}.
\newblock \showarticletitle{{Crafting the Image in Surgical Telemedicine}}. In
  \bibinfo{booktitle}{\emph{Proceedings of the 19th ACM Conference on
  Computer-Supported Cooperative Work \& Social Computing}}
  \emph{(\bibinfo{series}{CSCW '16})}. \bibinfo{publisher}{ACM}.
\newblock
\urldef\tempurl%
\url{https://doi.org/10.1145/2818048.2819978}
\showDOI{\tempurl}


\bibitem[Mhasawade et~al\mbox{.}(2021)]%
        {mhasawade2021machine}
\bibfield{author}{\bibinfo{person}{Vishwali Mhasawade}, \bibinfo{person}{Yuan
  Zhao}, {and} \bibinfo{person}{Rumi Chunara}.}
  \bibinfo{year}{2021}\natexlab{}.
\newblock \showarticletitle{{Machine learning and algorithmic fairness in
  public and population health}}.
\newblock \bibinfo{journal}{\emph{Nature Machine Intelligence}}
  (\bibinfo{year}{2021}).
\newblock
\urldef\tempurl%
\url{https://doi.org/10.1038/s42256-021-00373-4}
\showDOI{\tempurl}


\bibitem[Miceli et~al\mbox{.}(2022)]%
        {miceli2022documenting}
\bibfield{author}{\bibinfo{person}{Milagros Miceli}, \bibinfo{person}{Tianling
  Yang}, \bibinfo{person}{Adriana Alvarado~Garcia}, \bibinfo{person}{Julian
  Posada}, \bibinfo{person}{Sonja~Mei Wang}, \bibinfo{person}{Marc Pohl}, {and}
  \bibinfo{person}{Alex Hanna}.} \bibinfo{year}{2022}\natexlab{}.
\newblock \showarticletitle{{Documenting Data Production Processes: A
  Participatory Approach for Data Work}}.
\newblock \bibinfo{journal}{\emph{Proc. ACM Hum.-Comput. Interact.}}
  (\bibinfo{date}{Nov.} \bibinfo{year}{2022}).
\newblock
\urldef\tempurl%
\url{https://doi.org/10.1145/3555623}
\showDOI{\tempurl}


\bibitem[{Microsoft}(2022)]%
        {t2022responsible}
\bibfield{author}{\bibinfo{person}{{Microsoft}}.}
  \bibinfo{year}{2022}\natexlab{}.
\newblock \bibinfo{booktitle}{\emph{{Responsible AI}}}.
\newblock
\urldef\tempurl%
\url{https://www.microsoft.com/en-us/ai/responsible-ai}
\showURL{%
Retrieved February 2023 from \tempurl}


\bibitem[Mlynar et~al\mbox{.}(2022)]%
        {mlynar2022ai}
\bibfield{author}{\bibinfo{person}{Jakub Mlynar}, \bibinfo{person}{Farzaneh
  Bahrami}, \bibinfo{person}{André Ourednik}, \bibinfo{person}{Nico Mutzner},
  \bibinfo{person}{Himanshu Verma}, {and} \bibinfo{person}{Hamed Alavi}.}
  \bibinfo{year}{2022}\natexlab{}.
\newblock \showarticletitle{{AI beyond Deus Ex Machina – Reimagining
  Intelligence in Future Cities with Urban Experts}}. In
  \bibinfo{booktitle}{\emph{Proceedings of the 2022 {CHI} {Conference} on
  {Human} {Factors} in {Computing} {Systems}}} \emph{(\bibinfo{series}{{CHI}
  '22})}. \bibinfo{publisher}{ACM}.
\newblock
\urldef\tempurl%
\url{https://doi.org/10.1145/3491102.3517502}
\showDOI{\tempurl}


\bibitem[Moitra et~al\mbox{.}(2022)]%
        {moitra2022ai}
\bibfield{author}{\bibinfo{person}{Aparna Moitra}, \bibinfo{person}{Dennis
  Wagenaar}, \bibinfo{person}{Manveer Kalirai}, \bibinfo{person}{Syed~Ishtiaque
  Ahmed}, {and} \bibinfo{person}{Robert Soden}.}
  \bibinfo{year}{2022}\natexlab{}.
\newblock \showarticletitle{{AI and Disaster Risk: A Practitioner
  Perspective}}.
\newblock \bibinfo{journal}{\emph{Proc. ACM Hum.-Comput. Interact.}}
  (\bibinfo{date}{Nov.} \bibinfo{year}{2022}).
\newblock
\urldef\tempurl%
\url{https://doi.org/10.1145/3555163}
\showDOI{\tempurl}


\bibitem[Moore et~al\mbox{.}(2023)]%
        {moore2023failurenotes}
\bibfield{author}{\bibinfo{person}{Steven Moore}, \bibinfo{person}{Q.~Vera
  Liao}, {and} \bibinfo{person}{Hariharan Subramonyam}.}
  \bibinfo{year}{2023}\natexlab{}.
\newblock \showarticletitle{{FAIlureNotes: Supporting Designers in
  Understanding the Limits of AI Models for Computer Vision Tasks}}. In
  \bibinfo{booktitle}{\emph{Proceedings of the 2023 {CHI} {Conference} on
  {Human} {Factors} in {Computing} {Systems}}} \emph{(\bibinfo{series}{{CHI}
  '23})}. \bibinfo{publisher}{ACM}.
\newblock
\urldef\tempurl%
\url{https://doi.org/10.1145/3544548.3581242}
\showDOI{\tempurl}


\bibitem[Moreschi et~al\mbox{.}(2020)]%
        {moreschi2020brazilian}
\bibfield{author}{\bibinfo{person}{Bruno Moreschi}, \bibinfo{person}{Gabriel
  Pereira}, {and} \bibinfo{person}{Fabio~G Cozman}.}
  \bibinfo{year}{2020}\natexlab{}.
\newblock \showarticletitle{{The Brazilian Workers in Amazon Mechanical Turk:
  dreams and realities of ghost workers.}}
\newblock \bibinfo{journal}{\emph{Contracampo}} (\bibinfo{year}{2020}).
\newblock
\urldef\tempurl%
\url{https://doi.org/10.22409/contracampo.v39i1.38252}
\showDOI{\tempurl}


\bibitem[Morrison et~al\mbox{.}(2021)]%
        {morrison2021social}
\bibfield{author}{\bibinfo{person}{Cecily Morrison}, \bibinfo{person}{Edward
  Cutrell}, \bibinfo{person}{Martin Grayson}, \bibinfo{person}{Anja Thieme},
  \bibinfo{person}{Alex Taylor}, \bibinfo{person}{Geert Roumen},
  \bibinfo{person}{Camilla Longden}, \bibinfo{person}{Sebastian Tschiatschek},
  \bibinfo{person}{Rita Faia~Marques}, {and} \bibinfo{person}{Abigail Sellen}.}
  \bibinfo{year}{2021}\natexlab{}.
\newblock \showarticletitle{{Social Sensemaking with AI: Designing an
  Open-Ended AI Experience with a Blind Child}}. In
  \bibinfo{booktitle}{\emph{Proceedings of the 2021 {CHI} {Conference} on
  {Human} {Factors} in {Computing} {Systems}}} \emph{(\bibinfo{series}{{CHI}
  '21})}. \bibinfo{publisher}{ACM}.
\newblock
\urldef\tempurl%
\url{https://doi.org/10.1145/3411764.3445290}
\showDOI{\tempurl}


\bibitem[Muller et~al\mbox{.}(2022)]%
        {muller2022neurips}
\bibfield{author}{\bibinfo{person}{Michael Muller}, \bibinfo{person}{Plamen
  Agelov}, \bibinfo{person}{Shion Guha}, \bibinfo{person}{Marina Kogan},
  \bibinfo{person}{Gina Neff}, \bibinfo{person}{Nuria Oliver},
  \bibinfo{person}{Manuel~Gomez Rodriguez}, {and} \bibinfo{person}{Adrian
  Weller}.} \bibinfo{year}{2022}\natexlab{}.
\newblock \bibinfo{booktitle}{\emph{{NeurIPS 2021 Workshop Proposal: Human
  Centered AI}}}.
\newblock
\urldef\tempurl%
\url{https://sites.google.com/view/hcai-human-centered-ai-neurips/home}
\showURL{%
Retrieved March 2023 from \tempurl}


\bibitem[Muller and Strohmayer(2022)]%
        {muller2022forgetting}
\bibfield{author}{\bibinfo{person}{Michael Muller} {and}
  \bibinfo{person}{Angelika Strohmayer}.} \bibinfo{year}{2022}\natexlab{}.
\newblock \showarticletitle{{Forgetting Practices in the Data Sciences}}. In
  \bibinfo{booktitle}{\emph{Proceedings of the 2022 {CHI} {Conference} on
  {Human} {Factors} in {Computing} {Systems}}} \emph{(\bibinfo{series}{{CHI}
  '22})}. \bibinfo{publisher}{ACM}.
\newblock
\urldef\tempurl%
\url{https://doi.org/10.1145/3491102.3517644}
\showDOI{\tempurl}


\bibitem[Muller and Weisz(2022)]%
        {muller2022extending}
\bibfield{author}{\bibinfo{person}{Michael Muller} {and}
  \bibinfo{person}{Justin Weisz}.} \bibinfo{year}{2022}\natexlab{}.
\newblock \showarticletitle{{Extending a Human-AI Collaboration Framework with
  Dynamism and Sociality}}. In \bibinfo{booktitle}{\emph{2022 Symposium on
  Human-Computer Interaction for Work}} \emph{(\bibinfo{series}{CHIWORK
  2022})}. \bibinfo{publisher}{ACM}, Article \bibinfo{articleno}{10}.
\newblock
\urldef\tempurl%
\url{https://doi.org/10.1145/3533406.3533407}
\showDOI{\tempurl}


\bibitem[Muller et~al\mbox{.}(2021)]%
        {muller2021designing}
\bibfield{author}{\bibinfo{person}{Michael Muller},
  \bibinfo{person}{Christine~T. Wolf}, \bibinfo{person}{Josh Andres},
  \bibinfo{person}{Michael Desmond}, \bibinfo{person}{Narendra~Nath Joshi},
  \bibinfo{person}{Zahra Ashktorab}, \bibinfo{person}{Aabhas Sharma},
  \bibinfo{person}{Kristina Brimijoin}, \bibinfo{person}{Qian Pan},
  \bibinfo{person}{Evelyn Duesterwald}, {and} \bibinfo{person}{Casey Dugan}.}
  \bibinfo{year}{2021}\natexlab{}.
\newblock \showarticletitle{{Designing Ground Truth and the Social Life of
  Labels}}. In \bibinfo{booktitle}{\emph{Proceedings of the 2021 CHI Conference
  on Human Factors in Computing Systems}} \emph{(\bibinfo{series}{CHI '21})}.
  \bibinfo{publisher}{ACM}, Article \bibinfo{articleno}{94}.
\newblock
\urldef\tempurl%
\url{https://doi.org/10.1145/3411764.3445402}
\showDOI{\tempurl}


\bibitem[Nakao et~al\mbox{.}(2022)]%
        {nakao2022involving}
\bibfield{author}{\bibinfo{person}{Yuri Nakao}, \bibinfo{person}{Simone
  Stumpf}, \bibinfo{person}{Subeida Ahmed}, \bibinfo{person}{Aisha Naseer},
  {and} \bibinfo{person}{Lorenzo Strappelli}.} \bibinfo{year}{2022}\natexlab{}.
\newblock \showarticletitle{{Toward Involving End-Users in Interactive
  Human-in-the-Loop AI Fairness}}.
\newblock \bibinfo{journal}{\emph{ACM Trans. Interact. Intell. Syst.}}, Article
  \bibinfo{articleno}{18} (\bibinfo{date}{July} \bibinfo{year}{2022}).
\newblock
\showISSN{2160-6455}
\urldef\tempurl%
\url{https://doi.org/10.1145/3514258}
\showDOI{\tempurl}


\bibitem[Nashed et~al\mbox{.}(2021)]%
        {nashed2021ethically}
\bibfield{author}{\bibinfo{person}{Samer Nashed}, \bibinfo{person}{Justin
  Svegliato}, {and} \bibinfo{person}{Shlomo Zilberstein}.}
  \bibinfo{year}{2021}\natexlab{}.
\newblock \showarticletitle{{Ethically Compliant Planning within Moral
  Communities}}. In \bibinfo{booktitle}{\emph{Proceedings of the 2021
  {AAAI}/{ACM} {Conference} on {AI}, {Ethics}, and {Society}}}
  \emph{(\bibinfo{series}{{AIES} '21})}. \bibinfo{publisher}{ACM}.
\newblock
\urldef\tempurl%
\url{https://doi.org/10.1145/3461702.3462522}
\showDOI{\tempurl}


\bibitem[{National Institute of Standards and Technology}(2023)]%
        {standards2023ai}
\bibfield{author}{\bibinfo{person}{{National Institute of Standards and
  Technology}}.} \bibinfo{year}{2023}\natexlab{}.
\newblock \bibinfo{booktitle}{\emph{{AI Risk Management Framework}}}.
\newblock
\urldef\tempurl%
\url{https://www.nist.gov/itl/ai-risk-management-framework}
\showURL{%
Retrieved February 2023 from \tempurl}


\bibitem[Neuhauser and Kreps(2011)]%
        {neuhauser2011participatory}
\bibfield{author}{\bibinfo{person}{Linda Neuhauser} {and}
  \bibinfo{person}{Gary~L Kreps}.} \bibinfo{year}{2011}\natexlab{}.
\newblock \showarticletitle{{Participatory Design and Artificial Intelligence:
  Strategies to Improve Health Communication for Diverse Audiences}}. In
  \bibinfo{booktitle}{\emph{AAAI Spring Symposium: AI and Health
  Communication}}.
\newblock
\urldef\tempurl%
\url{https://researchers.mq.edu.au/en/publications/participatory-design-and-artificial-intelligence-strategies-to-im}
\showURL{%
\tempurl}


\bibitem[Nielsen(2021)]%
        {nielsen2021measuring}
\bibfield{author}{\bibinfo{person}{Aileen Nielsen}.}
  \bibinfo{year}{2021}\natexlab{}.
\newblock \showarticletitle{{Measuring Lay Reactions to Personal Data
  Markets}}. In \bibinfo{booktitle}{\emph{Proceedings of the 2021 {AAAI}/{ACM}
  {Conference} on {AI}, {Ethics}, and {Society}}}
  \emph{(\bibinfo{series}{{AIES} '21})}. \bibinfo{publisher}{ACM}.
\newblock
\urldef\tempurl%
\url{https://doi.org/10.1145/3461702.3462582}
\showDOI{\tempurl}


\bibitem[Niforatos et~al\mbox{.}(2020)]%
        {niforatos2020would}
\bibfield{author}{\bibinfo{person}{Evangelos Niforatos}, \bibinfo{person}{Adam
  Palma}, \bibinfo{person}{Roman Gluszny}, \bibinfo{person}{Athanasios
  Vourvopoulos}, {and} \bibinfo{person}{Fotis Liarokapis}.}
  \bibinfo{year}{2020}\natexlab{}.
\newblock \showarticletitle{{Would You Do It?: Enacting Moral Dilemmas in
  Virtual Reality for Understanding Ethical Decision-Making}}. In
  \bibinfo{booktitle}{\emph{Proceedings of the 2020 {CHI} {Conference} on
  {Human} {Factors} in {Computing} {Systems}}} \emph{(\bibinfo{series}{{CHI}
  '20})}. \bibinfo{publisher}{ACM}.
\newblock
\urldef\tempurl%
\url{https://doi.org/10.1145/3313831.3376788}
\showDOI{\tempurl}


\bibitem[{Nokia Bell Labs}(2022)]%
        {labs2022responsible}
\bibfield{author}{\bibinfo{person}{{Nokia Bell Labs}}.}
  \bibinfo{year}{2022}\natexlab{}.
\newblock \bibinfo{booktitle}{\emph{{Responsible AI}}}.
\newblock
\urldef\tempurl%
\url{https://www.bell-labs.com/research-innovation/responsible-ai/}
\showURL{%
Retrieved January 2023 from \tempurl}


\bibitem[Olteanu et~al\mbox{.}(2019)]%
        {olteanu2019social}
\bibfield{author}{\bibinfo{person}{Alexandra Olteanu}, \bibinfo{person}{Carlos
  Castillo}, \bibinfo{person}{Fernando Diaz}, {and} \bibinfo{person}{Emre
  Kıcıman}.} \bibinfo{year}{2019}\natexlab{}.
\newblock \showarticletitle{{Social Data: Biases, Methodological Pitfalls, and
  Ethical Boundaries}}.
\newblock \bibinfo{journal}{\emph{Frontiers in Big Data}}
  (\bibinfo{year}{2019}).
\newblock
\urldef\tempurl%
\url{https://doi.org/10.3389/fdata.2019.00013}
\showDOI{\tempurl}


\bibitem[Orphanou et~al\mbox{.}(2022)]%
        {orphanou2022mitigating}
\bibfield{author}{\bibinfo{person}{Kalia Orphanou}, \bibinfo{person}{Jahna
  Otterbacher}, \bibinfo{person}{Styliani Kleanthous},
  \bibinfo{person}{Khuyagbaatar Batsuren}, \bibinfo{person}{Fausto
  Giunchiglia}, \bibinfo{person}{Veronika Bogina},
  \bibinfo{person}{Avital~Shulner Tal}, \bibinfo{person}{Alan Hartman}, {and}
  \bibinfo{person}{Tsvi Kuflik}.} \bibinfo{year}{2022}\natexlab{}.
\newblock \showarticletitle{{Mitigating Bias in Algorithmic Systems--A Fish-Eye
  View}}.
\newblock \bibinfo{journal}{\emph{ACM Comput. Surv.}} (\bibinfo{date}{Dec.}
  \bibinfo{year}{2022}).
\newblock
\showISSN{0360-0300}
\urldef\tempurl%
\url{https://doi.org/10.1145/3527152}
\showDOI{\tempurl}


\bibitem[Page et~al\mbox{.}(2021)]%
        {page2021prisma}
\bibfield{author}{\bibinfo{person}{Matthew~J Page}, \bibinfo{person}{Joanne~E
  McKenzie}, \bibinfo{person}{Patrick~M Bossuyt}, \bibinfo{person}{Isabelle
  Boutron}, \bibinfo{person}{Tammy~C Hoffmann}, \bibinfo{person}{Cynthia~D
  Mulrow}, \bibinfo{person}{Larissa Shamseer}, \bibinfo{person}{Jennifer~M
  Tetzlaff}, \bibinfo{person}{Elie~A Akl}, \bibinfo{person}{Sue~E Brennan},
  \bibinfo{person}{Roger Chou}, \bibinfo{person}{Julie Glanville},
  \bibinfo{person}{Jeremy~M Grimshaw}, \bibinfo{person}{Asbj{\o}rn
  Hr{\'o}bjartsson}, \bibinfo{person}{Manoj~M Lalu}, \bibinfo{person}{Tianjing
  Li}, \bibinfo{person}{Elizabeth~W Loder}, \bibinfo{person}{Evan Mayo-Wilson},
  \bibinfo{person}{Steve McDonald}, \bibinfo{person}{Luke~A McGuinness},
  \bibinfo{person}{Lesley~A Stewart}, \bibinfo{person}{James Thomas},
  \bibinfo{person}{Andrea~C Tricco}, \bibinfo{person}{Vivian~A Welch},
  \bibinfo{person}{Penny Whiting}, {and} \bibinfo{person}{David Moher}.}
  \bibinfo{year}{2021}\natexlab{}.
\newblock \showarticletitle{{The PRISMA 2020 statement: an updated guideline
  for reporting systematic reviews}}.
\newblock \bibinfo{journal}{\emph{BMJ}} (\bibinfo{year}{2021}).
\newblock
\urldef\tempurl%
\url{https://doi.org/10.1136/bmj.n71}
\showDOI{\tempurl}


\bibitem[Park et~al\mbox{.}(2021)]%
        {park2021human}
\bibfield{author}{\bibinfo{person}{Hyanghee Park}, \bibinfo{person}{Daehwan
  Ahn}, \bibinfo{person}{Kartik Hosanagar}, {and} \bibinfo{person}{Joonhwan
  Lee}.} \bibinfo{year}{2021}\natexlab{}.
\newblock \showarticletitle{{Human-AI Interaction in Human Resource Management:
  Understanding Why Employees Resist Algorithmic Evaluation at Workplaces and
  How to Mitigate Burdens}}. In \bibinfo{booktitle}{\emph{Proceedings of the
  2021 {CHI} {Conference} on {Human} {Factors} in {Computing} {Systems}}}
  \emph{(\bibinfo{series}{{CHI} '21})}. \bibinfo{publisher}{ACM}.
\newblock
\urldef\tempurl%
\url{https://doi.org/10.1145/3411764.3445304}
\showDOI{\tempurl}


\bibitem[Park et~al\mbox{.}(2022)]%
        {park2022designing}
\bibfield{author}{\bibinfo{person}{Hyanghee Park}, \bibinfo{person}{Daehwan
  Ahn}, \bibinfo{person}{Kartik Hosanagar}, {and} \bibinfo{person}{Joonhwan
  Lee}.} \bibinfo{year}{2022}\natexlab{}.
\newblock \showarticletitle{{Designing Fair AI in Human Resource Management:
  Understanding Tensions Surrounding Algorithmic Evaluation and Envisioning
  Stakeholder-Centered Solutions}}. In \bibinfo{booktitle}{\emph{Proceedings of
  the 2022 {CHI} {Conference} on {Human} {Factors} in {Computing} {Systems}}}
  \emph{(\bibinfo{series}{{CHI} '22})}. \bibinfo{publisher}{ACM}.
\newblock
\urldef\tempurl%
\url{https://doi.org/10.1145/3491102.3517672}
\showDOI{\tempurl}


\bibitem[Passi and Jackson(2017)]%
        {passi2017data}
\bibfield{author}{\bibinfo{person}{Samir Passi} {and} \bibinfo{person}{Steven
  Jackson}.} \bibinfo{year}{2017}\natexlab{}.
\newblock \showarticletitle{{Data Vision: Learning to See Through Algorithmic
  Abstraction}}. In \bibinfo{booktitle}{\emph{Proceedings of the 2017 ACM
  Conference on Computer Supported Cooperative Work and Social Computing}}
  \emph{(\bibinfo{series}{CSCW '17})}. \bibinfo{publisher}{ACM}.
\newblock
\urldef\tempurl%
\url{https://doi.org/10.1145/2998181.2998331}
\showDOI{\tempurl}


\bibitem[Pessach and Shmueli(2022)]%
        {pessach2022review}
\bibfield{author}{\bibinfo{person}{Dana Pessach} {and} \bibinfo{person}{Erez
  Shmueli}.} \bibinfo{year}{2022}\natexlab{}.
\newblock \showarticletitle{{A Review on Fairness in Machine Learning}}.
\newblock \bibinfo{journal}{\emph{ACM Comput. Surv.}} (\bibinfo{date}{Feb.}
  \bibinfo{year}{2022}).
\newblock
\showISSN{0360-0300}
\urldef\tempurl%
\url{https://doi.org/10.1145/3494672}
\showDOI{\tempurl}


\bibitem[Peters et~al\mbox{.}(2013)]%
        {peters2013bridging}
\bibfield{author}{\bibinfo{person}{Anicia Peters}, \bibinfo{person}{Heike
  Winschiers-Theophilus}, {and} \bibinfo{person}{Brian Mennecke}.}
  \bibinfo{year}{2013}\natexlab{}.
\newblock \showarticletitle{{Bridging the Digital Divide through Facebook
  Friendships: A Cross-Cultural Study}}. In
  \bibinfo{booktitle}{\emph{Proceedings of the 2013 Conference on Computer
  Supported Cooperative Work Companion}} \emph{(\bibinfo{series}{CSCW '13})}.
  \bibinfo{publisher}{ACM}.
\newblock
\urldef\tempurl%
\url{https://doi.org/10.1145/2441955.2442014}
\showDOI{\tempurl}


\bibitem[Pine and Liboiron(2015)]%
        {pine2015politics}
\bibfield{author}{\bibinfo{person}{Kathleen~H. Pine} {and} \bibinfo{person}{Max
  Liboiron}.} \bibinfo{year}{2015}\natexlab{}.
\newblock \showarticletitle{{The Politics of Measurement and Action}}. In
  \bibinfo{booktitle}{\emph{Proceedings of the 33rd Annual ACM Conference on
  Human Factors in Computing Systems}} \emph{(\bibinfo{series}{CHI '15})}.
  \bibinfo{publisher}{ACM}.
\newblock
\urldef\tempurl%
\url{https://doi.org/10.1145/2702123.2702298}
\showDOI{\tempurl}


\bibitem[Pushkarna et~al\mbox{.}(2022)]%
        {pushkarna2022data}
\bibfield{author}{\bibinfo{person}{Mahima Pushkarna}, \bibinfo{person}{Andrew
  Zaldivar}, {and} \bibinfo{person}{Oddur Kjartansson}.}
  \bibinfo{year}{2022}\natexlab{}.
\newblock \showarticletitle{{Data Cards: Purposeful and Transparent Dataset
  Documentation for Responsible AI}}. In \bibinfo{booktitle}{\emph{2022 {ACM}
  {Conference} on {Fairness}, {Accountability}, and {Transparency}}}
  \emph{(\bibinfo{series}{{FAccT} '22})}. \bibinfo{publisher}{ACM}.
\newblock
\urldef\tempurl%
\url{https://doi.org/10.1145/3531146.3533231}
\showDOI{\tempurl}


\bibitem[{PwC}(2022)]%
        {c2022pwc}
\bibfield{author}{\bibinfo{person}{{PwC}}.} \bibinfo{year}{2022}\natexlab{}.
\newblock \bibinfo{booktitle}{\emph{{PwC's Responsible AI}}}.
\newblock
\urldef\tempurl%
\url{https://www.pwc.com/gx/en/issues/data-and-analytics/artificial-intelligence/what-is-responsible-ai/pwc-responsible-ai.pdf}
\showURL{%
Retrieved February 2023 from \tempurl}


\bibitem[Rakova et~al\mbox{.}(2021)]%
        {rakova2021where}
\bibfield{author}{\bibinfo{person}{Bogdana Rakova}, \bibinfo{person}{Jingying
  Yang}, \bibinfo{person}{Henriette Cramer}, {and} \bibinfo{person}{Rumman
  Chowdhury}.} \bibinfo{year}{2021}\natexlab{}.
\newblock \showarticletitle{{Where Responsible AI Meets Reality: Practitioner
  Perspectives on Enablers for Shifting Organizational Practices}}.
\newblock \bibinfo{journal}{\emph{Proc. ACM Hum.-Comput. Interact.}}
  (\bibinfo{date}{April} \bibinfo{year}{2021}).
\newblock
\urldef\tempurl%
\url{https://doi.org/10.1145/3449081}
\showDOI{\tempurl}


\bibitem[Ramesh et~al\mbox{.}(2022)]%
        {ramesh2022how}
\bibfield{author}{\bibinfo{person}{Divya Ramesh}, \bibinfo{person}{Vaishnav
  Kameswaran}, \bibinfo{person}{Ding Wang}, {and} \bibinfo{person}{Nithya
  Sambasivan}.} \bibinfo{year}{2022}\natexlab{}.
\newblock \showarticletitle{{How Platform-User Power Relations Shape
  Algorithmic Accountability: A Case Study of Instant Loan Platforms and
  Financially Stressed Users in India}}. In \bibinfo{booktitle}{\emph{2022
  {ACM} {Conference} on {Fairness}, {Accountability}, and {Transparency}}}
  \emph{(\bibinfo{series}{{FAccT} '22})}. \bibinfo{publisher}{ACM}.
\newblock
\urldef\tempurl%
\url{https://doi.org/10.1145/3531146.3533237}
\showDOI{\tempurl}


\bibitem[Raz et~al\mbox{.}(2021)]%
        {raz2021face}
\bibfield{author}{\bibinfo{person}{Daniella Raz}, \bibinfo{person}{Corinne
  Bintz}, \bibinfo{person}{Vivian Guetler}, \bibinfo{person}{Aaron Tam},
  \bibinfo{person}{Michael Katell}, \bibinfo{person}{Dharma Dailey},
  \bibinfo{person}{Bernease Herman}, \bibinfo{person}{P.~M. Krafft}, {and}
  \bibinfo{person}{Meg Young}.} \bibinfo{year}{2021}\natexlab{}.
\newblock \showarticletitle{{Face Mis-ID: An Interactive Pedagogical Tool
  Demonstrating Disparate Accuracy Rates in Facial Recognition}}. In
  \bibinfo{booktitle}{\emph{Proceedings of the 2021 {AAAI}/{ACM} {Conference}
  on {AI}, {Ethics}, and {Society}}} \emph{(\bibinfo{series}{{AIES} '21})}.
  \bibinfo{publisher}{ACM}.
\newblock
\urldef\tempurl%
\url{https://doi.org/10.1145/3461702.3462627}
\showDOI{\tempurl}


\bibitem[Reinecke et~al\mbox{.}(2013)]%
        {reinecke2013doodle}
\bibfield{author}{\bibinfo{person}{Katharina Reinecke},
  \bibinfo{person}{Minh~Khoa Nguyen}, \bibinfo{person}{Abraham Bernstein},
  \bibinfo{person}{Michael N\"{a}f}, {and} \bibinfo{person}{Krzysztof~Z.
  Gajos}.} \bibinfo{year}{2013}\natexlab{}.
\newblock \showarticletitle{{Doodle around the World: Online Scheduling
  Behavior Reflects Cultural Differences in Time Perception and Group
  Decision-Making}}. In \bibinfo{booktitle}{\emph{Proceedings of the 2013
  Conference on Computer Supported Cooperative Work}}
  \emph{(\bibinfo{series}{CSCW '13})}. \bibinfo{publisher}{ACM}.
\newblock
\urldef\tempurl%
\url{https://doi.org/10.1145/2441776.2441784}
\showDOI{\tempurl}


\bibitem[Reitmaier et~al\mbox{.}(2022)]%
        {reitmaier2022opportunities}
\bibfield{author}{\bibinfo{person}{Thomas Reitmaier}, \bibinfo{person}{Electra
  Wallington}, \bibinfo{person}{Dani Kalarikalayil~Raju},
  \bibinfo{person}{Ondrej Klejch}, \bibinfo{person}{Jennifer Pearson},
  \bibinfo{person}{Matt Jones}, \bibinfo{person}{Peter Bell}, {and}
  \bibinfo{person}{Simon Robinson}.} \bibinfo{year}{2022}\natexlab{}.
\newblock \showarticletitle{{Opportunities and Challenges of Automatic Speech
  Recognition Systems for Low-Resource Language Speakers}}. In
  \bibinfo{booktitle}{\emph{Proceedings of the 2022 {CHI} {Conference} on
  {Human} {Factors} in {Computing} {Systems}}} \emph{(\bibinfo{series}{{CHI}
  '22})}. \bibinfo{publisher}{ACM}.
\newblock
\urldef\tempurl%
\url{https://doi.org/10.1145/3491102.3517639}
\showDOI{\tempurl}


\bibitem[Richards et~al\mbox{.}(2019)]%
        {richards2019factsheets}
\bibfield{author}{\bibinfo{person}{John Richards}, \bibinfo{person}{David
  Piorkowski}, \bibinfo{person}{Michael Hind}, \bibinfo{person}{Stephanie
  Houde}, {and} \bibinfo{person}{Aleksandra Mojsilović}.}
  \bibinfo{year}{2019}\natexlab{}.
\newblock \showarticletitle{{FactSheets: Increasing trust in AI services
  through supplier's declarations of conformity}}.
\newblock \bibinfo{journal}{\emph{IBM Journal of Research and Development}}
  (\bibinfo{year}{2019}).
\newblock
\urldef\tempurl%
\url{https://doi.org/10.1147/JRD.2019.2942288}
\showDOI{\tempurl}


\bibitem[Richardson et~al\mbox{.}(2021)]%
        {richardson2021towards}
\bibfield{author}{\bibinfo{person}{Brianna Richardson}, \bibinfo{person}{Jean
  Garcia-Gathright}, \bibinfo{person}{Samuel~F. Way}, \bibinfo{person}{Jennifer
  Thom}, {and} \bibinfo{person}{Henriette Cramer}.}
  \bibinfo{year}{2021}\natexlab{}.
\newblock \showarticletitle{{Towards Fairness in Practice: A
  Practitioner-Oriented Rubric for Evaluating Fair ML Toolkits}}. In
  \bibinfo{booktitle}{\emph{Proceedings of the 2021 {CHI} {Conference} on
  {Human} {Factors} in {Computing} {Systems}}} \emph{(\bibinfo{series}{{CHI}
  '21})}. \bibinfo{publisher}{ACM}.
\newblock
\urldef\tempurl%
\url{https://doi.org/10.1145/3411764.3445604}
\showDOI{\tempurl}


\bibitem[Rismani et~al\mbox{.}(2023)]%
        {rismani2023plane}
\bibfield{author}{\bibinfo{person}{Shalaleh Rismani}, \bibinfo{person}{Renee
  Shelby}, \bibinfo{person}{Andrew Smart}, \bibinfo{person}{Edgar Jatho},
  \bibinfo{person}{Joshua Kroll}, \bibinfo{person}{AJung Moon}, {and}
  \bibinfo{person}{Negar Rostamzadeh}.} \bibinfo{year}{2023}\natexlab{}.
\newblock \showarticletitle{{From Plane Crashes to Algorithmic Harm:
  Applicability of Safety Engineering Frameworks for Responsible ML}}. In
  \bibinfo{booktitle}{\emph{Proceedings of the 2023 {CHI} {Conference} on
  {Human} {Factors} in {Computing} {Systems}}} \emph{(\bibinfo{series}{{CHI}
  '23})}. \bibinfo{publisher}{ACM}.
\newblock
\urldef\tempurl%
\url{https://doi.org/10.1145/3544548.3581407}
\showDOI{\tempurl}


\bibitem[Roemmich and Andalibi(2021)]%
        {roemmich2021data}
\bibfield{author}{\bibinfo{person}{Kat Roemmich} {and} \bibinfo{person}{Nazanin
  Andalibi}.} \bibinfo{year}{2021}\natexlab{}.
\newblock \showarticletitle{{Data Subjects' Conceptualizations of and Attitudes
  Toward Automatic Emotion Recognition-Enabled Wellbeing Interventions on
  Social Media}}.
\newblock \bibinfo{journal}{\emph{Proc. ACM Hum.-Comput. Interact.}}
  (\bibinfo{date}{Oct.} \bibinfo{year}{2021}).
\newblock
\urldef\tempurl%
\url{https://doi.org/10.1145/3476049}
\showDOI{\tempurl}


\bibitem[Sambasivan et~al\mbox{.}(2021a)]%
        {sambasivan2021re}
\bibfield{author}{\bibinfo{person}{Nithya Sambasivan}, \bibinfo{person}{Erin
  Arnesen}, \bibinfo{person}{Ben Hutchinson}, \bibinfo{person}{Tulsee Doshi},
  {and} \bibinfo{person}{Vinodkumar Prabhakaran}.}
  \bibinfo{year}{2021}\natexlab{a}.
\newblock \showarticletitle{{Re-Imagining Algorithmic Fairness in India and
  Beyond}}. In \bibinfo{booktitle}{\emph{Proceedings of the 2021 {ACM}
  {Conference} on {Fairness}, {Accountability}, and {Transparency}}}
  \emph{(\bibinfo{series}{{FAccT} '21})}. \bibinfo{publisher}{ACM}.
\newblock
\urldef\tempurl%
\url{https://doi.org/10.1145/3442188.3445896}
\showDOI{\tempurl}


\bibitem[Sambasivan et~al\mbox{.}(2021b)]%
        {sambasivan2021everyone}
\bibfield{author}{\bibinfo{person}{Nithya Sambasivan}, \bibinfo{person}{Shivani
  Kapania}, \bibinfo{person}{Hannah Highfill}, \bibinfo{person}{Diana Akrong},
  \bibinfo{person}{Praveen Paritosh}, {and} \bibinfo{person}{Lora~M Aroyo}.}
  \bibinfo{year}{2021}\natexlab{b}.
\newblock \showarticletitle{{“Everyone Wants to Do the Model Work, Not the
  Data Work”: Data Cascades in High-Stakes AI}}. In
  \bibinfo{booktitle}{\emph{Proceedings of the 2021 {CHI} {Conference} on
  {Human} {Factors} in {Computing} {Systems}}} \emph{(\bibinfo{series}{{CHI}
  '21})}. \bibinfo{publisher}{ACM}.
\newblock
\urldef\tempurl%
\url{https://doi.org/10.1145/3411764.3445518}
\showDOI{\tempurl}


\bibitem[Saxena et~al\mbox{.}(2020)]%
        {saxena2020human}
\bibfield{author}{\bibinfo{person}{Devansh Saxena}, \bibinfo{person}{Karla
  Badillo-Urquiola}, \bibinfo{person}{Pamela~J. Wisniewski}, {and}
  \bibinfo{person}{Shion Guha}.} \bibinfo{year}{2020}\natexlab{}.
\newblock \showarticletitle{{A Human-Centered Review of Algorithms Used within
  the U.S. Child Welfare System}}. In \bibinfo{booktitle}{\emph{Proceedings of
  the 2020 CHI Conference on Human Factors in Computing Systems}}
  \emph{(\bibinfo{series}{CHI '20})}. \bibinfo{publisher}{ACM}.
\newblock
\urldef\tempurl%
\url{https://doi.org/10.1145/3313831.3376229}
\showDOI{\tempurl}


\bibitem[Schiff et~al\mbox{.}(2020)]%
        {schiff2020what}
\bibfield{author}{\bibinfo{person}{Daniel Schiff}, \bibinfo{person}{Justin
  Biddle}, \bibinfo{person}{Jason Borenstein}, {and} \bibinfo{person}{Kelly
  Laas}.} \bibinfo{year}{2020}\natexlab{}.
\newblock \showarticletitle{{What's Next for AI Ethics, Policy, and Governance?
  A Global Overview}}. In \bibinfo{booktitle}{\emph{Proceedings of the
  {AAAI}/{ACM} {Conference} on {AI}, {Ethics}, and {Society}}}
  \emph{(\bibinfo{series}{{AIES} '20})}. \bibinfo{publisher}{ACM}.
\newblock
\urldef\tempurl%
\url{https://doi.org/10.1145/3375627.3375804}
\showDOI{\tempurl}


\bibitem[Septiandri et~al\mbox{.}(2023)]%
        {septiandri2023weird}
\bibfield{author}{\bibinfo{person}{Ali~Akbar Septiandri},
  \bibinfo{person}{Marios Constantinides}, \bibinfo{person}{Mohammad Tahaei},
  {and} \bibinfo{person}{Daniele Quercia}.} \bibinfo{year}{2023}\natexlab{}.
\newblock \showarticletitle{WEIRD FAccTs: How Western, Educated,
  Industrialized, Rich, and Democratic is FAccT?}. In
  \bibinfo{booktitle}{\emph{Proceedings of the 2023 ACM Conference on Fairness,
  Accountability, and Transparency}} \emph{(\bibinfo{series}{FAccT '23})}.
  \bibinfo{publisher}{ACM}.
\newblock
\urldef\tempurl%
\url{https://doi.org/10.1145/3593013.3593985}
\showDOI{\tempurl}


\bibitem[Shahid and Vashistha(2023)]%
        {shahid2023decolonizing}
\bibfield{author}{\bibinfo{person}{Farhana Shahid} {and}
  \bibinfo{person}{Aditya Vashistha}.} \bibinfo{year}{2023}\natexlab{}.
\newblock \showarticletitle{{Decolonizing Content Moderation: Does Uniform
  Global Community Standard Resemble Utopian Equality or Western Power
  Hegemony?}}. In \bibinfo{booktitle}{\emph{Proceedings of the 2023 {CHI}
  {Conference} on {Human} {Factors} in {Computing} {Systems}}}
  \emph{(\bibinfo{series}{{CHI} '23})}. \bibinfo{publisher}{ACM}.
\newblock
\urldef\tempurl%
\url{https://doi.org/10.1145/3544548.3581538}
\showDOI{\tempurl}


\bibitem[Sharma et~al\mbox{.}(2021)]%
        {sharma2021fair}
\bibfield{author}{\bibinfo{person}{Shubham Sharma}, \bibinfo{person}{Alan~H.
  Gee}, \bibinfo{person}{David Paydarfar}, {and} \bibinfo{person}{Joydeep
  Ghosh}.} \bibinfo{year}{2021}\natexlab{}.
\newblock \showarticletitle{{FaiR-N: Fair and Robust Neural Networks for
  Structured Data}}. In \bibinfo{booktitle}{\emph{Proceedings of the 2021
  {AAAI}/{ACM} {Conference} on {AI}, {Ethics}, and {Society}}}
  \emph{(\bibinfo{series}{{AIES} '21})}. \bibinfo{publisher}{ACM}.
\newblock
\urldef\tempurl%
\url{https://doi.org/10.1145/3461702.3462559}
\showDOI{\tempurl}


\bibitem[Shaw et~al\mbox{.}(2018)]%
        {shaw2018towards}
\bibfield{author}{\bibinfo{person}{Nolan~P. Shaw}, \bibinfo{person}{Andreas
  Stöckel}, \bibinfo{person}{Ryan~W. Orr}, \bibinfo{person}{Thomas~F.
  Lidbetter}, {and} \bibinfo{person}{Robin Cohen}.}
  \bibinfo{year}{2018}\natexlab{}.
\newblock \showarticletitle{{Towards Provably Moral AI Agents in Bottom-up
  Learning Frameworks}}. In \bibinfo{booktitle}{\emph{Proceedings of the 2018
  {AAAI}/{ACM} {Conference} on {AI}, {Ethics}, and {Society}}}
  \emph{(\bibinfo{series}{{AIES} '18})}. \bibinfo{publisher}{ACM}.
\newblock
\urldef\tempurl%
\url{https://doi.org/10.1145/3278721.3278728}
\showDOI{\tempurl}


\bibitem[Shen et~al\mbox{.}(2021)]%
        {shen2021everyday}
\bibfield{author}{\bibinfo{person}{Hong Shen}, \bibinfo{person}{Alicia DeVos},
  \bibinfo{person}{Motahhare Eslami}, {and} \bibinfo{person}{Kenneth
  Holstein}.} \bibinfo{year}{2021}\natexlab{}.
\newblock \showarticletitle{{Everyday Algorithm Auditing: Understanding the
  Power of Everyday Users in Surfacing Harmful Algorithmic Behaviors}}.
\newblock \bibinfo{journal}{\emph{Proc. ACM Hum.-Comput. Interact.}}
  (\bibinfo{date}{Oct.} \bibinfo{year}{2021}).
\newblock
\urldef\tempurl%
\url{https://doi.org/10.1145/3479577}
\showDOI{\tempurl}


\bibitem[Shen et~al\mbox{.}(2020)]%
        {shen2020designing}
\bibfield{author}{\bibinfo{person}{Hong Shen}, \bibinfo{person}{Haojian Jin},
  \bibinfo{person}{Ángel~Alexander Cabrera}, \bibinfo{person}{Adam Perer},
  \bibinfo{person}{Haiyi Zhu}, {and} \bibinfo{person}{Jason~I. Hong}.}
  \bibinfo{year}{2020}\natexlab{}.
\newblock \showarticletitle{{Designing Alternative Representations of Confusion
  Matrices to Support Non-Expert Public Understanding of Algorithm
  Performance}}.
\newblock \bibinfo{journal}{\emph{Proc. ACM Hum.-Comput. Interact.}}
  (\bibinfo{date}{Oct.} \bibinfo{year}{2020}).
\newblock
\urldef\tempurl%
\url{https://doi.org/10.1145/3415224}
\showDOI{\tempurl}


\bibitem[Shneiderman(2020)]%
        {shneiderman2020bridging}
\bibfield{author}{\bibinfo{person}{Ben Shneiderman}.}
  \bibinfo{year}{2020}\natexlab{}.
\newblock \showarticletitle{{Bridging the Gap Between Ethics and Practice:
  Guidelines for Reliable, Safe, and Trustworthy Human-Centered AI Systems}}.
\newblock \bibinfo{journal}{\emph{ACM Trans. Interact. Intell. Syst.}}
  (\bibinfo{date}{Oct.} \bibinfo{year}{2020}).
\newblock
\showISSN{2160-6455}
\urldef\tempurl%
\url{https://doi.org/10.1145/3419764}
\showDOI{\tempurl}


\bibitem[Shneiderman(2022)]%
        {shneiderman2022human}
\bibfield{author}{\bibinfo{person}{Ben Shneiderman}.}
  \bibinfo{year}{2022}\natexlab{}.
\newblock \bibinfo{booktitle}{\emph{{Human-centered AI}}}.
\newblock \bibinfo{publisher}{Oxford University Press}.
\newblock


\bibitem[Siala and Wang(2022)]%
        {siala2022shifting}
\bibfield{author}{\bibinfo{person}{Haytham Siala} {and}
  \bibinfo{person}{Yichuan Wang}.} \bibinfo{year}{2022}\natexlab{}.
\newblock \showarticletitle{{SHIFTing artificial intelligence to be responsible
  in healthcare: A systematic review}}.
\newblock \bibinfo{journal}{\emph{Social Science \& Medicine}}
  (\bibinfo{year}{2022}).
\newblock
\showISSN{0277-9536}
\urldef\tempurl%
\url{https://doi.org/10.1016/j.socscimed.2022.114782}
\showDOI{\tempurl}


\bibitem[Siapka(2022)]%
        {siapka2022towards}
\bibfield{author}{\bibinfo{person}{Anastasia Siapka}.}
  \bibinfo{year}{2022}\natexlab{}.
\newblock \showarticletitle{{Towards a Feminist Metaethics of AI}}. In
  \bibinfo{booktitle}{\emph{Proceedings of the 2022 {AAAI}/{ACM} {Conference}
  on {AI}, {Ethics}, and {Society}}} \emph{(\bibinfo{series}{{AIES} '22})}.
  \bibinfo{publisher}{ACM}.
\newblock
\urldef\tempurl%
\url{https://doi.org/10.1145/3514094.3534197}
\showDOI{\tempurl}


\bibitem[Silva and Kenney(2018)]%
        {silva2018algorithms}
\bibfield{author}{\bibinfo{person}{Selena Silva} {and} \bibinfo{person}{Martin
  Kenney}.} \bibinfo{year}{2018}\natexlab{}.
\newblock \showarticletitle{{Algorithms, platforms, and ethnic bias: An
  integrative essay}}.
\newblock \bibinfo{journal}{\emph{Phylon (1960-)}} (\bibinfo{year}{2018}).
\newblock
\urldef\tempurl%
\url{https://www.jstor.org/stable/26545017}
\showURL{%
\tempurl}


\bibitem[Simons et~al\mbox{.}(2021)]%
        {simons2021machine}
\bibfield{author}{\bibinfo{person}{Joshua Simons}, \bibinfo{person}{Sophia
  Adams~Bhatti}, {and} \bibinfo{person}{Adrian Weller}.}
  \bibinfo{year}{2021}\natexlab{}.
\newblock \showarticletitle{{Machine Learning and the Meaning of Equal
  Treatment}}. In \bibinfo{booktitle}{\emph{Proceedings of the 2021
  {AAAI}/{ACM} {Conference} on {AI}, {Ethics}, and {Society}}}
  \emph{(\bibinfo{series}{{AIES} '21})}. \bibinfo{publisher}{ACM}.
\newblock
\urldef\tempurl%
\url{https://doi.org/10.1145/3461702.3462556}
\showDOI{\tempurl}


\bibitem[Sloane and Zakrzewski(2022)]%
        {sloane2022german}
\bibfield{author}{\bibinfo{person}{Mona Sloane} {and} \bibinfo{person}{Janina
  Zakrzewski}.} \bibinfo{year}{2022}\natexlab{}.
\newblock \showarticletitle{{German AI Start-Ups and “AI Ethics”: Using A
  Social Practice Lens for Assessing and Implementing Socio-Technical
  Innovation}}. In \bibinfo{booktitle}{\emph{2022 {ACM} {Conference} on
  {Fairness}, {Accountability}, and {Transparency}}}
  \emph{(\bibinfo{series}{{FAccT} '22})}. \bibinfo{publisher}{ACM}.
\newblock
\urldef\tempurl%
\url{https://doi.org/10.1145/3531146.3533156}
\showDOI{\tempurl}


\bibitem[Smith et~al\mbox{.}(2022)]%
        {smith2022real}
\bibfield{author}{\bibinfo{person}{Jessie~J. Smith}, \bibinfo{person}{Saleema
  Amershi}, \bibinfo{person}{Solon Barocas}, \bibinfo{person}{Hanna Wallach},
  {and} \bibinfo{person}{Jennifer Wortman~Vaughan}.}
  \bibinfo{year}{2022}\natexlab{}.
\newblock \showarticletitle{{REAL ML: Recognizing, Exploring, and Articulating
  Limitations of Machine Learning Research}}. In \bibinfo{booktitle}{\emph{2022
  {ACM} {Conference} on {Fairness}, {Accountability}, and {Transparency}}}
  \emph{(\bibinfo{series}{{FAccT} '22})}. \bibinfo{publisher}{ACM}.
\newblock
\urldef\tempurl%
\url{https://doi.org/10.1145/3531146.3533122}
\showDOI{\tempurl}


\bibitem[Stapleton et~al\mbox{.}(2022)]%
        {stapleton2022imagining}
\bibfield{author}{\bibinfo{person}{Logan Stapleton}, \bibinfo{person}{Min~Hun
  Lee}, \bibinfo{person}{Diana Qing}, \bibinfo{person}{Marya Wright},
  \bibinfo{person}{Alexandra Chouldechova}, \bibinfo{person}{Ken Holstein},
  \bibinfo{person}{Zhiwei~Steven Wu}, {and} \bibinfo{person}{Haiyi Zhu}.}
  \bibinfo{year}{2022}\natexlab{}.
\newblock \showarticletitle{{Imagining New Futures beyond Predictive Systems in
  Child Welfare: A Qualitative Study with Impacted Stakeholders}}. In
  \bibinfo{booktitle}{\emph{2022 {ACM} {Conference} on {Fairness},
  {Accountability}, and {Transparency}}} \emph{(\bibinfo{series}{{FAccT}
  '22})}. \bibinfo{publisher}{ACM}.
\newblock
\urldef\tempurl%
\url{https://doi.org/10.1145/3531146.3533177}
\showDOI{\tempurl}


\bibitem[Stark and Hoey(2021)]%
        {stark2021ethics}
\bibfield{author}{\bibinfo{person}{Luke Stark} {and} \bibinfo{person}{Jesse
  Hoey}.} \bibinfo{year}{2021}\natexlab{}.
\newblock \showarticletitle{{The Ethics of Emotion in Artificial Intelligence
  Systems}}. In \bibinfo{booktitle}{\emph{Proceedings of the 2021 {ACM}
  {Conference} on {Fairness}, {Accountability}, and {Transparency}}}
  \emph{(\bibinfo{series}{{FAccT} '21})}. \bibinfo{publisher}{ACM}.
\newblock
\urldef\tempurl%
\url{https://doi.org/10.1145/3442188.3445939}
\showDOI{\tempurl}


\bibitem[Steiger et~al\mbox{.}(2021)]%
        {steiger2021psychological}
\bibfield{author}{\bibinfo{person}{Miriah Steiger}, \bibinfo{person}{Timir~J
  Bharucha}, \bibinfo{person}{Sukrit Venkatagiri}, \bibinfo{person}{Martin~J.
  Riedl}, {and} \bibinfo{person}{Matthew Lease}.}
  \bibinfo{year}{2021}\natexlab{}.
\newblock \showarticletitle{{The Psychological Well-Being of Content
  Moderators: The Emotional Labor of Commercial Moderation and Avenues for
  Improving Support}}. In \bibinfo{booktitle}{\emph{Proceedings of the 2021
  {CHI} {Conference} on {Human} {Factors} in {Computing} {Systems}}}
  \emph{(\bibinfo{series}{{CHI} '21})}. \bibinfo{publisher}{ACM}.
\newblock
\urldef\tempurl%
\url{https://doi.org/10.1145/3411764.3445092}
\showDOI{\tempurl}


\bibitem[Subramonyam et~al\mbox{.}(2022)]%
        {subramonyam2022solving}
\bibfield{author}{\bibinfo{person}{Hariharan Subramonyam},
  \bibinfo{person}{Jane Im}, \bibinfo{person}{Colleen Seifert}, {and}
  \bibinfo{person}{Eytan Adar}.} \bibinfo{year}{2022}\natexlab{}.
\newblock \showarticletitle{{Solving Separation-of-Concerns Problems in
  Collaborative Design of Human-AI Systems through Leaky Abstractions}}. In
  \bibinfo{booktitle}{\emph{{CHI} {Conference} on {Human} {Factors} in
  {Computing} {Systems}}} \emph{(\bibinfo{series}{CHI '22})}.
  \bibinfo{publisher}{ACM}.
\newblock
\urldef\tempurl%
\url{https://doi.org/10.1145/3491102.3517537}
\showDOI{\tempurl}


\bibitem[Sun et~al\mbox{.}(2019)]%
        {sun2019mitigating}
\bibfield{author}{\bibinfo{person}{Tony Sun}, \bibinfo{person}{Andrew Gaut},
  \bibinfo{person}{Shirlyn Tang}, \bibinfo{person}{Yuxin Huang},
  \bibinfo{person}{Mai ElSherief}, \bibinfo{person}{Jieyu Zhao},
  \bibinfo{person}{Diba Mirza}, \bibinfo{person}{Elizabeth Belding},
  \bibinfo{person}{Kai-Wei Chang}, {and} \bibinfo{person}{William~Yang Wang}.}
  \bibinfo{year}{2019}\natexlab{}.
\newblock \showarticletitle{{Mitigating Gender Bias in Natural Language
  Processing: Literature Review}}. In \bibinfo{booktitle}{\emph{Proceedings of
  the 57th Annual Meeting of the Association for Computational Linguistics}}.
  \bibinfo{publisher}{Association for Computational Linguistics}.
\newblock
\urldef\tempurl%
\url{https://doi.org/10.18653/v1/P19-1159}
\showDOI{\tempurl}


\bibitem[Tahaei et~al\mbox{.}(2023)]%
        {tahaei2023human}
\bibfield{author}{\bibinfo{person}{Mohammad Tahaei}, \bibinfo{person}{Marios
  Constantinides}, \bibinfo{person}{Daniele Quercia}, \bibinfo{person}{Sean
  Kennedy}, \bibinfo{person}{Michael Muller}, \bibinfo{person}{Simone Stumpf},
  \bibinfo{person}{Q.~Vera Liao}, \bibinfo{person}{Ricardo Baeza-Yates},
  \bibinfo{person}{Lora Aroyo}, \bibinfo{person}{Jess Holbrook},
  \bibinfo{person}{Ewa Luger}, \bibinfo{person}{Michael Madaio},
  \bibinfo{person}{Ilana~Golbin Blumenfeld}, \bibinfo{person}{Maria
  De-Arteaga}, \bibinfo{person}{Jessica Vitak}, {and}
  \bibinfo{person}{Alexandra Olteanu}.} \bibinfo{year}{2023}\natexlab{}.
\newblock \showarticletitle{{Human-Centered Responsible Artificial
  Intelligence: Current \& Future Trends}}. In
  \bibinfo{booktitle}{\emph{Extended Abstracts of the 2023 CHI Conference on
  Human Factors in Computing Systems}} \emph{(\bibinfo{series}{CHI EA '23})}.
  \bibinfo{publisher}{ACM}.
\newblock
\urldef\tempurl%
\url{https://doi.org/10.1145/3544549.3583178}
\showDOI{\tempurl}


\bibitem[Tahaei and Vaniea(2019)]%
        {tahaei2019survey}
\bibfield{author}{\bibinfo{person}{Mohammad Tahaei} {and} \bibinfo{person}{Kami
  Vaniea}.} \bibinfo{year}{2019}\natexlab{}.
\newblock \showarticletitle{{A Survey on Developer-Centred Security}}. In
  \bibinfo{booktitle}{\emph{{2019 IEEE European Symposium on Security and
  Privacy Workshops (EuroS\&PW)}}}. \bibinfo{publisher}{{IEEE}}.
\newblock
\urldef\tempurl%
\url{https://doi.org/10.1109/EuroSPW.2019.00021}
\showDOI{\tempurl}


\bibitem[Tahaei et~al\mbox{.}(2020)]%
        {tahaei2020so}
\bibfield{author}{\bibinfo{person}{Mohammad Tahaei}, \bibinfo{person}{Kami
  Vaniea}, {and} \bibinfo{person}{Naomi Saphra}.}
  \bibinfo{year}{2020}\natexlab{}.
\newblock \showarticletitle{{Understanding Privacy-Related Questions on Stack
  Overflow}}. In \bibinfo{booktitle}{\emph{{Proceedings of the 2020 CHI
  Conference on Human Factors in Computing Systems}}}
  \emph{(\bibinfo{series}{CHI '20})}. \bibinfo{publisher}{{ACM}},
  \bibinfo{numpages}{14}~pages.
\newblock
\urldef\tempurl%
\url{https://doi.org/10.1145/3313831.3376768}
\showDOI{\tempurl}


\bibitem[Terzis(2020)]%
        {terzis2020onward}
\bibfield{author}{\bibinfo{person}{Petros Terzis}.}
  \bibinfo{year}{2020}\natexlab{}.
\newblock \showarticletitle{{Onward for the Freedom of Others: Marching beyond
  the AI Ethics}}. In \bibinfo{booktitle}{\emph{Proceedings of the 2020
  {Conference} on {Fairness}, {Accountability}, and {Transparency}}}
  \emph{(\bibinfo{series}{{FAT}* '20})}. \bibinfo{publisher}{ACM}.
\newblock
\urldef\tempurl%
\url{https://doi.org/10.1145/3351095.3373152}
\showDOI{\tempurl}


\bibitem[{The European Parliament and the Council of the European
  Union}(2018)]%
        {parliament2018general}
\bibfield{author}{\bibinfo{person}{{The European Parliament and the Council of
  the European Union}}.} \bibinfo{year}{2018}\natexlab{}.
\newblock \bibinfo{booktitle}{\emph{{General Data Protection Regulation
  (GDPR)}}}.
\newblock {The European Parliament and the Council of the European Union}.
\newblock
\urldef\tempurl%
\url{https://eur-lex.europa.eu/legal-content/EN/TXT/PDF/?uri=CELEX:32016R0679}
\showURL{%
Retrieved January 2023 from \tempurl}


\bibitem[{The Organisation for Economic Co-operation and Development
  (OECD)}(2019)]%
        {co_operation2019recommendation}
\bibfield{author}{\bibinfo{person}{{The Organisation for Economic Co-operation
  and Development (OECD)}}.} \bibinfo{year}{2019}\natexlab{}.
\newblock \bibinfo{booktitle}{\emph{{Recommendation of the Council on
  Artificial Intelligence}}}.
\newblock
\urldef\tempurl%
\url{https://legalinstruments.oecd.org/en/instruments/oecd-legal-0449}
\showURL{%
Retrieved February 2023 from \tempurl}


\bibitem[To et~al\mbox{.}(2021)]%
        {to2021reducing}
\bibfield{author}{\bibinfo{person}{Alexandra To}, \bibinfo{person}{Hillary
  Carey}, \bibinfo{person}{Geoff Kaufman}, {and} \bibinfo{person}{Jessica
  Hammer}.} \bibinfo{year}{2021}\natexlab{}.
\newblock \showarticletitle{{Reducing Uncertainty and Offering Comfort:
  Designing Technology for Coping with Interpersonal Racism}}. In
  \bibinfo{booktitle}{\emph{Proceedings of the 2021 {CHI} {Conference} on
  {Human} {Factors} in {Computing} {Systems}}} \emph{(\bibinfo{series}{{CHI}
  '21})}. \bibinfo{publisher}{ACM}.
\newblock
\urldef\tempurl%
\url{https://doi.org/10.1145/3411764.3445590}
\showDOI{\tempurl}


\bibitem[Tolmeijer et~al\mbox{.}(2022)]%
        {tolmeijer2022capable}
\bibfield{author}{\bibinfo{person}{Suzanne Tolmeijer}, \bibinfo{person}{Markus
  Christen}, \bibinfo{person}{Serhiy Kandul}, \bibinfo{person}{Markus Kneer},
  {and} \bibinfo{person}{Abraham Bernstein}.} \bibinfo{year}{2022}\natexlab{}.
\newblock \showarticletitle{{Capable but Amoral? Comparing AI and Human Expert
  Collaboration in Ethical Decision Making}}. In
  \bibinfo{booktitle}{\emph{Proceedings of the 2022 {CHI} {Conference} on
  {Human} {Factors} in {Computing} {Systems}}} \emph{(\bibinfo{series}{{CHI}
  '22})}. \bibinfo{publisher}{ACM}.
\newblock
\urldef\tempurl%
\url{https://doi.org/10.1145/3491102.3517732}
\showDOI{\tempurl}


\bibitem[Toreini et~al\mbox{.}(2020)]%
        {toreini2020relationship}
\bibfield{author}{\bibinfo{person}{Ehsan Toreini}, \bibinfo{person}{Mhairi
  Aitken}, \bibinfo{person}{Kovila Coopamootoo}, \bibinfo{person}{Karen
  Elliott}, \bibinfo{person}{Carlos~Gonzalez Zelaya}, {and}
  \bibinfo{person}{Aad van Moorsel}.} \bibinfo{year}{2020}\natexlab{}.
\newblock \showarticletitle{{The Relationship between Trust in AI and
  Trustworthy Machine Learning Technologies}}. In
  \bibinfo{booktitle}{\emph{Proceedings of the 2020 {Conference} on {Fairness},
  {Accountability}, and {Transparency}}} \emph{(\bibinfo{series}{{FAT}* '20})}.
  \bibinfo{publisher}{ACM}.
\newblock
\urldef\tempurl%
\url{https://doi.org/10.1145/3351095.3372834}
\showDOI{\tempurl}


\bibitem[Tsosie(2019)]%
        {tsosie2019tribal}
\bibfield{author}{\bibinfo{person}{Rebecca Tsosie}.}
  \bibinfo{year}{2019}\natexlab{}.
\newblock \showarticletitle{Tribal Data Governance and informational privacy:
  constructing indigenous data sovereignty}.
\newblock \bibinfo{journal}{\emph{Mont. L. Rev.}} (\bibinfo{year}{2019}).
\newblock


\bibitem[{United States Patent and Trademark Office}(2023)]%
        {patents2023}
\bibfield{author}{\bibinfo{person}{{United States Patent and Trademark
  Office}}.} \bibinfo{year}{2023}\natexlab{}.
\newblock \bibinfo{booktitle}{\emph{{United States Patent and Trademark
  Office}}}.
\newblock
\urldef\tempurl%
\url{https://www.uspto.gov}
\showURL{%
Retrieved June 2023 from \tempurl}


\bibitem[Vaccaro et~al\mbox{.}(2021)]%
        {vaccaro2021contestability}
\bibfield{author}{\bibinfo{person}{Kristen Vaccaro}, \bibinfo{person}{Ziang
  Xiao}, \bibinfo{person}{Kevin Hamilton}, {and} \bibinfo{person}{Karrie
  Karahalios}.} \bibinfo{year}{2021}\natexlab{}.
\newblock \showarticletitle{{Contestability For Content Moderation}}.
\newblock \bibinfo{journal}{\emph{Proc. ACM Hum.-Comput. Interact.}}
  (\bibinfo{date}{Oct.} \bibinfo{year}{2021}).
\newblock
\urldef\tempurl%
\url{https://doi.org/10.1145/3476059}
\showDOI{\tempurl}


\bibitem[Vaisman(2021)]%
        {vaisman2021your}
\bibfield{author}{\bibinfo{person}{Carmel Vaisman}.}
  \bibinfo{year}{2021}\natexlab{}.
\newblock \showarticletitle{{Your Next Robotic Boy/Girlfriend}}.
\newblock \bibinfo{journal}{\emph{Afeka Journal of Engineering and Science}}
  (\bibinfo{date}{Oct.} \bibinfo{year}{2021}).
\newblock
Issue 3.


\bibitem[Valencia et~al\mbox{.}(2023)]%
        {valencia2023less}
\bibfield{author}{\bibinfo{person}{Stephanie Valencia},
  \bibinfo{person}{Richard Cave}, \bibinfo{person}{Krystal Kallarackal},
  \bibinfo{person}{Katie Seaver}, \bibinfo{person}{Michael Terry}, {and}
  \bibinfo{person}{Shaun~K. Kane}.} \bibinfo{year}{2023}\natexlab{}.
\newblock \showarticletitle{{“The Less I Type, the Better”: How AI Language
  Models Can Enhance or Impede Communication for AAC Users}}. In
  \bibinfo{booktitle}{\emph{Proceedings of the 2023 {CHI} {Conference} on
  {Human} {Factors} in {Computing} {Systems}}} \emph{(\bibinfo{series}{{CHI}
  '23})}. \bibinfo{publisher}{ACM}.
\newblock
\urldef\tempurl%
\url{https://doi.org/10.1145/3544548.3581560}
\showDOI{\tempurl}


\bibitem[van Berkel et~al\mbox{.}(2023)]%
        {van_berkel2023methodology}
\bibfield{author}{\bibinfo{person}{Niels van Berkel}, \bibinfo{person}{Zhanna
  Sarsenbayeva}, {and} \bibinfo{person}{Jorge Goncalves}.}
  \bibinfo{year}{2023}\natexlab{}.
\newblock \showarticletitle{{The methodology of studying fairness perceptions
  in Artificial Intelligence: Contrasting CHI and FAccT}}.
\newblock \bibinfo{journal}{\emph{International Journal of Human-Computer
  Studies}} (\bibinfo{year}{2023}).
\newblock


\bibitem[Varanasi and Goyal(2023)]%
        {varanasi2023it}
\bibfield{author}{\bibinfo{person}{Rama~Adithya Varanasi} {and}
  \bibinfo{person}{Nitesh Goyal}.} \bibinfo{year}{2023}\natexlab{}.
\newblock \showarticletitle{{“It is Currently Hodgepodge”: Examining AI/ML
  Practitioners’ Challenges during Co-Production of Responsible AI Values}}.
  In \bibinfo{booktitle}{\emph{Proceedings of the 2023 {CHI} {Conference} on
  {Human} {Factors} in {Computing} {Systems}}} \emph{(\bibinfo{series}{{CHI}
  '23})}. \bibinfo{publisher}{ACM}.
\newblock
\urldef\tempurl%
\url{https://doi.org/10.1145/3544548.3580903}
\showDOI{\tempurl}


\bibitem[Verma et~al\mbox{.}(2023)]%
        {verma2023rethinking}
\bibfield{author}{\bibinfo{person}{Himanshu Verma}, \bibinfo{person}{Jakub
  Mlynar}, \bibinfo{person}{Roger Schaer}, \bibinfo{person}{Julien
  Reichenbach}, \bibinfo{person}{Mario Jreige}, \bibinfo{person}{John Prior},
  \bibinfo{person}{Florian Evéquoz}, {and} \bibinfo{person}{Adrien
  Depeursinge}.} \bibinfo{year}{2023}\natexlab{}.
\newblock \showarticletitle{{Rethinking the Role of AI with Physicians in
  Oncology: Revealing Perspectives from Clinical and Research Workflows}}. In
  \bibinfo{booktitle}{\emph{Proceedings of the 2023 {CHI} {Conference} on
  {Human} {Factors} in {Computing} {Systems}}} \emph{(\bibinfo{series}{{CHI}
  '23})}. \bibinfo{publisher}{ACM}.
\newblock
\urldef\tempurl%
\url{https://doi.org/10.1145/3544548.3581506}
\showDOI{\tempurl}


\bibitem[Vigil-Hayes et~al\mbox{.}(2017)]%
        {vigil_hayes2017indigenous}
\bibfield{author}{\bibinfo{person}{Morgan Vigil-Hayes}, \bibinfo{person}{Marisa
  Duarte}, \bibinfo{person}{Nicholet~Deschine Parkhurst}, {and}
  \bibinfo{person}{Elizabeth Belding}.} \bibinfo{year}{2017}\natexlab{}.
\newblock \showarticletitle{{\#Indigenous: Tracking the Connective Actions of
  Native American Advocates on Twitter}}. In
  \bibinfo{booktitle}{\emph{Proceedings of the 2017 ACM Conference on Computer
  Supported Cooperative Work and Social Computing}}
  \emph{(\bibinfo{series}{CSCW '17})}. \bibinfo{publisher}{ACM}.
\newblock
\urldef\tempurl%
\url{https://doi.org/10.1145/2998181.2998194}
\showDOI{\tempurl}


\bibitem[Viswanathan et~al\mbox{.}(2022)]%
        {viswanathan2022situational}
\bibfield{author}{\bibinfo{person}{Sruthi Viswanathan}, \bibinfo{person}{Cecile
  Boulard}, \bibinfo{person}{Adrien Bruyat}, {and} \bibinfo{person}{Antonietta
  Maria~Grasso}.} \bibinfo{year}{2022}\natexlab{}.
\newblock \showarticletitle{{Situational Recommender: Are You On the Spot,
  Refining Plans, or Just Bored?}}. In \bibinfo{booktitle}{\emph{Proceedings of
  the 2022 {CHI} {Conference} on {Human} {Factors} in {Computing} {Systems}}}
  \emph{(\bibinfo{series}{{CHI} '22})}. \bibinfo{publisher}{ACM}.
\newblock
\urldef\tempurl%
\url{https://doi.org/10.1145/3491102.3501909}
\showDOI{\tempurl}


\bibitem[Vitos et~al\mbox{.}(2017)]%
        {vitos2017supporting}
\bibfield{author}{\bibinfo{person}{Michalis Vitos}, \bibinfo{person}{Julia
  Altenbuchner}, \bibinfo{person}{Matthias Stevens}, \bibinfo{person}{Gillian
  Conquest}, \bibinfo{person}{Jerome Lewis}, {and} \bibinfo{person}{Muki
  Haklay}.} \bibinfo{year}{2017}\natexlab{}.
\newblock \showarticletitle{{Supporting Collaboration with Non-Literate Forest
  Communities in the Congo-Basin}}. In \bibinfo{booktitle}{\emph{Proceedings of
  the 2017 ACM Conference on Computer Supported Cooperative Work and Social
  Computing}} \emph{(\bibinfo{series}{CSCW '17})}. \bibinfo{publisher}{ACM}.
\newblock
\urldef\tempurl%
\url{https://doi.org/10.1145/2998181.2998242}
\showDOI{\tempurl}


\bibitem[Waldman(2021)]%
        {waldman2021industry}
\bibfield{author}{\bibinfo{person}{Ari~Ezra Waldman}.}
  \bibinfo{year}{2021}\natexlab{}.
\newblock \bibinfo{booktitle}{\emph{{Industry Unbound: The Inside Story of
  Privacy, Data, and Corporate Power}}}.
\newblock \bibinfo{publisher}{Cambridge University Press}.
\newblock
\urldef\tempurl%
\url{https://doi.org/10.1017/9781108591386}
\showDOI{\tempurl}


\bibitem[Walter and Suina(2019)]%
        {walter2019indigenous}
\bibfield{author}{\bibinfo{person}{Maggie Walter} {and}
  \bibinfo{person}{Michele Suina}.} \bibinfo{year}{2019}\natexlab{}.
\newblock \showarticletitle{Indigenous data, indigenous methodologies and
  indigenous data sovereignty}.
\newblock \bibinfo{journal}{\emph{International Journal of Social Research
  Methodology}} (\bibinfo{year}{2019}).
\newblock


\bibitem[Wang et~al\mbox{.}(2022b)]%
        {wang2022whose}
\bibfield{author}{\bibinfo{person}{Ding Wang}, \bibinfo{person}{Shantanu
  Prabhat}, {and} \bibinfo{person}{Nithya Sambasivan}.}
  \bibinfo{year}{2022}\natexlab{b}.
\newblock \showarticletitle{{Whose AI Dream? In Search of the Aspiration in
  Data Annotation.}}. In \bibinfo{booktitle}{\emph{Proceedings of the 2022
  {CHI} {Conference} on {Human} {Factors} in {Computing} {Systems}}}
  \emph{(\bibinfo{series}{{CHI} '22})}. \bibinfo{publisher}{ACM}.
\newblock
\urldef\tempurl%
\url{https://doi.org/10.1145/3491102.3502121}
\showDOI{\tempurl}


\bibitem[Wang et~al\mbox{.}(2019a)]%
        {wang2019human}
\bibfield{author}{\bibinfo{person}{Dakuo Wang}, \bibinfo{person}{Justin~D.
  Weisz}, \bibinfo{person}{Michael Muller}, \bibinfo{person}{Parikshit Ram},
  \bibinfo{person}{Werner Geyer}, \bibinfo{person}{Casey Dugan},
  \bibinfo{person}{Yla Tausczik}, \bibinfo{person}{Horst Samulowitz}, {and}
  \bibinfo{person}{Alexander Gray}.} \bibinfo{year}{2019}\natexlab{a}.
\newblock \showarticletitle{{Human-AI Collaboration in Data Science: Exploring
  Data Scientists' Perceptions of Automated AI}}.
\newblock \bibinfo{journal}{\emph{Proc. ACM Hum.-Comput. Interact.}}
  (\bibinfo{date}{Nov.} \bibinfo{year}{2019}).
\newblock
\urldef\tempurl%
\url{https://doi.org/10.1145/3359313}
\showDOI{\tempurl}


\bibitem[Wang et~al\mbox{.}(2019b)]%
        {wang2019designing}
\bibfield{author}{\bibinfo{person}{Danding Wang}, \bibinfo{person}{Qian Yang},
  \bibinfo{person}{Ashraf Abdul}, {and} \bibinfo{person}{Brian~Y. Lim}.}
  \bibinfo{year}{2019}\natexlab{b}.
\newblock \showarticletitle{{Designing Theory-Driven User-Centric Explainable
  AI}}. In \bibinfo{booktitle}{\emph{Proceedings of the 2019 CHI Conference on
  Human Factors in Computing Systems}} \emph{(\bibinfo{series}{CHI '19})}.
  \bibinfo{publisher}{ACM}.
\newblock
\urldef\tempurl%
\url{https://doi.org/10.1145/3290605.3300831}
\showDOI{\tempurl}


\bibitem[Wang et~al\mbox{.}(2022c)]%
        {wang2022informing}
\bibfield{author}{\bibinfo{person}{Ge Wang}, \bibinfo{person}{Jun Zhao},
  \bibinfo{person}{Max Van~Kleek}, {and} \bibinfo{person}{Nigel Shadbolt}.}
  \bibinfo{year}{2022}\natexlab{c}.
\newblock \showarticletitle{{Informing Age-Appropriate AI: Examining Principles
  and Practices of AI for Children}}. In \bibinfo{booktitle}{\emph{Proceedings
  of the 2022 {CHI} {Conference} on {Human} {Factors} in {Computing}
  {Systems}}} \emph{(\bibinfo{series}{{CHI} '22})}. \bibinfo{publisher}{ACM}.
\newblock
\urldef\tempurl%
\url{https://doi.org/10.1145/3491102.3502057}
\showDOI{\tempurl}


\bibitem[Wang et~al\mbox{.}(2021)]%
        {wang2021cass}
\bibfield{author}{\bibinfo{person}{Liuping Wang}, \bibinfo{person}{Dakuo Wang},
  \bibinfo{person}{Feng Tian}, \bibinfo{person}{Zhenhui Peng},
  \bibinfo{person}{Xiangmin Fan}, \bibinfo{person}{Zhan Zhang},
  \bibinfo{person}{Mo Yu}, \bibinfo{person}{Xiaojuan Ma}, {and}
  \bibinfo{person}{Hongan Wang}.} \bibinfo{year}{2021}\natexlab{}.
\newblock \showarticletitle{{CASS: Towards Building a Social-Support Chatbot
  for Online Health Community}}.
\newblock \bibinfo{journal}{\emph{Proc. ACM Hum.-Comput. Interact.}}
  (\bibinfo{date}{April} \bibinfo{year}{2021}).
\newblock
\urldef\tempurl%
\url{https://doi.org/10.1145/3449083}
\showDOI{\tempurl}


\bibitem[Wang et~al\mbox{.}(2022a)]%
        {wang2022understanding}
\bibfield{author}{\bibinfo{person}{Qiaosi Wang}, \bibinfo{person}{Ida Camacho},
  \bibinfo{person}{Shan Jing}, {and} \bibinfo{person}{Ashok~K. Goel}.}
  \bibinfo{year}{2022}\natexlab{a}.
\newblock \showarticletitle{{Understanding the Design Space of AI-Mediated
  Social Interaction in Online Learning: Challenges and Opportunities}}.
\newblock \bibinfo{journal}{\emph{Proc. ACM Hum.-Comput. Interact.}}
  (\bibinfo{date}{April} \bibinfo{year}{2022}).
\newblock
\urldef\tempurl%
\url{https://doi.org/10.1145/3512977}
\showDOI{\tempurl}


\bibitem[Wang et~al\mbox{.}(2023)]%
        {wang2023designing}
\bibfield{author}{\bibinfo{person}{Qiaosi Wang}, \bibinfo{person}{Michael
  Madaio}, \bibinfo{person}{Shaun Kane}, \bibinfo{person}{Shivani Kapania},
  \bibinfo{person}{Michael Terry}, {and} \bibinfo{person}{Lauren Wilcox}.}
  \bibinfo{year}{2023}\natexlab{}.
\newblock \showarticletitle{{Designing Responsible AI: Adaptations of UX
  Practice to Meet Responsible AI Challenges}}. In
  \bibinfo{booktitle}{\emph{Proceedings of the 2023 {CHI} {Conference} on
  {Human} {Factors} in {Computing} {Systems}}} \emph{(\bibinfo{series}{{CHI}
  '23})}. \bibinfo{publisher}{ACM}.
\newblock
\urldef\tempurl%
\url{https://doi.org/10.1145/3544548.3581278}
\showDOI{\tempurl}


\bibitem[Washington and Kuo(2020)]%
        {washington2020whose}
\bibfield{author}{\bibinfo{person}{Anne~L. Washington} {and}
  \bibinfo{person}{Rachel Kuo}.} \bibinfo{year}{2020}\natexlab{}.
\newblock \showarticletitle{{Whose Side Are Ethics Codes on? Power,
  Responsibility and the Social Good}}. In
  \bibinfo{booktitle}{\emph{Proceedings of the 2020 {Conference} on {Fairness},
  {Accountability}, and {Transparency}}} \emph{(\bibinfo{series}{{FAT}* '20})}.
  \bibinfo{publisher}{ACM}.
\newblock
\urldef\tempurl%
\url{https://doi.org/10.1145/3351095.3372844}
\showDOI{\tempurl}


\bibitem[Watkins(2023)]%
        {watkins2023face}
\bibfield{author}{\bibinfo{person}{Elizabeth~Anne Watkins}.}
  \bibinfo{year}{2023}\natexlab{}.
\newblock \showarticletitle{{Face Work: A Human-Centered Investigation into
  Facial Verification in Gig Work}}.
\newblock \bibinfo{journal}{\emph{Proc. ACM Hum.-Comput. Interact.}}
  (\bibinfo{date}{April} \bibinfo{year}{2023}).
\newblock
\urldef\tempurl%
\url{https://doi.org/10.1145/3579485}
\showDOI{\tempurl}


\bibitem[Weick(1995)]%
        {weick1995sensemaking}
\bibfield{author}{\bibinfo{person}{Karl~E Weick}.}
  \bibinfo{year}{1995}\natexlab{}.
\newblock \bibinfo{booktitle}{\emph{{Sensemaking in organizations}}}.
\newblock \bibinfo{publisher}{Sage}.
\newblock
\urldef\tempurl%
\url{https://us.sagepub.com/en-us/nam/sensemaking-in-organizations/book4988}
\showURL{%
\tempurl}


\bibitem[Weidinger et~al\mbox{.}(2022)]%
        {weidinger2022taxonomy}
\bibfield{author}{\bibinfo{person}{Laura Weidinger}, \bibinfo{person}{Jonathan
  Uesato}, \bibinfo{person}{Maribeth Rauh}, \bibinfo{person}{Conor Griffin},
  \bibinfo{person}{Po-Sen Huang}, \bibinfo{person}{John Mellor},
  \bibinfo{person}{Amelia Glaese}, \bibinfo{person}{Myra Cheng},
  \bibinfo{person}{Borja Balle}, \bibinfo{person}{Atoosa Kasirzadeh},
  \bibinfo{person}{Courtney Biles}, \bibinfo{person}{Sasha Brown},
  \bibinfo{person}{Zac Kenton}, \bibinfo{person}{Will Hawkins},
  \bibinfo{person}{Tom Stepleton}, \bibinfo{person}{Abeba Birhane},
  \bibinfo{person}{Lisa~Anne Hendricks}, \bibinfo{person}{Laura Rimell},
  \bibinfo{person}{William Isaac}, \bibinfo{person}{Julia Haas},
  \bibinfo{person}{Sean Legassick}, \bibinfo{person}{Geoffrey Irving}, {and}
  \bibinfo{person}{Iason Gabriel}.} \bibinfo{year}{2022}\natexlab{}.
\newblock \showarticletitle{{Taxonomy of Risks Posed by Language Models}}. In
  \bibinfo{booktitle}{\emph{2022 {ACM} {Conference} on {Fairness},
  {Accountability}, and {Transparency}}} \emph{(\bibinfo{series}{{FAccT}
  '22})}. \bibinfo{publisher}{ACM}.
\newblock
\urldef\tempurl%
\url{https://doi.org/10.1145/3531146.3533088}
\showDOI{\tempurl}


\bibitem[Whittlestone et~al\mbox{.}(2019)]%
        {whittlestone2019role}
\bibfield{author}{\bibinfo{person}{Jess Whittlestone}, \bibinfo{person}{Rune
  Nyrup}, \bibinfo{person}{Anna Alexandrova}, {and} \bibinfo{person}{Stephen
  Cave}.} \bibinfo{year}{2019}\natexlab{}.
\newblock \showarticletitle{{The Role and Limits of Principles in AI Ethics:
  Towards a Focus on Tensions}}. In \bibinfo{booktitle}{\emph{Proceedings of
  the 2019 {AAAI}/{ACM} {Conference} on {AI}, {Ethics}, and {Society}}}
  \emph{(\bibinfo{series}{{AIES} '19})}. \bibinfo{publisher}{ACM}.
\newblock
\urldef\tempurl%
\url{https://doi.org/10.1145/3306618.3314289}
\showDOI{\tempurl}


\bibitem[Widder et~al\mbox{.}(2022)]%
        {widder2022limits}
\bibfield{author}{\bibinfo{person}{David~Gray Widder}, \bibinfo{person}{Dawn
  Nafus}, \bibinfo{person}{Laura Dabbish}, {and} \bibinfo{person}{James
  Herbsleb}.} \bibinfo{year}{2022}\natexlab{}.
\newblock \showarticletitle{{Limits and Possibilities for “Ethical AI” in
  Open Source: A Study of Deepfakes}}. In \bibinfo{booktitle}{\emph{2022 {ACM}
  {Conference} on {Fairness}, {Accountability}, and {Transparency}}}
  \emph{(\bibinfo{series}{{FAccT} '22})}. \bibinfo{publisher}{ACM}.
\newblock
\urldef\tempurl%
\url{https://doi.org/10.1145/3531146.3533779}
\showDOI{\tempurl}


\bibitem[Wilkinson et~al\mbox{.}(2020)]%
        {wilkinson2020which}
\bibfield{author}{\bibinfo{person}{Dominic Wilkinson}, \bibinfo{person}{Hazem
  Zohny}, \bibinfo{person}{Andreas Kappes}, \bibinfo{person}{Walter
  Sinnott-Armstrong}, {and} \bibinfo{person}{Julian Savulescu}.}
  \bibinfo{year}{2020}\natexlab{}.
\newblock \showarticletitle{{Which factors should be included in triage? An
  online survey of the attitudes of the UK general public to pandemic triage
  dilemmas}}.
\newblock \bibinfo{journal}{\emph{BMJ open}} (\bibinfo{year}{2020}).
\newblock
\urldef\tempurl%
\url{https://doi.org/10.1136/bmjopen-2020-045593}
\showDOI{\tempurl}


\bibitem[Willen et~al\mbox{.}(2022)]%
        {willen2022rethinking}
\bibfield{author}{\bibinfo{person}{Sarah~S. Willen},
  \bibinfo{person}{Abigail~Fisher Williamson}, \bibinfo{person}{Colleen~C.
  Walsh}, \bibinfo{person}{Mikayla Hyman}, {and} \bibinfo{person}{William
  Tootle}.} \bibinfo{year}{2022}\natexlab{}.
\newblock \showarticletitle{{Rethinking flourishing: Critical insights and
  qualitative perspectives from the U.S. Midwest}}.
\newblock \bibinfo{journal}{\emph{SSM - Mental Health}} (\bibinfo{year}{2022}).
\newblock
\showISSN{2666-5603}
\urldef\tempurl%
\url{https://doi.org/10.1016/j.ssmmh.2021.100057}
\showDOI{\tempurl}


\bibitem[Winecoff and Watkins(2022)]%
        {winecoff2022artificial}
\bibfield{author}{\bibinfo{person}{Amy~A. Winecoff} {and}
  \bibinfo{person}{Elizabeth~Anne Watkins}.} \bibinfo{year}{2022}\natexlab{}.
\newblock \showarticletitle{{Artificial Concepts of Artificial Intelligence:
  Institutional Compliance and Resistance in AI Startups}}. In
  \bibinfo{booktitle}{\emph{Proceedings of the 2022 {AAAI}/{ACM} {Conference}
  on {AI}, {Ethics}, and {Society}}} \emph{(\bibinfo{series}{{AIES} '22})}.
  \bibinfo{publisher}{ACM}.
\newblock
\urldef\tempurl%
\url{https://doi.org/10.1145/3514094.3534138}
\showDOI{\tempurl}


\bibitem[Wong et~al\mbox{.}(2023a)]%
        {wong2023privacy}
\bibfield{author}{\bibinfo{person}{Richmond~Y. Wong}, \bibinfo{person}{Andrew
  Chong}, {and} \bibinfo{person}{R.~Cooper Aspegren}.}
  \bibinfo{year}{2023}\natexlab{a}.
\newblock \showarticletitle{{Privacy Legislation as Business Risks: How GDPR
  and CCPA Are Represented in Technology Companies' Investment Risk
  Disclosures}}.
\newblock \bibinfo{journal}{\emph{Proc. ACM Hum.-Comput. Interact.}}
  (\bibinfo{date}{April} \bibinfo{year}{2023}).
\newblock
\urldef\tempurl%
\url{https://doi.org/10.1145/3579515}
\showDOI{\tempurl}


\bibitem[Wong et~al\mbox{.}(2023b)]%
        {wong2023seeing}
\bibfield{author}{\bibinfo{person}{Richmond~Y. Wong},
  \bibinfo{person}{Michael~A. Madaio}, {and} \bibinfo{person}{Nick Merrill}.}
  \bibinfo{year}{2023}\natexlab{b}.
\newblock \showarticletitle{{Seeing Like a Toolkit: How Toolkits Envision the
  Work of AI Ethics}}.
\newblock \bibinfo{journal}{\emph{Proc. ACM Hum.-Comput. Interact.}}
  (\bibinfo{date}{April} \bibinfo{year}{2023}).
\newblock
\urldef\tempurl%
\url{https://doi.org/10.1145/3579621}
\showDOI{\tempurl}


\bibitem[Xu et~al\mbox{.}(2022)]%
        {xu2022algorithmic}
\bibfield{author}{\bibinfo{person}{Jie Xu}, \bibinfo{person}{Yunyu Xiao},
  \bibinfo{person}{Wendy~Hui Wang}, \bibinfo{person}{Yue Ning},
  \bibinfo{person}{Elizabeth~A Shenkman}, \bibinfo{person}{Jiang Bian}, {and}
  \bibinfo{person}{Fei Wang}.} \bibinfo{year}{2022}\natexlab{}.
\newblock \showarticletitle{{Algorithmic fairness in computational medicine}}.
\newblock \bibinfo{journal}{\emph{EBioMedicine}} (\bibinfo{year}{2022}).
\newblock
\urldef\tempurl%
\url{https://doi.org/10.1016/j.ebiom.2022.104250}
\showDOI{\tempurl}


\bibitem[Yang et~al\mbox{.}(2023)]%
        {yang2023harnessing}
\bibfield{author}{\bibinfo{person}{Qian Yang}, \bibinfo{person}{Yuexing Hao},
  \bibinfo{person}{Kexin Quan}, \bibinfo{person}{Stephen Yang},
  \bibinfo{person}{Yiran Zhao}, \bibinfo{person}{Volodymyr Kuleshov}, {and}
  \bibinfo{person}{Fei Wang}.} \bibinfo{year}{2023}\natexlab{}.
\newblock \showarticletitle{{Harnessing Biomedical Literature to Calibrate
  Clinicians’ Trust in AI Decision Support Systems}}. In
  \bibinfo{booktitle}{\emph{Proceedings of the 2023 {CHI} {Conference} on
  {Human} {Factors} in {Computing} {Systems}}} \emph{(\bibinfo{series}{{CHI}
  '23})}. \bibinfo{publisher}{ACM}.
\newblock
\urldef\tempurl%
\url{https://doi.org/10.1145/3544548.3581393}
\showDOI{\tempurl}


\bibitem[Yang et~al\mbox{.}(2022)]%
        {yang2022enhancing}
\bibfield{author}{\bibinfo{person}{Yu Yang}, \bibinfo{person}{Aayush Gupta},
  \bibinfo{person}{Jianwei Feng}, \bibinfo{person}{Prateek Singhal},
  \bibinfo{person}{Vivek Yadav}, \bibinfo{person}{Yue Wu},
  \bibinfo{person}{Pradeep Natarajan}, \bibinfo{person}{Varsha Hedau}, {and}
  \bibinfo{person}{Jungseock Joo}.} \bibinfo{year}{2022}\natexlab{}.
\newblock \showarticletitle{{Enhancing Fairness in Face Detection in Computer
  Vision Systems by Demographic Bias Mitigation}}. In
  \bibinfo{booktitle}{\emph{Proceedings of the 2022 {AAAI}/{ACM} {Conference}
  on {AI}, {Ethics}, and {Society}}} \emph{(\bibinfo{series}{{AIES} '22})}.
  \bibinfo{publisher}{ACM}.
\newblock
\urldef\tempurl%
\url{https://doi.org/10.1145/3514094.3534153}
\showDOI{\tempurl}


\bibitem[Yfantidou et~al\mbox{.}(2023)]%
        {yfantidou2023beyond}
\bibfield{author}{\bibinfo{person}{Sofia Yfantidou}, \bibinfo{person}{Marios
  Constantinides}, \bibinfo{person}{Dimitris Spathis}, \bibinfo{person}{Athena
  Vakali}, \bibinfo{person}{Daniele Quercia}, {and} \bibinfo{person}{Fahim
  Kawsar}.} \bibinfo{year}{2023}\natexlab{}.
\newblock \showarticletitle{Beyond Accuracy: A Critical Review of Fairness in
  Machine Learning for Mobile and Wearable Computing}.
\newblock \bibinfo{journal}{\emph{arXiv preprint arXiv:2303.15585}}
  (\bibinfo{year}{2023}).
\newblock


\bibitem[Yildirim et~al\mbox{.}(2022)]%
        {yildirim2022how}
\bibfield{author}{\bibinfo{person}{Nur Yildirim}, \bibinfo{person}{Alex Kass},
  \bibinfo{person}{Teresa Tung}, \bibinfo{person}{Connor Upton},
  \bibinfo{person}{Donnacha Costello}, \bibinfo{person}{Robert Giusti},
  \bibinfo{person}{Sinem Lacin}, \bibinfo{person}{Sara Lovic},
  \bibinfo{person}{James~M O'Neill}, \bibinfo{person}{Rudi~O'Reilly Meehan},
  \bibinfo{person}{Eoin Ó~Loideáin}, \bibinfo{person}{Azzurra Pini},
  \bibinfo{person}{Medb Corcoran}, \bibinfo{person}{Jeremiah Hayes},
  \bibinfo{person}{Diarmuid~J Cahalane}, \bibinfo{person}{Gaurav Shivhare},
  \bibinfo{person}{Luigi Castoro}, \bibinfo{person}{Giovanni Caruso},
  \bibinfo{person}{Changhoon Oh}, \bibinfo{person}{James McCann},
  \bibinfo{person}{Jodi Forlizzi}, {and} \bibinfo{person}{John Zimmerman}.}
  \bibinfo{year}{2022}\natexlab{}.
\newblock \showarticletitle{{How Experienced Designers of Enterprise
  Applications Engage AI as a Design Material}}. In
  \bibinfo{booktitle}{\emph{Proceedings of the 2022 {CHI} {Conference} on
  {Human} {Factors} in {Computing} {Systems}}} \emph{(\bibinfo{series}{{CHI}
  '22})}. \bibinfo{publisher}{ACM}.
\newblock
\urldef\tempurl%
\url{https://doi.org/10.1145/3491102.3517491}
\showDOI{\tempurl}


\bibitem[Yildirim et~al\mbox{.}(2023)]%
        {yildirim2023investigating}
\bibfield{author}{\bibinfo{person}{Nur Yildirim}, \bibinfo{person}{Mahima
  Pushkarna}, \bibinfo{person}{Nitesh Goyal}, \bibinfo{person}{Martin
  Wattenberg}, {and} \bibinfo{person}{Fernanda Viégas}.}
  \bibinfo{year}{2023}\natexlab{}.
\newblock \showarticletitle{{Investigating How Practitioners Use Human-AI
  Guidelines: A Case Study on the People + AI Guidebook}}. In
  \bibinfo{booktitle}{\emph{Proceedings of the 2023 {CHI} {Conference} on
  {Human} {Factors} in {Computing} {Systems}}} \emph{(\bibinfo{series}{{CHI}
  '23})}. \bibinfo{publisher}{ACM}.
\newblock
\urldef\tempurl%
\url{https://doi.org/10.1145/3544548.3580900}
\showDOI{\tempurl}


\bibitem[Young(2021)]%
        {young2021danger}
\bibfield{author}{\bibinfo{person}{James Young}.}
  \bibinfo{year}{2021}\natexlab{}.
\newblock \showarticletitle{{Danger! This robot may be trying to manipulate
  you}}.
\newblock \bibinfo{journal}{\emph{Science Robotics}} (\bibinfo{year}{2021}).
\newblock
\urldef\tempurl%
\url{https://doi.org/10.1126/scirobotics.abk3479}
\showDOI{\tempurl}


\bibitem[Young et~al\mbox{.}(2022)]%
        {young2022confronting}
\bibfield{author}{\bibinfo{person}{Meg Young}, \bibinfo{person}{Michael
  Katell}, {and} \bibinfo{person}{P.M. Krafft}.}
  \bibinfo{year}{2022}\natexlab{}.
\newblock \showarticletitle{{Confronting Power and Corporate Capture at the
  FAccT Conference}}. In \bibinfo{booktitle}{\emph{2022 {ACM} {Conference} on
  {Fairness}, {Accountability}, and {Transparency}}}
  \emph{(\bibinfo{series}{{FAccT} '22})}. \bibinfo{publisher}{ACM}.
\newblock
\urldef\tempurl%
\url{https://doi.org/10.1145/3531146.3533194}
\showDOI{\tempurl}


\bibitem[Yuan et~al\mbox{.}(2023)]%
        {yuan2023contextualizing}
\bibfield{author}{\bibinfo{person}{Chien Wen~(Tina) Yuan},
  \bibinfo{person}{Nanyi Bi}, \bibinfo{person}{Ya-Fang Lin}, {and}
  \bibinfo{person}{Yuen-Hsien Tseng}.} \bibinfo{year}{2023}\natexlab{}.
\newblock \showarticletitle{{Contextualizing User Perceptions about Biases for
  Human-Centered Explainable Artificial Intelligence}}. In
  \bibinfo{booktitle}{\emph{Proceedings of the 2023 {CHI} {Conference} on
  {Human} {Factors} in {Computing} {Systems}}} \emph{(\bibinfo{series}{{CHI}
  '23})}. \bibinfo{publisher}{ACM}.
\newblock
\urldef\tempurl%
\url{https://doi.org/10.1145/3544548.3580945}
\showDOI{\tempurl}


\bibitem[Yurrita et~al\mbox{.}(2022)]%
        {yurrita2022towards}
\bibfield{author}{\bibinfo{person}{Mireia Yurrita}, \bibinfo{person}{Dave
  Murray-Rust}, \bibinfo{person}{Agathe Balayn}, {and}
  \bibinfo{person}{Alessandro Bozzon}.} \bibinfo{year}{2022}\natexlab{}.
\newblock \showarticletitle{{Towards a Multi-Stakeholder Value-Based Assessment
  Framework for Algorithmic Systems}}. In \bibinfo{booktitle}{\emph{2022 {ACM}
  {Conference} on {Fairness}, {Accountability}, and {Transparency}}}
  \emph{(\bibinfo{series}{{FAccT} '22})}. \bibinfo{publisher}{ACM}.
\newblock
\urldef\tempurl%
\url{https://doi.org/10.1145/3531146.3533118}
\showDOI{\tempurl}


\bibitem[{Yuval Noah Harari}(2021)]%
        {harari2021lessons}
\bibfield{author}{\bibinfo{person}{{Yuval Noah Harari}}.}
  \bibinfo{year}{2021}\natexlab{}.
\newblock \bibinfo{booktitle}{\emph{{Lessons from a year of Covid}}}.
\newblock {Financial Times}.
\newblock
\urldef\tempurl%
\url{https://www.ft.com/content/f1b30f2c-84aa-4595-84f2-7816796d6841}
\showURL{%
Retrieved January 2023 from \tempurl}


\bibitem[Zhang et~al\mbox{.}(2023a)]%
        {zhang2023stakeholder}
\bibfield{author}{\bibinfo{person}{Angie Zhang}, \bibinfo{person}{Alexander
  Boltz}, \bibinfo{person}{Jonathan Lynn}, \bibinfo{person}{Chun-Wei Wang},
  {and} \bibinfo{person}{Min~Kyung Lee}.} \bibinfo{year}{2023}\natexlab{a}.
\newblock \showarticletitle{{Stakeholder-Centered AI Design: Co-Designing
  Worker Tools with Gig Workers through Data Probes}}. In
  \bibinfo{booktitle}{\emph{Proceedings of the 2023 {CHI} {Conference} on
  {Human} {Factors} in {Computing} {Systems}}} \emph{(\bibinfo{series}{{CHI}
  '23})}. \bibinfo{publisher}{ACM}.
\newblock
\urldef\tempurl%
\url{https://doi.org/10.1145/3544548.3581354}
\showDOI{\tempurl}


\bibitem[Zhang et~al\mbox{.}(2023b)]%
        {zhang2023deliberating}
\bibfield{author}{\bibinfo{person}{Angie Zhang}, \bibinfo{person}{Olympia
  Walker}, \bibinfo{person}{Kaci Nguyen}, \bibinfo{person}{Jiajun Dai},
  \bibinfo{person}{Anqing Chen}, {and} \bibinfo{person}{Min~Kyung Lee}.}
  \bibinfo{year}{2023}\natexlab{b}.
\newblock \showarticletitle{{Deliberating with AI: Improving Decision-Making
  for the Future through Participatory AI Design and Stakeholder
  Deliberation}}.
\newblock \bibinfo{journal}{\emph{Proc. ACM Hum.-Comput. Interact.}}
  (\bibinfo{date}{April} \bibinfo{year}{2023}).
\newblock
\urldef\tempurl%
\url{https://doi.org/10.1145/3579601}
\showDOI{\tempurl}


\bibitem[Zhang et~al\mbox{.}(2022)]%
        {zhang2022debiased}
\bibfield{author}{\bibinfo{person}{Wencan Zhang}, \bibinfo{person}{Mariella
  Dimiccoli}, {and} \bibinfo{person}{Brian~Y Lim}.}
  \bibinfo{year}{2022}\natexlab{}.
\newblock \showarticletitle{{Debiased-CAM to Mitigate Image Perturbations with
  Faithful Visual Explanations of Machine Learning}}. In
  \bibinfo{booktitle}{\emph{Proceedings of the 2022 {CHI} {Conference} on
  {Human} {Factors} in {Computing} {Systems}}} \emph{(\bibinfo{series}{{CHI}
  '22})}. \bibinfo{publisher}{ACM}.
\newblock
\urldef\tempurl%
\url{https://doi.org/10.1145/3491102.3517522}
\showDOI{\tempurl}


\bibitem[Zheng et~al\mbox{.}(2023)]%
        {zheng2023competent}
\bibfield{author}{\bibinfo{person}{Chengbo Zheng}, \bibinfo{person}{Yuheng Wu},
  \bibinfo{person}{Chuhan Shi}, \bibinfo{person}{Shuai Ma},
  \bibinfo{person}{Jiehui Luo}, {and} \bibinfo{person}{Xiaojuan Ma}.}
  \bibinfo{year}{2023}\natexlab{}.
\newblock \showarticletitle{{Competent but Rigid: Identifying the Gap in
  Empowering AI to Participate Equally in Group Decision-Making}}. In
  \bibinfo{booktitle}{\emph{Proceedings of the 2023 {CHI} {Conference} on
  {Human} {Factors} in {Computing} {Systems}}} \emph{(\bibinfo{series}{{CHI}
  '23})}. \bibinfo{publisher}{ACM}.
\newblock
\urldef\tempurl%
\url{https://doi.org/10.1145/3544548.3581131}
\showDOI{\tempurl}


\bibitem[Zhou et~al\mbox{.}(2023)]%
        {zhou2023synthetic}
\bibfield{author}{\bibinfo{person}{Jiawei Zhou}, \bibinfo{person}{Yixuan
  Zhang}, \bibinfo{person}{Qianni Luo}, \bibinfo{person}{Andrea~G Parker},
  {and} \bibinfo{person}{Munmun De~Choudhury}.}
  \bibinfo{year}{2023}\natexlab{}.
\newblock \showarticletitle{{Synthetic Lies: Understanding AI-Generated
  Misinformation and Evaluating Algorithmic and Human Solutions}}. In
  \bibinfo{booktitle}{\emph{Proceedings of the 2023 {CHI} {Conference} on
  {Human} {Factors} in {Computing} {Systems}}} \emph{(\bibinfo{series}{{CHI}
  '23})}. \bibinfo{publisher}{ACM}.
\newblock
\urldef\tempurl%
\url{https://doi.org/10.1145/3544548.3581318}
\showDOI{\tempurl}


\bibitem[Zytko et~al\mbox{.}(2022)]%
        {zytko2022participatory}
\bibfield{author}{\bibinfo{person}{Douglas Zytko}, \bibinfo{person}{Pamela
  J.~Wisniewski}, \bibinfo{person}{Shion Guha}, \bibinfo{person}{Eric
  P.~S.~Baumer}, {and} \bibinfo{person}{Min~Kyung Lee}.}
  \bibinfo{year}{2022}\natexlab{}.
\newblock \showarticletitle{{Participatory Design of AI Systems: Opportunities
  and Challenges Across Diverse Users, Relationships, and Application
  Domains}}. In \bibinfo{booktitle}{\emph{Extended Abstracts of the 2022 CHI
  Conference on Human Factors in Computing Systems}}
  \emph{(\bibinfo{series}{CHI EA '22})}. \bibinfo{publisher}{ACM}, Article
  \bibinfo{articleno}{154}.
\newblock
\urldef\tempurl%
\url{https://doi.org/10.1145/3491101.3516506}
\showDOI{\tempurl}


\end{thebibliography}

\appendix
\clearpage
\section{Appendix: Summary of All Reviewed HCER-AI Research Papers}

\begin{center}
\scriptsize
\begin{longtable}{@{}llllcccccc@{}}
\caption{A list of all HCER-AI reviewed papers with their research methods and themes (n=164).}
\label{tab:all-papers}\\
\toprule
Citation & Research method & Year & Venue & Explainability & Human flourishing & Privacy & Security & Fairness & Governance \\* \midrule
\endfirsthead
\multicolumn{10}{c}%
{{\bfseries Table \thetable\ continued from previous page}} \\
\toprule
Citation & Research method & Year & Venue & Explainability & Human flourishing & Privacy & Security & Fairness & Governance \\* \midrule
\endhead
%
\citet{maas2018regulating} & Theory & 2018 & AIES &  &  &  & \fullcircle &  & \fullcircle \\
\citet{whittlestone2019role} & Review \& Theory & 2019 & AIES &  &  &  &  &  & \fullcircle \\
\citet{schiff2020what} & Review \& Theory & 2020 & AIES &  &  &  &  &  & \fullcircle \\
\citet{erdélyi2020ai} & Review & 2020 & AIES &  &  &  &  &  & \fullcircle \\
\citet{hopkins2021machine} & Qualitative & 2021 & AIES &  &  &  &  &  & \fullcircle \\
\citet{henriksen2021situated} & Qualitative & 2021 & AIES & \fullcircle &  & \fullcircle &  & \fullcircle &  \\
\citet{nielsen2021measuring} & Qualitative & 2021 & AIES &  &  & \fullcircle &  &  &  \\
\citet{klinova2021ai} & Theory & 2021 & AIES &  & \fullcircle &  &  & \fullcircle &  \\
\citet{raz2021face} & Qualitative & 2021 & AIES & \fullcircle &  &  &  & \fullcircle &  \\
\citet{abdalla2021grey} & Quantitative & 2021 & AIES &  &  &  &  &  & \fullcircle \\
\citet{simons2021machine} & Qualitative & 2021 & AIES &  & \fullcircle &  &  & \fullcircle &  \\
\citet{lee2021participatory} & Theory & 2021 & AIES &  &  &  &  & \fullcircle & \fullcircle \\
\citet{deshpande2022responsible} & Review & 2022 & AIES &  &  &  &  &  & \fullcircle \\
\citet{bertrand2022how} & Review & 2022 & AIES & \fullcircle &  &  &  &  & \fullcircle \\
\citet{bessen2022cost} & Qualitative & 2022 & AIES &  &  &  &  &  & \fullcircle \\
\citet{aka2021measuring} & Quantitative & 2022 & AIES &  &  & \fullcircle &  &  & \fullcircle \\
\citet{sharma2021fair} & Quantitative & 2021 & AIES &  &  &  &  & \fullcircle &  \\
\citet{winecoff2022artificial} & Quantitative & 2021 & AIES &  &  &  &  & \fullcircle &  \\
\citet{nashed2021ethically} & Qualitative & 2021 & AIES &  &  &  &  &  & \fullcircle \\
\citet{leavy2021ethical} & Qualitative & 2021 & AIES &  &  &  &  & \fullcircle &  \\
\citet{cruz_cortés2020invitation} & Quantitative & 2020 & AIES &  &  &  &  & \fullcircle &  \\
\citet{kasirzadeh2021ethical} & Theory & 2021 & AIES &  &  &  &  & \fullcircle & \fullcircle \\
\citet{yang2022enhancing} & Theory & 2021 & AIES &  &  &  &  &  & \fullcircle \\
\citet{flathmann2021modeling} & Quantitative & 2022 & AIES &  &  &  &  & \fullcircle &  \\
\citet{shaw2018towards} & Quantitative & 2018 & AIES &  &  &  &  &  & \fullcircle \\
\citet{benthall2021artificial} & Qualitative & 2021 & AIES &  &  &  &  & \fullcircle & \fullcircle \\
\citet{kim2018computational} & Theory & 2021 & AIES &  &  & \fullcircle & \fullcircle & \fullcircle & \fullcircle \\
\citet{eicher2018jill} & Quantitative & 2018 & AIES &  &  &  &  &  & \fullcircle \\
\citet{chi2021reconfiguring} & Theory & 2018 & AIES &  &  &  &  &  & \fullcircle \\
\citet{siapka2022towards} & Theory & 2022 & AIES &  &  &  &  &  & \fullcircle \\
\citet{beede2020human} & Qualitative & 2020 & CHI &  & \fullcircle &  &  &  &  \\
\citet{liao2020questioning} & Qualitative & 2020 & CHI & \fullcircle &  &  &  &  &  \\
\citet{cheng2021soliciting} & Qualitative & 2021 & CHI &  & \fullcircle &  &  & \fullcircle &  \\
\citet{morrison2021social} & Qualitative & 2021 & CHI &  & \fullcircle &  &  & \fullcircle &  \\
\citet{park2021human} & Quantitative & 2021 & CHI & \fullcircle &  &  &  &  &  \\
\citet{bansal2021does} & Qualitative & 2021 & CHI & \fullcircle &  & \fullcircle & \fullcircle &  &  \\
\citet{ehsan2021expanding} & Qualitative & 2021 & CHI & \fullcircle &  & \fullcircle & \fullcircle & \fullcircle &  \\
\citet{sambasivan2021everyone} & Mixed & 2021 & CHI &  &  & \fullcircle &  & \fullcircle &  \\
\citet{langer2022look} & Review & 2021 & CHI &  & \fullcircle &  &  &  &  \\
\citet{lee2021who} & Qualitative & 2021 & CHI &  & \fullcircle &  &  &  &  \\
\citet{steiger2021psychological} & Theory & 2022 & CHI & \fullcircle &  &  &  & \fullcircle & \fullcircle \\
\citet{muller2022forgetting} & Quantitative & 2022 & CHI &  &  &  &  & \fullcircle & \fullcircle \\
\citet{cheng2022how} & Mixed & 2022 & CHI &  & \fullcircle &  &  & \fullcircle &  \\
\citet{park2022designing} & Qualitative & 2022 & CHI &  &  & \fullcircle &  &  &  \\
\citet{mlynar2022ai} & Qualitative & 2022 & CHI & \fullcircle & \fullcircle & \fullcircle & \fullcircle & \fullcircle &  \\
\citet{wang2022informing} & Mixed & 2022 & CHI &  &  &  &  & \fullcircle &  \\
\citet{kapania2022because} & Review & 2022 & CHI &  & \fullcircle & \fullcircle & \fullcircle & \fullcircle &  \\
\citet{wang2022whose} & Qualitative & 2022 & CHI &  & \fullcircle &  &  &  &  \\
\citet{reitmaier2022opportunities} & Mixed & 2022 & CHI &  & \fullcircle &  &  &  &  \\
\citet{yildirim2022how} & Qualitative & 2022 & CHI & \fullcircle &  & \fullcircle &  &  &  \\
\citet{hall2022supporting} & Qualitative & 2022 & CHI &  &  &  &  & \fullcircle &  \\
\citet{long2022family} & Qualitative & 2022 & CHI & \fullcircle &  &  &  &  &  \\
\citet{docherty2022re} & Qualitative & 2022 & CHI &  &  & \fullcircle &  & \fullcircle &  \\
\citet{kang2022how} & Theory & 2022 & CHI &  & \fullcircle &  &  & \fullcircle &  \\
\citet{zhang2022debiased} & Qualitative & 2022 & CHI & \fullcircle &  &  &  &  &  \\
\citet{viswanathan2022situational} & Mixed & 2022 & CHI & \fullcircle &  & \fullcircle &  & \fullcircle &  \\
\citet{tolmeijer2022capable} & Review & 2022 & CHI &  &  & \fullcircle &  & \fullcircle &  \\
\citet{subramonyam2022solving} & Quantitative & 2022 & CHI &  &  &  &  & \fullcircle & \fullcircle \\
\citet{lin2021engaging} & Quantitative & 2020 & CHI &  &  &  &  & \fullcircle & \fullcircle \\
\citet{niforatos2020would} & Qualitative & 2021 & CHI &  &  &  &  &  & \fullcircle \\
\citet{elsayed_ali2023responsible} & Mixed & 2021 & CHI &  &  &  &  & \fullcircle &  \\
\citet{varanasi2023it} & Quantitative & 2021 & CHI &  &  &  &  & \fullcircle & \fullcircle \\
\citet{to2021reducing} & Qualitative & 2021 & CHI &  &  &  &  & \fullcircle &  \\
\citet{wang2023designing} & Qualitative & 2023 & CHI &  &  &  &  &  & \fullcircle \\
\citet{lima2021human} & Qualitative & 2023 & CHI &  &  &  &  &  & \fullcircle \\
\citet{richardson2021towards} & Qualitative & 2023 & CHI & \fullcircle &  &  &  & \fullcircle & \fullcircle \\
\citet{kamikubo2023contributing} & Qualitative & 2023 & CHI &  &  &  &  & \fullcircle & \fullcircle \\
\citet{moore2023failurenotes} & Qualitative & 2023 & CHI &  &  &  &  & \fullcircle & \fullcircle \\
\citet{verma2023rethinking} & Mixed & 2023 & CHI &  &  & \fullcircle & \fullcircle & \fullcircle & \fullcircle \\
\citet{deng2023understanding} & Qualitative & 2023 & CHI &  & \fullcircle &  &  &  &  \\
\citet{chu2023work} & Qualitative & 2023 & CHI &  &  &  &  & \fullcircle &  \\
\citet{zhou2023synthetic} & Quantitative & 2023 & CHI & \fullcircle &  &  &  &  & \fullcircle \\
\citet{kapania2023hunt} & Quantitative & 2023 & CHI &  &  &  &  & \fullcircle &  \\
\citet{chen2023is} & Qualitative & 2023 & CHI & \fullcircle &  &  &  &  &  \\
\citet{yildirim2023investigating} & Mixed & 2023 & CHI &  &  &  &  & \fullcircle & \fullcircle \\
\citet{ayobi2023computational} & Qualitative & 2023 & CHI &  &  &  &  &  & \fullcircle \\
\citet{alfrink2023contestable} & Qualitative & 2023 & CHI & \fullcircle &  &  &  & \fullcircle & \fullcircle \\
\citet{liao2023designerly} & Qualitative & 2023 & CHI &  &  &  &  & \fullcircle & \fullcircle \\
\citet{zhang2023stakeholder} & Qualitative & 2023 & CHI &  & \fullcircle &  &  & \fullcircle & \fullcircle \\
\citet{lee2023fostering} & Mixed & 2023 & CHI &  &  &  &  &  & \fullcircle \\
\citet{cabitza2023ai} & Quantitative & 2023 & CHI & \fullcircle &  &  &  &  &  \\
\citet{valencia2023less} & Qualitative & 2023 & CHI &  & \fullcircle &  &  &  &  \\
\citet{hsieh2023what} & Qualitative & 2023 & CHI &  &  &  &  & \fullcircle &  \\
\citet{ashktorab2023fairness} & Qualitative & 2023 & CHI &  &  &  &  &  & \fullcircle \\
\citet{lam2023model} & Mixed & 2023 & CHI &  &  &  &  &  & \fullcircle \\
\citet{baughan2023mixed} & Mixed & 2023 & CHI &  &  &  &  &  & \fullcircle \\
\citet{zheng2023competent} & Qualitative & 2023 & CHI &  &  &  &  & \fullcircle &  \\
\citet{rismani2023plane} & Qualitative & 2023 & CHI &  &  &  &  &  & \fullcircle \\
\citet{shahid2023decolonizing} & Qualitative & 2023 & CHI &  &  &  &  &  & \fullcircle \\
\citet{li2023participation} & Mixed & 2023 & CHI &  &  &  &  &  & \fullcircle \\
\citet{lima2023blaming} & Quantitative & 2023 & CHI & \fullcircle &  &  &  & \fullcircle & \fullcircle \\
\citet{yang2023harnessing} & Qualitative & 2023 & CHI & \fullcircle &  &  &  &  &  \\
\citet{kim2023bubbleu} & Mixed & 2023 & CHI & \fullcircle &  &  &  &  &  \\
\citet{yuan2023contextualizing} & Qualitative & 2023 & CHI &  &  &  &  &  & \fullcircle \\
\citet{burgess2023healthcare} & Qualitative & 2023 & CHI & \fullcircle &  &  &  & \fullcircle &  \\
\citet{lewicki2023out} & Qualitative & 2023 & CHI & \fullcircle &  &  &  & \fullcircle &  \\
\citet{choi2023creator} & Qualitative & 2023 & CHI & \fullcircle &  &  &  & \fullcircle &  \\
\citet{capel2023what} & Qualitative & 2023 & CHI &  &  &  &  &  & \fullcircle \\
\citet{rakova2021where} & Qualitative & 2021 & CSCW &  &  &  &  &  & \fullcircle \\
\citet{gu2021lessons} & Qualitative & 2022 & CSCW &  &  &  &  & \fullcircle & \fullcircle \\
\citet{heger2022understanding} & Qualitative & 2021 & CSCW &  &  &  &  &  & \fullcircle \\
\citet{lam2022end} & Qualitative & 2022 & CSCW &  &  &  &  & \fullcircle &  \\
\citet{lee2019webuildai} & Qualitative & 2019 & CSCW &  &  &  &  & \fullcircle & \fullcircle \\
\citet{bawa2020do} & Mixed & 2020 & CSCW &  & \fullcircle &  &  & \fullcircle &  \\
\citet{shen2020designing} & Mixed & 2020 & CSCW & \fullcircle &  &  &  & \fullcircle &  \\
\citet{lee2023speculating} & Qualitative & 2023 & CSCW &  &  &  &  &  & \fullcircle \\
\citet{wang2019human} & Qualitative & 2019 & CSCW &  &  &  &  &  & \fullcircle \\
\citet{ehsan2023charting} & Qualitative & 2023 & CSCW & \fullcircle &  &  &  &  & \fullcircle \\
\citet{roemmich2021data} & Qualitative & 2021 & CSCW &  & \fullcircle &  &  & \fullcircle & \fullcircle \\
\citet{banovic2023being} & Quantitative & 2023 & CSCW & \fullcircle &  &  &  & \fullcircle &  \\
\citet{watkins2023face} & Qualitative & 2023 & CSCW &  & \fullcircle &  &  &  &  \\
\citet{feuston2021putting} & Qualitative & 2021 & CSCW &  &  &  &  &  & \fullcircle \\
\citet{wang2021cass} & Mixed & 2021 & CSCW &  & \fullcircle &  &  &  &  \\
\citet{chancellor2019who} & Qualitative & 2019 & CSCW &  &  &  &  &  & \fullcircle \\
\citet{wang2022understanding} & Qualitative & 2022 & CSCW &  & \fullcircle &  &  &  &  \\
\citet{wong2023seeing} & Qualitative & 2023 & CSCW &  &  &  &  &  & \fullcircle \\
\citet{long2021co} & Qualitative & 2021 & CSCW & \fullcircle &  &  &  &  &  \\
\citet{mcdonald2020intersectional} & Qualitative & 2020 & CSCW &  &  &  &  &  & \fullcircle \\
\citet{holstein2023supporting} & Mixed & 2023 & CSCW & \fullcircle &  &  &  &  &  \\
\citet{cabrera2021discovering} & Qualitative & 2021 & CSCW &  &  &  &  &  & \fullcircle \\
\citet{moitra2022ai} & Qualitative & 2022 & CSCW & \fullcircle &  & \fullcircle & \fullcircle & \fullcircle &  \\
\citet{jia2022understanding} & Quantitative & 2022 & CSCW & \fullcircle &  &  &  &  &  \\
\citet{shen2021everyday} & Qualitative & 2021 & CSCW &  &  &  &  &  & \fullcircle \\
\citet{zhang2023deliberating} & Qualitative & 2023 & CSCW &  &  &  &  &  & \fullcircle \\
\citet{burrell2019when} & Mixed & 2019 & CSCW &  &  &  &  & \fullcircle & \fullcircle \\
\citet{lyons2021conceptualising} & Qualitative & 2021 & CSCW &  &  &  &  & \fullcircle & \fullcircle \\
\citet{buçinca2021trust} & Quantitative & 2021 & CSCW & \fullcircle &  &  &  &  &  \\
\citet{fox2023patchwork} & Qualitative & 2023 & CSCW &  & \fullcircle &  &  &  & \fullcircle \\
\citet{wong2023privacy} & Qualitative & 2023 & CSCW &  &  & \fullcircle &  &  &  \\
\citet{greiffenhagen2023work} & Quantitative & 2023 & CSCW &  &  & \fullcircle &  &  & \fullcircle \\
\citet{vaccaro2021contestability} & Qualitative & 2021 & CSCW &  &  &  &  & \fullcircle & \fullcircle \\
\citet{miceli2022documenting} & Qualitative & 2022 & CSCW &  &  &  &  &  & \fullcircle \\
\citet{kroll2021outlining} & Qualitative & 2021 & FAccT &  &  &  &  & \fullcircle & \fullcircle \\
\citet{krafft2021action} & Theory & 2021 & FAccT &  &  &  &  &  & \fullcircle \\
\citet{sambasivan2021re} & Qualitative & 2021 & FAccT &  &  &  &  & \fullcircle & \fullcircle \\
\citet{knowles2021sanction} & Theory & 2021 & FAccT &  &  &  &  &  & \fullcircle \\
\citet{jakesch2022how} & Quantitative & 2022 & FAccT & \fullcircle &  & \fullcircle & \fullcircle & \fullcircle &  \\
\citet{pushkarna2022data} & Qualitative & 2022 & FAccT & \fullcircle &  &  &  &  & \fullcircle \\
\citet{liao2022designing} & Theory & 2022 & FAccT & \fullcircle &  &  &  &  &  \\
\citet{bell2022it} & Qualitative & 2022 & FAccT &  & \fullcircle &  &  & \fullcircle & \fullcircle \\
\citet{stapleton2022imagining} & Quantitative & 2022 & FAccT & \fullcircle &  &  &  &  &  \\
\citet{weidinger2022taxonomy} & Review & 2022 & FAccT &  &  &  &  &  & \fullcircle \\
\citet{ramesh2022how} & Qualitative & 2022 & FAccT &  &  &  &  & \fullcircle & \fullcircle \\
\citet{ha2022south} & Qualitative & 2022 & FAccT &  &  &  &  &  & \fullcircle \\
\citet{smith2022real} & Qualitative & 2022 & FAccT &  &  &  &  &  & \fullcircle \\
\citet{donahue2022human} & Mixed & 2022 & FAccT &  &  &  &  &  & \fullcircle \\
\citet{crisan2022interactive} & Theory & 2022 & FAccT &  &  &  &  & \fullcircle & \fullcircle \\
\citet{huang2022social} & Quantitative & 2022 & FAccT &  &  &  &  & \fullcircle & \fullcircle \\
\citet{black2022algorithmic} & Review \& Theory & 2022 & FAccT &  &  &  &  & \fullcircle &  \\
\citet{sloane2022german} & Qualitative & 2022 & FAccT &  &  &  &  &  & \fullcircle \\
\citet{ashurst2022disentangling} & Theory & 2022 & FAccT &  &  &  &  &  & \fullcircle \\
\citet{kaur2022sensible} & Theory & 2022 & FAccT & \fullcircle &  &  &  &  &  \\
\citet{costanza_chock2022who} & Mixed & 2022 & FAccT &  &  &  &  &  & \fullcircle \\
\citet{widder2022limits} & Qualitative & 2022 & FAccT & \fullcircle &  &  &  &  & \fullcircle \\
\citet{kasirzadeh2021use} & Review & 2021 & FAccT & \fullcircle &  &  &  & \fullcircle &  \\
\citet{engelmann2022what} & Quantitative & 2022 & FAccT &  &  &  &  & \fullcircle &  \\
\citet{terzis2020onward} & Theory & 2020 & FAccT &  &  &  &  &  & \fullcircle \\
\citet{toreini2020relationship} & Review \& Theory & 2020 & FAccT & \fullcircle &  &  & \fullcircle & \fullcircle & \fullcircle \\
\citet{boag2022tech} & Review \& Theory & 2022 & FAccT &  &  &  &  &  & \fullcircle \\
\citet{bietti2020ethics} & Theory & 2020 & FAccT &  &  &  &  &  & \fullcircle \\
\citet{washington2020whose} & Qualitative & 2020 & FAccT &  &  &  &  &  & \fullcircle \\
\citet{lima2022conflict} & Theory & 2022 & FAccT & \fullcircle &  &  &  &  & \fullcircle \\
\citet{stark2021ethics} & Theory & 2021 & FAccT &  &  &  &  &  & \fullcircle \\
\citet{young2022confronting} & Theory & 2022 & FAccT &  &  &  &  &  & \fullcircle \\* \bottomrule
\end{longtable}
\end{center}
\section{Appendix: Interactions With ChatGPT 4.0}
\label{app:chatgpt-report}

We first let ChatGPT 4.0 know that our content is larger than its text limit with the following prompt:

``The total length of the content that I want to send you is too large to send in only one piece.      
For sending you that content, I will follow this rule:
[START PART 1/15]
this is the content of the part 1 out of 15 in total
[END PART 1/15]
Then you just answer: `Received part 1/15'
        
And when I tell you `ALL PARTS SENT,' then you can continue processing the data and answering my requests.''

\subsection{Thematic Analysis for Research Papers}
ChatGPT generated the following text with minor stylist changes using all the abstracts of the 164 papers in our review. We first asked for ``apply thematic analysis  to all of the previous 15 prompts based on the method proposed by Braun and Clarke (2008) and generate themes,'' which resulted in step-by-step instructions for how to do the thematic analysis. In a follow-up prompt, we asked to ``produce a report based on the thematic analysis proposed by Braun and Clarke (2008).'' ChatGPT produced the following paragraphs with this prompt:


\textit{This approach led to the identification of four main themes, each underscoring key areas of interest and investigation among researchers in the domain of AI ethics. The themes are as follows: Ethical Implications of AI, Practical Applications of AI, Understanding and Documenting Data, and User Engagement.}

\paragraph{Theme 1: Ethical Implications of AI}
\textit{This theme emerges prominently, signaling that ethical considerations are of primary concern in AI research and practice. The studies under this theme scrutinize ethics codes, considering how they may conflate consumers with society and ignore the political aspects of shared resources. The analyses highlight a gulf between the ideals presented in ethics codes and lived experiences, particularly for marginalized communities.}

\textit{Moreover, this theme reflects the power of collective action among tech workers as a countervailing force against corporate impulses to prioritize growth or profit over ethical considerations. Successful campaigns are discussed, along with frameworks for future action. This suggests a recognition of the potential for internal dynamics within tech companies to act as a transformative lever in the pursuit of ethical AI.}

\paragraph{Theme 2: Practical Applications of AI}
\textit{This theme examines the deployment of AI in real-world contexts, such as healthcare and disaster risk management. AI's practical applications can offer significant benefits, from improving clinician workflows and patient outcomes in the detection of diabetic eye disease to supporting disaster risk management by making data collection or analysis practices faster, less costly, or more accurate.}

\textit{However, these studies also acknowledge the potential for negative consequences. In healthcare, socio-environmental factors can impact model performance and patient experience, emphasizing the need for human-centered evaluative research. In disaster risk management, discussions with technical experts reveal the need for careful consideration of potential negative impacts and the identification of mitigation strategies.}

\paragraph{Theme 3: Understanding and Documenting Data}
\textit{This theme addresses the challenges and importance of understanding and documenting the data that fuel AI models. Recognizing that the origins, development, intent, and evolution of datasets influence the resultant models, researchers propose a shift from merely documenting datasets towards documenting data production.}

\textit{One notable contribution is the proposal of ``Data Cards,'' structured summaries of essential facts about ML datasets, designed to support transparent, purposeful, and human-centered documentation of datasets. The goal is to make data documentation a user-centric product, providing stakeholders with the critical information necessary for responsible AI development.}

\paragraph{Theme 4: User Engagement}
\textit{The final theme highlights the importance of user engagement in AI systems. A growing interest among researchers and practitioners lies in user-engaged approaches to algorithm auditing. The challenges faced by industry practitioners, such as recruiting and incentivizing user auditors, scaffolding user audits, and deriving actionable insights from user-engaged audit reports, are thoroughly discussed.}

\textit{Further, the studies suggest a complex relationship between practitioners and user auditors, which represents organizational obstacles to user-engaged auditing. Addressing these obstacles could lead to more robust and user-responsive AI systems, amplifying the potential for effective AI auditing and user-focused AI implementation.}

\textit{In conclusion, the themes identified through the thematic analysis reveal the multifaceted nature of AI ethics research, which is as concerned with technical aspects as it is with the ethical implications, practical applications, understanding and documenting data, and engaging users in the AI systems lifecycle. The interplay of these themes will undoubtedly continue to shape the discourse in the field.}


\subsection{Research Methods of Research Papers}
Here we asked ChatGPT to about the research methods used in the abstract of the 164 papers using this prompt: ``what are the research methods used the all the previous prompts?'' Below is the answer:

\begin{itemize}
    \item \textit{Interviews: Semi-structured interviews were a common method across several summaries. They were used to gather information about experiences and perspectives from various stakeholders such as data workers, industry practitioners, community organizers, and clinic staff.}
    \item \textit{Participatory Design and Co-Design Workshops: These methods involve users or stakeholders in the design process to ensure that the resulting system or product meets their needs and is usable in the intended context.}
    \item \textit{Prototyping: This approach involves creating early samples, models, or releases of a product to test a concept or process.}
    \item \textit{Observation: Some of the summaries mentioned direct observation as a way to understand current workflows and contexts, such as in the case of observing eye-screening workflows in Thai clinics.}
    \item \textit{Critical Discourse Analysis: This is a qualitative research method used to analyze written or spoken language within its social context. One study used this method to scrutinize data ethics codes.}
    \item \textit{Iterative Co-Design Activities: This approach combines co-design, which involves stakeholders in the design process, with iteration, repeating a process to refine and improve the outcome.}
    \item \textit{Characterization of Collective Actions: This method was used to analyze a collection of documented actions to identify patterns, themes, or trends.}
    \item \textit{Case Studies: Several summaries reported the use of case studies, which involve in-depth investigations of a single individual, group, or event. This approach can provide a detailed context to understand complex issues.}
    \item \textit{Prospective Evaluations of Model Accuracy: These evaluations involve testing a model's performance against a pre-established benchmark or goal.}
    \item \textit{Convening Technical Experts for Discussion: In one summary, experts were brought together to discuss potential negative impacts of AI in disaster risk management. }
\end{itemize}

\subsection{Future Research Recommendations Based on the Research Papers}
Here we asked ChatGPT to about the research recommendations and future research directions based on the abstract of the 164 papers using this prompt: ``what are some the research recommendation and future research directions based on the thematic analysis ?'' Below is the answer:

\begin{itemize}
    \item \textit{User-Centric Design: It's evident that user-focused design needs to be central in the development of AI systems. Future research can focus on understanding the needs of different user groups, developing better tools to engage users in the design process, and exploring ways to make AI systems more transparent and understandable.}
    \item \textit{Greater Transparency in AI: The need for more transparency in AI systems was highlighted in the data. Future research can explore methods and techniques to improve transparency, including the development of comprehensive documentation practices and standards.}
    \item \textit{AI in Marginalized Communities: The impact of AI and data technologies on marginalized communities was a key concern. Future work can focus on understanding these impacts in greater detail, developing technologies that support these communities, and exploring methods to ensure that their needs and concerns are considered in the design of AI systems.}
    \item \textit{Collective Action in Tech Industry: The potential power of collective action within the tech industry to address ethical issues was highlighted. Future research can explore this further, looking at how to support and facilitate such actions and understanding their impacts and effectiveness.}
    \item \textit{Real-World AI Evaluations: The need for more real-world evaluation of AI systems was evident. Future research can focus on developing methodologies for these evaluations, exploring the impacts of real-world contexts on AI performance, and understanding how AI systems can be effectively integrated into existing workflows.}
    \item \textit{AI in Disaster Risk Management: AI has potential to assist in disaster risk management, but more research is needed to understand its possible negative consequences. Future work could focus on investigating these consequences, developing strategies to mitigate potential harms, and building frameworks for responsible use of AI in this context.}
    \item \textit{Effective AI Communication: Developing effective communication strategies for AI, particularly for non-experts, is critical. Future research should investigate how to best communicate about AI, its capabilities and limitations, and the ethical issues it raises.}
    \item \textit{AI Ethics Integration in Education: The integration of AI ethics into educational programs is a key area for future exploration. Research could focus on developing and evaluating curricula, teaching methods, and assessments that effectively engage students in these topics.}
    \item \textit{Work-Integrated Learning for AI: Exploring work-integrated learning in the context of AI is a promising area for future research. This could involve studying the effectiveness of different models of work-integrated learning, and how they can be used to prepare students for ethical challenges in the AI industry.}
    \item \textit{AI Governance Practices: Future work should delve into establishing effective AI governance practices. This includes studying how AI can be regulated, what forms of oversight are most effective, and how organizations can implement ethical AI practices.}
\end{itemize}

\subsection{Thematic Analysis for Patents}

ChatGPT generated the following text with minor stylist changes using all the abstracts of the 28 patents in our review. We first asked for ``produce a report based on the thematic analysis proposed by Braun and Clarke (2008),'' ChatGPT produced the following paragraphs with this prompt:


\begin{itemize}
    \item  \textit{AI-driven decision making and predictions: This theme appeared as a prevalent aspect in many of the patent abstracts. In these cases, AI technologies are being utilized to predict and inform decisions in various areas, such as media content, environmental data, brain state, and security provisioning. For example, one of the inventions predicts moral insight scores from media content using an AI model. This theme demonstrates the value and potential of AI in shaping future decision-making processes in diverse fields.}

    \item  \textit{Transparency, Interpretability, and Explainability in AI: This theme encapsulates the patents that aim to make AI models more understandable and transparent. Given the often 'black-box' nature of AI systems, efforts are being made to develop systems that can provide explanations for the output they produce, thus increasing their reliability, credibility, and usability. For instance, one patent describes a system that couples interpretable and non-interpretable systems to provide explanations for AI-driven decisions. This theme underlines the importance of accountability and interpretability in the development and application of AI systems.}

    \item  \textit{Machine Learning Model Performance \& Enhancement: A subset of patents deal with improving the performance and capabilities of machine learning models, ranging from accuracy and efficiency to robustness and fairness. For instance, one patent discusses using a stochastic process and a novel hypothesis test to compare the performance of black-box models, providing a method to derive global and local explanations. This theme underscores the ongoing innovation in the development and refinement of machine learning models, which is critical to maximizing their utility and application in different fields.}
\end{itemize}

\end{document}